\documentclass{aa}  

\usepackage{graphicx}
\usepackage{txfonts}
\usepackage[]{hyperref} % To add links in your PDF file
\usepackage{booktabs} % for much better looking tables
\usepackage{array} % for better arrays (eg matrices) in maths
\usepackage{paralist} % very flexible & customisable lists (eg. enumerate/itemize, etc.)
\usepackage{verbatim} % adds environment for commenting out blocks of text & for better verbatim
\usepackage{ulem}
\usepackage{xspace}
\usepackage{siunitx}
\usepackage{amssymb}
\usepackage{float}
\usepackage{comment}
\usepackage{tabularx}
\usepackage{array}
\usepackage[dvipsnames]{xcolor}
\usepackage{textgreek}
\usepackage{rotating}
\usepackage{multirow}
\usepackage{hyperref}

\usepackage[version=4]{mhchem}

\usepackage[
    starfontserif % comment for sans glyphs
    ]{starfont}

\usepackage{natbib}
\bibpunct{(}{)}{;}{a}{}{,} % to follow the A&A style
\graphicspath{{./img/}}

\hypersetup{
    colorlinks=true,
    linkcolor=blue,
    filecolor=blue,
    urlcolor=blue,
    citecolor=blue
}

\makeatletter
\renewcommand*\aa@pageof{, page \thepage{} of \pageref*{LastPage}}
\makeatother

\addto\extrasenglish{%
}

\DeclareSymbolFont{starfontsym}{OT1}{sts}{m}{n}
\DeclareMathSymbol{\mathSun}{\mathord}{starfontsym}{115}
\DeclareMathSymbol{\mathMercury}{\mathord}{starfontsym}{102}
\DeclareMathSymbol{\mathVenus}{\mathord}{starfontsym}{103}
\DeclareMathSymbol{\mathTerra}{\mathord}{starfontsym}{76}
\DeclareMathSymbol{\mathMoon}{\mathord}{starfontsym}{100}
\DeclareMathSymbol{\mathMars}{\mathord}{starfontsym}{104}
\DeclareMathSymbol{\mathJupiter}{\mathord}{starfontsym}{106}
\DeclareMathSymbol{\mathSaturn}{\mathord}{starfontsym}{83}
\DeclareMathSymbol{\mathUranus}{\mathord}{starfontsym}{70}
\DeclareMathSymbol{\mathNeptune}{\mathord}{starfontsym}{71}
\DeclareMathSymbol{\mathPluto}{\mathord}{starfontsym}{74}

\newcommand{\wasp}{\mbox{WASP-107 b}\xspace}

\newcommand{\criresp}{CRIRES$^+$\xspace}
\newcommand{\citepeg}{\citep[e.g.][]}
\newcommand{\about}{\ensuremath{{\sim}}}
\newcommand{\um}{\textmu m\xspace}
\newcommand{\kpvsys}{$K_{\mathrm{p}}-v$\xspace} %_{\mathrm{sys}}
\newcommand{\kp}{$K_{\mathrm{p}}$\xspace}
\newcommand{\vsys}{$v$\xspace}
\newcommand{\pRT}{\texttt{petitRADTRANS}\xspace}
\newcommand{\sig}{$\sigma$\xspace}

\newcommand{\Msol}{$M_{\odot}$\xspace}
\newcommand{\Rsol}{$R_{\odot}$\xspace}
\newcommand{\Mearth}{$M_{\oplus}$\xspace}

\newcommand{\Mjup}{$M_\mathrm{J}$\xspace}
\newcommand{\Rjup}{$R_\mathrm{J}$\xspace}

\newcommand{\hto}{H$_2$O\xspace}
\newcommand{\sot}{SO$_2$\xspace}
\newcommand{\chf}{CH$_4$\xspace}
\newcommand{\hts}{H$_2$S\xspace}
\newcommand{\nht}{NH$_3$\xspace}

\newcommand{\changeo}[1]{\textcolor{black}{#1}}

\newcommand{\change}[1]{\textcolor{black}{#1}}%{\textbf{#1}}

\begin{document} 

   \title{VLT/CRIRES+ observations of warm Neptune WASP-107 b:}

   \subtitle{Challenges in detecting molecules with ground-based transmission spectroscopy of cooler and cloudy exoplanets}

   \author{ % actively involved in discussion (order of contribution)
            \mbox{L. Boldt-Christmas} \inst{1} 
            \and \mbox{A. D. Rains} \inst{1,2} 
            \and \mbox{N. Piskunov} \inst{1} 
            \and \mbox{L. Nortmann} \inst{3} 
            \and \mbox{F. Lesjak} \inst{3,4} 
            \and \mbox{D. Cont} \inst{5,6}
            % gto programme repsonsible
            \and \mbox{O. Kochukhov} \inst{1} 
            % data reduction (alphabetical)
            \and \mbox{A. Hahlin} \inst{1} 
            \and \mbox{A. Lavail} \inst{7} 
            \and \mbox{T. Marquart} \inst{1}
            % others (alphabetical)
            \and \mbox{U. Heiter} \inst{1}
            \and \mbox{M. Rengel} \inst{8} 
            \and \mbox{D. Shulyak} \inst{9}
            \and \mbox{F. Yan} \inst{10} 
            % CST (alphabetical)
            \and \mbox{A. Hatzes} \inst{11}
            \and \mbox{E. Nagel} \inst{3}
            \and \mbox{A. Reiners} \inst{3} 
            \and \mbox{U. Seemann} \inst{12}
            }

   \institute{
            %1 
            Observational Astrophysics, Department of Physics and Astronomy, Uppsala University, Sweden
            \and
            %2
            Instituto de Astrofísica, Pontificia Universidad Católica de Chile, Av. Vicuña Mackenna 4860, 782-0436 Macul, Santiago, Chile
            \and
            %3
            Institut f\"ur Astrophysik und Geophysik, Georg-August-Universit\"at, Friedrich-Hund-Platz 1, 37077 G\"ottingen, Germany 
            \and 
            %4
            Leibniz Institute for Astrophysics Potsdam (AIP), An der Sternwarte 16, 14482 Potsdam, Germany
            \and
            %5
            Universitäts-Sternwarte, Ludwig-Maximilians-Universität München, Scheinerstrasse 1, 81679 München, Germany 
            \and 
            %6
            Exzellenzcluster Origins, Boltzmannstrasse 2, 85748 Garching bei München, Germany 
            \and 
            %7
            Namzitu Astro, 31130 Quint-Fonsegrives, France
            \and
            %8
            Max-Planck-Institut für Sonnensystemforschung, Justus-von-Liebig-Weg 3, 37077 Göttingen, Germany
            \and
            %9
            Instituto de Astrof\'{\i}sica de Andaluc\'{\i}a - CSIC, c/ Glorieta de la Astronom\'{\i}a s/n, 18008 Granada, Spain
            \and
            %10
            Department of Astronomy, University of Science and Technology of China, Hefei 230026, PR China
            \and
            %11
            Thüringer Landessternwarte Tautenburg, Sternwarte 5, 07778 Tautenburg, Germany
            \and
            %12
            European Southern Observatory, Karl-Schwarzschild-Str. 2, 85748 Garching bei München, Germany}

   \date{Received 31 July 2025 / Accepted 24 April 2026}

% \abstract{}{}{}{}{} 
% 5 {} token are mandatory
 
  \abstract
   {The atmospheres of transiting exoplanets can be studied spectroscopically using space-based or ground-based observations. Each has its own set of strengths and weaknesses, so there are benefits to both approaches. This is especially true for more challenging targets such as cooler, smaller exoplanets whose atmospheres most likely contain many molecular species and cloud decks.} 
   {We aim to study the atmosphere of the warm Neptune-like exoplanet WASP-107 b ($T_\mathrm{eq} \approx$ 740 K). Several molecular species have been detected in this exoplanet in recent studies using the space-based JWST, and we aim to confirm and expand upon these detections using the ground-based VLT, evaluating how well our findings agree with previously retrieved atmospheric parameters.}
   {We observe two transits of \wasp with VLT/\criresp and create cross-correlation templates of the target atmosphere based on retrieval results from JWST studies. We create different templates to investigate the impact of varying volume mixing ratios of species and the inclusion or exclusion of clouds. Considering this target's observational challenges, we create simulated observations prior to evaluating our real data in order to assess our expected detection significances with the cross-correlation technique.}% to create \kpvsys diagrams.}
   {We \change{report cross-correlation signals} of two molecular species, CO (\about6\sig) and \ce{H2O} (\about4.5\sig), \change{in our \kpvsys maps}. This \change{supports} previous space-based detections and demonstrates, for the first time, the capability of VLT/\criresp to detect species in targets cooler than hot Jupiters using transmission spectroscopy. We show that our analysis is sensitive to the inclusion of clouds, but less so to different volume mixing ratios. Interestingly, our \change{signal maximum} deviates \change{notably but consistently from its expected location in the \kp direction of our maps}, and we speculate on the possible reasons for this effect. We demonstrate that the error budget for these relatively cooler exoplanets is severely reduced in comparison to hotter exoplanets, and underline the need for further work in the context of high-resolution spectroscopy.}
   {} 


   \maketitle

%----------------------------------------------------
\section{Introduction}
    \label{section:Introduction}
 As an exoplanet passes in front of its host star, its chemical composition may be studied by observing the fraction of starlight that is transmitted through the planet's upper atmosphere as a function of wavelength. This simple premise is what enables the field of transmission spectroscopy, and it has facilitated the growth of exoplanetary characterisation from a hypothetical possibility into a thriving and interdisciplinary research topic. Today, out of the near 6\,000 exoplanets\footnote{NASA Exoplanet Archive – \url{https://exoplanetarchive.ipac.caltech.edu/}} that have now been confirmed, over 250 planets have been chemically characterised by the methods of transmission, emission, and/or reflectance spectroscopy.\footnote{IAC ExoAtmospheres database – \url{https://research.iac.es/proyecto/exoatmospheres/}}\label{fn:IAC}

Spectroscopic studies of exoplanetary atmospheres can be broadly divided into two different branches: space-based observations of high photometric precision but lower spectral resolution, and ground-based observations of higher spectral resolution that suffer from Earth's atmospheric effects. Each of these two avenues come with their own benefits and drawbacks, and as such, it is difficult to argue that either approach is superior to the other. Ground-based spectrographs can generally be of higher resolving power $R$, enabling them to measure strong contrasts between the cores and the wings of individual spectral lines. Meanwhile, instruments of lower $R$ can only detect the combined effects of groups of lines that result in increase or decrease (emission or absorption) of local flux. Observations of high resolution therefore provide spectra of greater detail, allowing robust identification of chemical species. In recent years, ground-based transmission spectroscopy has been crucial for the study of detailed atmospheric effects such as winds, jet streams, and vertical stratification \citepeg{ehrenreich_nightside_2020,seidel_wind_2020,gandhi_spatially_2022,prinoth_titanium_2022,lesjak_retrieval_2023,cont_exploring_2024,nortmann_crires_2025,seidel_vertical_2025}. 
 
 However, instruments of higher spectral resolution currently remain prohibitively complex to use in space. For space-based telescopes such as JWST, the on-board spectrographs have a much lower spectral resolution ($R \lesssim$ 3\,000), meaning such instruments are not able to resolve individual spectral lines. Nonetheless, space-based telescopes and their spectrographs maintain a number of notable benefits including broader wavelength coverage as well as a lack of turbulent distortions, telluric absorption and emission, and daytime interruptions. The field of space-based atmospheric characterisation is more active than ever, and a small selection of recent successes include e.g. \citet{jwst_transiting_exoplanet_community_early_release_science_team_identification_2023}, \citet{tsai_photochemically_2023}, \citet{smith_combined_2024}, \citet{mukherjee_jwst_2025}, \citet{teske_jwst_2025}, and many more.

Thanks to these fundamental differences, space-based and ground-based transmission spectroscopy bring forth distinct but complementary contributions to the general research goal of exoplanet atmospheric characterisation. In the pursuit of recovering an atmospheric transmission spectrum of a planet and deriving as many details as possible about its physical conditions, both types of observations fulfil an important role that cannot be achieved by the other. This becomes increasingly important as the exoplanet community continues to move further towards the study of planets that are cooler and more chemically complex than the evergreen case studies of hot and ultra-hot Jupiters. With benefits from both types of data, in an ideal situation, one could capitalise on this by observing the same target with both a low-resolution space-based instrument as well as a high-resolution ground-based instrument in the same wavelength regime, ensuring that one may support the detections of the other. 

It is this ideal situation, and the comparison of conclusions derived from the two observing techniques, that is being explored in this paper. We analyse transit observations from two different nights of exoplanet \wasp \citep{anderson_discoveries_2017}, observed using the ground-based spectrograph \criresp on the Very Large Telescope (VLT) of Paranal Observatory, and compare these high-resolution observations of $R \approx$ 140\,000 covering K-band wavelengths $\sim$2.0–2.5 \um with the analysis results of space-based observations of the same target from a growing gallery of JWST observations. Previous studies have already resulted in a number of molecular detections in \wasp, and this work has been able to confirm some of these results at a wavelength range that no previous ground-based studies of this target have explored. 

We use the high-resolution cross-correlation spectroscopy technique (HRCCS) to report \change{the signals} of two molecular species, CO at \about\,6\sig and \ce{H2O} at \about\,4.5\sig, and that cross-correlation using a multi-species template of the atmosphere as retrieved by a previous JWST study \citep{welbanks_high_2024} produces a \about\,6\sig cross-correlation peak. Considering only ground-based transmission spectroscopy studies, these detections are the first ever to be made for a target of $T_\mathrm{eq} <$\,800 K, highlighting both the capability of instruments like VLT/\criresp and also the challenges that will face observers as they proceed towards studying targets of even lower temperatures.

As the exoplanet community continues to reap the fruits of JWST data, the number of exoplanetary atmosphere studies using space-based observations will continue to grow; in parallel, the Extremely Large Telescope (ELT) nears completion, ushering in a new age for ground-based observations very soon. In this context, there is a need for assessing the extent to which the two types of observations can be combined, as whatever parameters we obtain from either type of data and incorporate into our HRCCS template will affect our final result. To explore this, we used simulated observations together with real VLT/\criresp data to further explore (i) how sensitive our HRCCS detection significance is to the changes in parameters from space-based retrievals by cross-correlating templates derived from two different JWST retrievals; (ii) how the uncertainties in other parameters of our system can affect our results, exploring possible effects of cooler, smaller, and cloudier planets such as \wasp. 

\changeo{Considering these properties mean that a system such as that of \wasp becomes more challenging to analyse using HRCCS, there are many instances where the properties of that analysis are not fully understood, or where independence between conflating effects cannot be fully justified. As a result, this paper is above all an attempt to study the consequences of complexity and cross-talk between various components of the problem for the case of a relatively cooler target than a nominal hot Jupiter, based on good quality data and multiple transits. For cooler planets, the interplay between the dominant features in the transmission spectrum, underlying stellar (esp. cool dwarf) spectrum, and the telluric contamination becomes a very difficult task for more simplistic algorithms; and withal, more subtle effects such as presence of clouds in the planetary atmosphere or minor errors in the wavelength solution may result in major reduction of detection significance due to appearance of ``systematic features'' in the \kpvsys maps. Thus the overarching question of this paper is: to what extent may we fully account for artefacts in our cross-correlation maps, and bring the characterisation of cooler exoplanets such as WASP-107 b to the same level of significance that was achieved for hotter Jupiters with our traditional methods? This paper uses traditional HRCSS methodology -- but as we demonstrate, non-traditional targets clearly require non-traditional methodology, which is yet to be fully developed.} As such, the hope of the authors is that this paper will serve more as an exploration of \changeo{such} methodology rather than a reporting of chemical detections and/or non-detections only \changeo{using the tools available to us today}.

In Sect. \ref{section:PreviousStudies}, we provide context for our study by exploring previous atmospheric studies of this target that have been conducted since its discovery. In Sect. \ref{section:ObsDR}, we provide the details of how our observational data were obtained and prepared for interpretation. In Sect. \ref{section:Method}, we describe our methods for removing the stellar and telluric signal and our cross-correlation analysis. In Sect. \ref{section:SimDataAnalysis}, we use simulated observations to establish some context, explore the anticipated impact of certain parameters, and set expectations for what can be considered a reasonable detection significance for this target and instrument. In Sect. \ref{section:RealDataAnalysis}, we present the analysis of our VLT/\criresp observations and investigate possible interpretations of the \kpvsys plots. Our results are presented and discussed in Sect. \ref{section:ResultsDiscussion} before the final conclusions in Sect. \ref{section:Conclusions}.

\section{Previous studies}
    \label{section:PreviousStudies}

    \subsection{Discovery and earlier observational studies}
        \label{subsection:PreviousStudies:preJWST}

\wasp is a warm ($T_\mathrm{eq}$\,$<$\,800\,K) Neptune-like exoplanet with a radius of 0.924 \Rjup and mass of 0.096 \Mjup \citep{mocnik_starspots_2017,piaulet_wasp-107bs_2021}. Its host star, WASP-107, is a cool dwarf of solar metallicity located 65 pc away in the constellation Virgo. It is of spectral type K7V with an effective temperature of 4\,358 K, a radius of 0.656 \Rsol, and a mass of 0.696 \Msol \citep{anderson_discoveries_2017,dressing_characterizing_2019} with near-solar abundances and a carbon-to-oxygen ratio (C/O) of 0.50, compared to solar C/O = 0.54 \citep{hejazi_elemental_2023}. For a full overview of the stellar and planetary parameters of this system, see Table \ref{tab:WASP107_s+p_params}.

\begin{table}\small
    \centering
        \caption{Host star and planetary parameters for \wasp.}
    \begin{tabular}{ lrr } 
     \toprule
    \textbf{Star (WASP-107)} & & \\
        \midrule
       Parameter & Symbol/Units & Value \\
     \midrule
      Mass$^a$ & $M_* [M_\odot]$ & 0.683 $^{+0.017} _{-0.016}$ \\
      Radius$^a$ & $R_* [R_\odot]$ & 	0.67 $\pm$ 0.02\\
      Age$^a$ & [Gyr] & 3.4 $\pm$ 0.7 \\
      Distance$^b$ & [pc] & 64.401 $\pm$ 0.1078 \\
      Spectral type$^c$ & & K7V \\
      Luminosity$^a$ & $L_* [L_\odot]$ & 	0.132 $\pm$ 0.003 \\ 
      Magnitude (K-band)$^d$ & $m_{K_s}$ & 8.637 $\pm$ 0.023 \\
      Effective temperature$^a$ & $T_\mathrm{eff}$ [K] & 4425 $\pm$ 70 \\
      Metallicity$^a$ & [Fe/H] & 0.02 $\pm$ 0.09 \\
      Surface gravity (log $g$)$^a$ & log$_{10}$ [cm/s$^2$] & 	4.633 $\pm$ 0.012 \\
      Carbon-to-oxygen ratio$^e$ & C/O & 0.50 $\pm$ 0.10 \\
      $v$ sin $i^f$ & [km/s] & 0.507$^{+0.072}_{-0.086}$ \\
      %Systemic radial velocity$^?$ & [km/s] & ? $\pm$ ? \\ 
    \midrule
    \midrule
    \textbf{Planet (WASP-107 b)} & & \\
    \midrule
    Parameter & Symbol/Units & Value \\
     \midrule
     Mass$^a$ & $M_p [M_\mathrm{J}]$  & 0.0960 $\pm$ 0.0050 \\
     Radius$^g$ & $R_p [R_\mathrm{J}]$ & 0.94 $\pm$ 0.02 \\
     Density$^a$ & $\rho$ [g/cm$^3$] & 0.134 $^{+0.015}_{-0.013}$ \\
     Eq. temperature$^h$ & $T_\mathrm{eq}$ [K] & 736 $\pm$ 17\\
     Int. temperature$^i$ & $T_\mathrm{int}$ [K] & 460 $\pm$ 40 \\
     Orbital period$^a$ & $P$ [days] & 5.7214742 \\
     Orbital eccentricity$^a$ & $e$ & 0.06 $\pm$ 0.04 \\
     Orbital inclination$^g$ & $i$ (deg) & 	89.56 $\pm$ 0.08  \\
     Sky-projected inclination$^j$ &  $\lvert \lambda \rvert $ (deg) & 118 $^{+38}_{-19}$ \\
     %Radial velocity amplitude$^a$ & $K$ [m/s] & 	14.1 $\pm$ 0.8 \\
     Planet RV semi-amplitude$^k$ & $K_p$ [km/s] & 105.2 $\pm$ 2.5 \\
     Argument of periastron$^l$ & $\omega$ [deg] & -2.3 $\pm$ 6.1 \\
     Semi-major axis$^m$ & $a$ [au] & 0.055 $\pm$ 0.001 \\
     Transit depth$^m$ & $\delta$  $[\%]$  & 	2.17 $\pm$ 0.02  \\
    Transit duration$^f$ & $t_{14}$ [hours] & 2.7528 $\pm$ 0.0072\\

     \bottomrule
      & & \\
      & & \\

    \end{tabular}
    \textbf{References:} 
    
    $^a$\citet{piaulet_wasp-107bs_2021},
    $^b$Gaia EDR3 \citep{gaia_collaboration_gaia_2021}, %https://simbad.cds.unistra.fr/simbad/sim-ref?bibcode=2020yCat.1350....0G
    $^c$\citet{dressing_characterizing_2019},
    $^d$2MASS All-Sky Catalog \citep{cutri_vizier_2003}, $^e$\citet{hejazi_elemental_2023}, $^f$\citet{bourrier_dream_2023}, $^g$\citet{kokori_exoclock_2023}, $^h$\citet{mocnik_starspots_2017},
    $^i$\citet{sing_warm_2024}, $^j$\citet{rubenzahl_tess-keck_2021}, $^k$\citet{guilluy_gaps_2024},
    $^l$\citet{murphy_evidence_2024},
    $^m$\citet{anderson_discoveries_2017}
    \label{tab:WASP107_s+p_params}
    \vspace{-0.2cm}
\end{table}

The matter of this planet's atmospheric conditions was identified as a point of possible scientific interest already at the time of its discovery by \citet{anderson_discoveries_2017}. In that work, its mass was established to be 0.12 \Mjup, which placed the planet in a transition region between gas giants where the planetary mass is less than half that of Saturn (0.5 $M_{\mathSaturn} = 0.15$ \Mjup) but above twice that of Neptune (2 $M_{\mathNeptune} = 0.11$ \Mjup). This is a regime where our Solar System has no analogues, especially not at short orbital distances, and so speculations about the possible climate of \wasp were difficult to justify at that time. It was noted that the planet could be characterised more accurately if one was to determine e.g. its atmospheric metallicity, considering this is a parameter that varies between the ice giants (higher metallicity) and gas giants (lower) in our Solar System. 

Due to the potential scientific value of studying its atmosphere and the observational suitability of the target, \wasp was suggested in the discovery paper as an excellent candidate for future transmission spectroscopy studies of its atmospheric composition. The observational advantages of \wasp are twofold: (i) its large atmospheric scale height, and (ii) its small, bright host star. \wasp has a notably large radius for its mass, and was found in \citet{piaulet_wasp-107bs_2021} to have an even lower density (0.134 g/cm$^3$) than previously measured, placing its mass at 30.5 \Mearth or 0.096 \Mjup (which makes it even more exotic to our Solar System, at \about 10\% of Jupiter's mass and \about 90\% of its radius). As such, \wasp is a very ``puffy'' planet with a significantly inflated atmosphere – i.e. larger scale height – which is beneficial for transit studies. Furthermore, the host star is particularly bright in the infrared (K-band magnitude of 8.637 mag; see Table \ref{tab:WASP107_s+p_params}), making it highly suitable for obtaining high signal-to-noise (S/N) observations in the infrared and near-infrared regime, and thus for atmospheric spectroscopy studies as molecular lines appear at these wavelengths.

The first two atmospheric studies of \wasp were published in May 2018, both using space-based observational data of transits from the Hubble Space Telescope (HST). The work by \citet{kreidberg_water_2018} used observations with HST/WFC3 (with the G141 grism, which covers the wavelength range of 1.1–1.7 \um) of a single \wasp transit to detect water in its upper atmosphere through atmospheric retrievals. This work constrained atmospheric metallicity to an upper limit of $\times30$ solar metallicity, and noted a depletion of methane. The work by \citet{spake_helium_2018} also used HST/WFC3 observations (this time with the G102 grism of wavelength range 0.8–1.1 \um) to detect a post-transit tail at 10,833 Å (He I triplet), suggesting that this planet's atmosphere was highly extended. In this work, the authors also noted the possibility of studying this tail via observations of this triplet using high-resolution spectrographs in the infrared, as had been done at the time for e.g. WASP-69 b by \citet{nortmann_ground-based_2018} and HAT-P-11 b by \citep{allart_spectrally_2018} using the CARMENES instrument ($R\approx$ 80,000) on the 3.5 m telescope at the Calar Alto Observatory. 

The first ground-based, high resolution study of \wasp was published by \citet{allart_high-resolution_2019}, which confirmed the target's extended helium atmosphere using Calar Alto Observatory/CARMENES to confirm the space-based HST/WCF3 findings whose helium feature was poorly resolved. The data obtained during these observations were also used in a study by \citet{kesseli_search_2020} who used archival CARMENES data to search for the molecule FeH in a number of exoplanetary atmospheres, including that of \wasp, but no FeH was found in this target. Two subsequent follow-up surveys were made using more ground-based observations, both studies using Keck II/NIRSPEC. \citet{kirk_confirmation_2020} confirmed that the signature of the escaping helium saw no significant temporal variation in the two years since the previous result, while \citet{spake_posttransit_2021} managed to obtain significant post-transit coverage that confirmed the tail's length to be the equivalent of \about 7 planet radii, corresponding to approximately twice the planet's Roche lobe radius. More recently, studies have confirmed and further characterised \wasp's escaping helium tail with both ground-based observations – namely \citet{guilluy_gaps_2024}, this time using the GIARPS (GIANO-B + HARPS-N) observing mode of the Telescopio Nazionale Galileo – and also with a very recent space-based observational study by JWST that included pre-transit coverage \citep{krishnamurthy_continuous_2025}.

In 2021, a radial velocity study by \citet{piaulet_wasp-107bs_2021} using an extensive data set from Keck I/HIRES and archival data (overall spanning observations from 2011–2020) confirmed that the density of \wasp was significantly lower than previously measured and also the detection of another more massive planet in the WASP-107 system (\mbox{WASP-107 c}) at a much longer, eccentric orbit. Also using Keck I/HIRES, following up TESS data as part of the TESS-Keck survey collaboration, a study by \citet{rubenzahl_tess-keck_2021} measured the misalignment of \wasp's orbit, i.e. its obliquity, based on observations of the target's Rossiter-McLaughlin effect. A polar/retrograde orbit had already been suspected by \citet{dai_oblique_2017} who constrained the likely obliquity to be in the range of 40–140$^{\circ}\xspace$ based on the finding of a lower number of starspot-crossings by \wasp in $K2$ data than expected. The TESS-Keck study successfully confirmed this anticipated obliquity, with a final calculation of sky-projected inclination being $\lvert \lambda \rvert = 118^{\circ}$. This result was further supported by \citet{bourrier_dream_2023} who confirmed its polar, retrograde orbit by analysis of its Rossiter-McLaughlin effect, and was complemented by \citet{dholakia_general_2025} who placed an upper bound on the planet's oblateness of $f<0.23$. 

\subsection{Transmission studies using JWST}
    \label{subsection:PreviousStudies:JWST}

Considering the diversity of these previous studies, the benefit of observing a target using both space-based and ground-based telescopes is clear for many science cases. By repeatedly studying the same target across overlapping wavelength ranges with both types of observations, one hopes to reach more robust conclusions about physical properties of the atmosphere in question. 

While all the transmission studies listed so far of \wasp have remained within the wavelength range between $\sim$0.8-1.7 \um, more studies featuring space-based results further into the infrared have started to be published in the last few years. In particular, JWST has been able to study \wasp using transmission spectroscopy across a redder wavelength range that had not previously been probed for this target. \citet{dyrek_so2_2024} observed \wasp with JWST on 19–20 January 2023 under the JWST MIRI GTO programme. %(programme identifier (PID) 1280; PI P. O. Lagage). 
The transit was observed using the MIRI spectrometer, which provides a spectral resolution of 30 to 100 across 4.61 to 11.83 \um (mid-infrared). Together with archival HST data from \citet{kreidberg_water_2018} of wavelength region 1.121–1.629 \um, the authors of this work performed atmospheric retrievals whose best fits to these data sets resulted in detections of \sot ($\sim$9\sig), \hto ($\sim$12\sig), and silicate clouds ($\sim$7\sig) alongside a non-detection of \chf and tentative detections of \hts ($\sim$4\sig), \nht ($\sim$2–3\sig), and CO ($\sim$2–3\sig). From these retrievals, the team also obtained volume mixing ratios for which \sot came out several orders of magnitude higher than equilibrium chemistry predicts. The interpretation advanced in their paper is that the atmosphere of the planet must therefore be in chemical disequilibrium, caused by photochemical reactions through (i) photodissociation of \hto in the uppermost atmospheric layers, generating atomic H and OH radicals that in turn create \sot by oxidising \hts, and (ii) further introduction of more OH radicals through photodissociation of other molecules beyond \hto in the lower atmospheric layers. 

Another recent publication of this target also used a combination of HST and JWST observation to further confirm and expand upon these detections. \citet{welbanks_high_2024} observed \wasp over two transits on 14 January 2023 and 4 July 2023 as part of the MANATEE NIRCam+MIRI GTO program using NIRCam F322W2 covering 2.4–4.0 \um and F444W covering 3.9–5.0 \um. Combining these data sets with the MIRI results of \citet{dyrek_so2_2024} plus the archival HST WFC3 data from \citet{kreidberg_water_2018} and \citet{spake_helium_2018} resulted in an impressive sequence of spectra covering a total wavelength region of 0.8–1.7 \um and 2.4–12.2 \um. In their atmospheric retrieval, the best fit confirmed the \citet{dyrek_so2_2024} detections (\ce{CO2} at 27\sig, \ce{H2O}  at 18\sig, and \ce{SO2} at 8\sig), as well as two tentative detections of \ce{CO} and \ce{NH3} (both at 5\sig). This work also provided the first detection of \ce{CH4} (at 8\sig), found at wavelengths that had not been previously studied for this target, i.e. $\sim$3.2–3.8 \um. Furthermore, this work managed to constrain internal temperatures to $>$345 K, suggesting that the notable inflation of this exoplanet may be explained by a Neptune-like internal structure of tidally-induced heating – speculated to possibly be due to the planet's non-circular orbit of $e=0.06\pm0.04$ \citep{piaulet_wasp-107bs_2021}. 

Other studies have recently been published using transmission spectra from JWST of this target to study two very relevant atmospheric factors: core interior and limb asymmetry. The first study by \citet{sing_warm_2024} analysed a JWST-NIRSpec transmission spectrum using the G395H grating, which provides a wavelength range of 2.7–5.18 \um, and again detected \ce{SO2}, \ce{CH4}, \ce{H2O}, \ce{CO2}, and CO through retrievals. Once abundances for these species were established – with special attention to the confirmation of previously detected methane depletion – the authors were able to run a grid of forward models in order to investigate vertical mixing, metallicity, and temperature structure of \wasp. They establish a relatively hot intrinsic temperature for \wasp of $T_\mathrm{int} = 460 \pm 40$ K, which is presumably responsible for the planet's inflated atmosphere; and they infer a core mass of 11.5 \Mearth (i.e. an approximate third of the planet's total mass) which is significantly higher than previously established upper limits.

The other study by \citet{murphy_evidence_2024} uses spectra taken at 2.5–4.0 \um with the JWST/NIRCam F210M filter and F322W2 grism in order to study the morning and evening limbs of \wasp with the ambition to search for potential asymmetries between the morning and evening terminators. For tidally-locked exoplanets of equilibrium temperatures above $\sim$1\,200 K, atmospheric models predict that limbs should become heterogeneous due to day-to-night circulation \citepeg{kataria_atmospheric_2016,powell_transit_2019} while remaining more homogeneous for exoplanets of lower temperatures such as \wasp ($T_{\mathrm{eq}} =$ 736 K from Table \ref{tab:WASP107_s+p_params}). However, this work indicates that the planet's morning limb is cooler by approximately 100~K, resulting in a scale height difference between the limbs for the wavelengths studied. These findings were confirmed and expanded upon by a follow-up study led by the same team, \citet{murphy_panchromatic_2025}, now using JWST data from all of JWST’s science intruments (NIRISS, NIRCam, NIRSpec, and MIRI) and covering for the first time the entire range of $\sim1-12$ \um. In this study, they find further evidence for strong variation between its morning and evening limb, specifically in abundances of \ce{SO2} and \ce{CO2}, and that clouds appear to only form on the morning limb, leaving the evening limb clear.

These studies, together with many other modelling studies not discussed here \citepeg{schlawin_clear_2018, millholland_tidal_2020, wang_metastable_2021, khodachenko_simulation_2021, linssen_constraining_2022}, demonstrate that the formation history, atmospheric structure, and dynamics of \wasp are not yet fully understood. Establishing specific pathways for the formation of exoplanets – especially those without analogues in our Solar System, such as \wasp – is a very active field of research that is currently making good progress in providing evidence that complicated formation histories may manifest in both an exoplanet's orbit \citep[see e.g.][and references therein]{maire_workshop_2023} and interiors \citep[see e.g. reviews by][]{nettelmann_exoplanetary_2021, guillot_giant_2022,foley_exoplanet_2024}. The studies listed in this section therefore shine light on the uniqueness of this target, and it has become increasingly clear that \wasp appears to be a highly unusual planet both atmospherically, compositionally, and dynamically, which is an important piece of context for the reader moving forward. Thus far, the vast majority of these results have not been confirmed at high spectral resolution through any ground-based observations, which served as a key motivator for this study.

%----------------------------------------------------
\section{Observations and data reduction}\label{section:ObsDR}

\begin{table}\small
    \centering
        \caption{VLT/CRIRES+ observations used in this work. Relative humidity and seeing are given by their minimum and maximum values of the night. Airmass is given by three values corresponding to the start of the observation; the minimum value (occurring close to mid-transit); and the end of the observation.}
    \vspace{-0.4cm}
    \begin{tabularx}\columnwidth{Xrr} 
    
     \\
     
          \toprule
     \textbf{Night 1} & & \\
        \midrule
       Parameter & Symbol/Units & Value \\
     \midrule
    Date & YYYY-MM-DD & 2022-03-11 \\
    Obs. start & UTC & 03:54:23 \\
    Transit start & '' & 05:32 \\
    Transit mid-point & '' & 06:55 \\
    Transit end & '' & 08:17 \\
    Obs. end & '' & 09:30:11 \\
    No. exposures (in/out) &  & 64 (34/30) \\
    Exposure length & [s] & 300 \\
    Avg. S/N per pixel &  & 132 \\
    Airmass &  & 1.19-1.03-1.59 \\
    Relative humidity & \% & 15–27 \\
    Seeing & $\arcsec$ & 0.32–1.04 \\
    Avg. resolution & $R$ & \changeo{143\,000} \\
    
    \midrule
    \textbf{Night 2} & & \\
    \midrule
       Parameter & Symbol/Units & Value \\
     \midrule
    Date & YYYY-MM-DD & 2023-02-23 \\
    Obs. start & UTC & 04:55:26 \\
    Transit start & '' & 05:48 \\
    Transit mid-point & '' & 07:11  \\
    Transit end & '' & 08:33 \\
    Obs. end & '' &  09:16:53\\
    No. exposures (in/out) &  & 50 (34/16) \\
    Exposure length & [s] & 300 \\
    Avg. S/N per pixel &  & 130 \\
    Airmass & & 1.20-1.03-1.21 \\
    Relative humidity & \% & 27–38 \\
    Seeing & $\arcsec$ & 0.58–1.26 \\
    Avg. resolution & $R$ & 125\,300  \\

        \midrule
    \textbf{General settings} & & \\
    \midrule
    Parameter & Symbol/Units & Value \\
        \midrule
    Wavelength setting & & \texttt{K2148} \\
    Wavelength coverage & \um & 1.972–2.452 \\ %#2.007–2.491 \\
    No. echelle orders &  & 6 \\
    Exposure time & [s] & 300 \\
    Readout time & [s] & 14 \\
    Nodding pattern &  & ABBA \\
    
     \bottomrule
      & & \\
    %\textbf{References:} $^X$ (for wl coverage) \citet{dorn_crires_2023} & & \\
     
    \end{tabularx}
    \vspace{-0.2cm}
    \label{tab:Observations}
\end{table}

In this section, the observations and the treatment of the data are described in detail. Sect. \ref{section:Observations} describes VLT/\criresp observations, Sect. \ref{subsection:Method:DataReduction} summarises data reduction, Sect. \ref{subsection:Method:WavelengthMorphing} details the alignment of the wavelength scales, Sect. \ref{subsection:Method:psf} discusses stability and variability, and Sect. \ref{subsection:Method:cleaning} covers cleaning and pre-processing necessary before our exoplanet detection methodology outlined in the next section.

\subsection{Observations}\label{section:Observations}

In this work, two primary transits of \wasp were observed as part of ESO programme IDs 108.C-0267(D) and 110.C-4127(D) \changeo{(PI: L. Nortmann)}. A third transit was scheduled to be observed in May 2023, but this night was lost due to bad weather and no observations were taken. The two successfully observed transits were obtained using the \criresp instrument that is installed on UT3 of the VLT at Paranal Observatory, Chile. \criresp is the upgrade project of the CRIRES instrument (CRyogenic InfraRed Echelle Spectrograph), which was previously in use at the VLT until 2014. It is a cross-dispersed echelle spectrograph operating in the near-infrared and mid-infrared regions of 0.95 \um to 5.3 \um (YJHKLM bands) at a nominal spectral resolution of $R \sim 100\,000$ with the 0.2 arcsecond slit \citep{dorn_crires_2023}. 

This work is the first ground-based study of \wasp at the K-band wavelengths of $\sim$2.0–2.5 \um, which is significantly redder than previous ground-based studies at $\sim$1 \um (Y-band). While previous studies have either covered this target at several different wavelength ranges from both space and ground, this is the only ground-based study so far to cover this particular wavelength regime that crucially straddles the relatively unexplored area between the shorter $\sim$1 \um regime and the longer $>$3 \um regime that is populated by several important molecular spectral features. All details regarding the parameters of the observations used in this work can be found in Table \ref{tab:Observations}. 

The first observation, henceforth referred to as Night 1 (N1), was obtained on the night to 11 March 2022. The second observation, henceforth Night 2 (N2), was obtained on the night to 23 February 2023. For both nights, observations consisted of 34 in-transit and the remainder of out-of-transit exposures.\footnote{\changeo{Following recommendations supported by e.g. \citet{cheverall_feasibility_2024} and \citet{palle_andes_2025}, the number of out-of-transit exposures taken was intentionally as high as our observing schedule allowed.}} All exposures were 300 seconds long, taken in an ABBA nodding pattern, \changeo{where the change between two exposures during the transit is 370 m/s (or approx. 25\% of the planetary diameter)}. This exposure cadence resulted in an average S/N of 130 per exposure, selected intentionally to optimise the trade-off between good S/N per exposure (in long exposures) and minimal smearing due to the planetary movement (in short exposures). Excessive smearing has been proven to have a significant impact on the HRCCS analysis used in this work (see Sect. \ref{subsection:Method:CCF_templates}), and so our observing strategy followed the recommendations of \citet{boldt-christmas_optimising_2024} to maximise the detection probability.

% ==== Figure for airmass + SNR/exp ====
\begin{figure*}
    \centering
    \includegraphics[width=\textwidth]{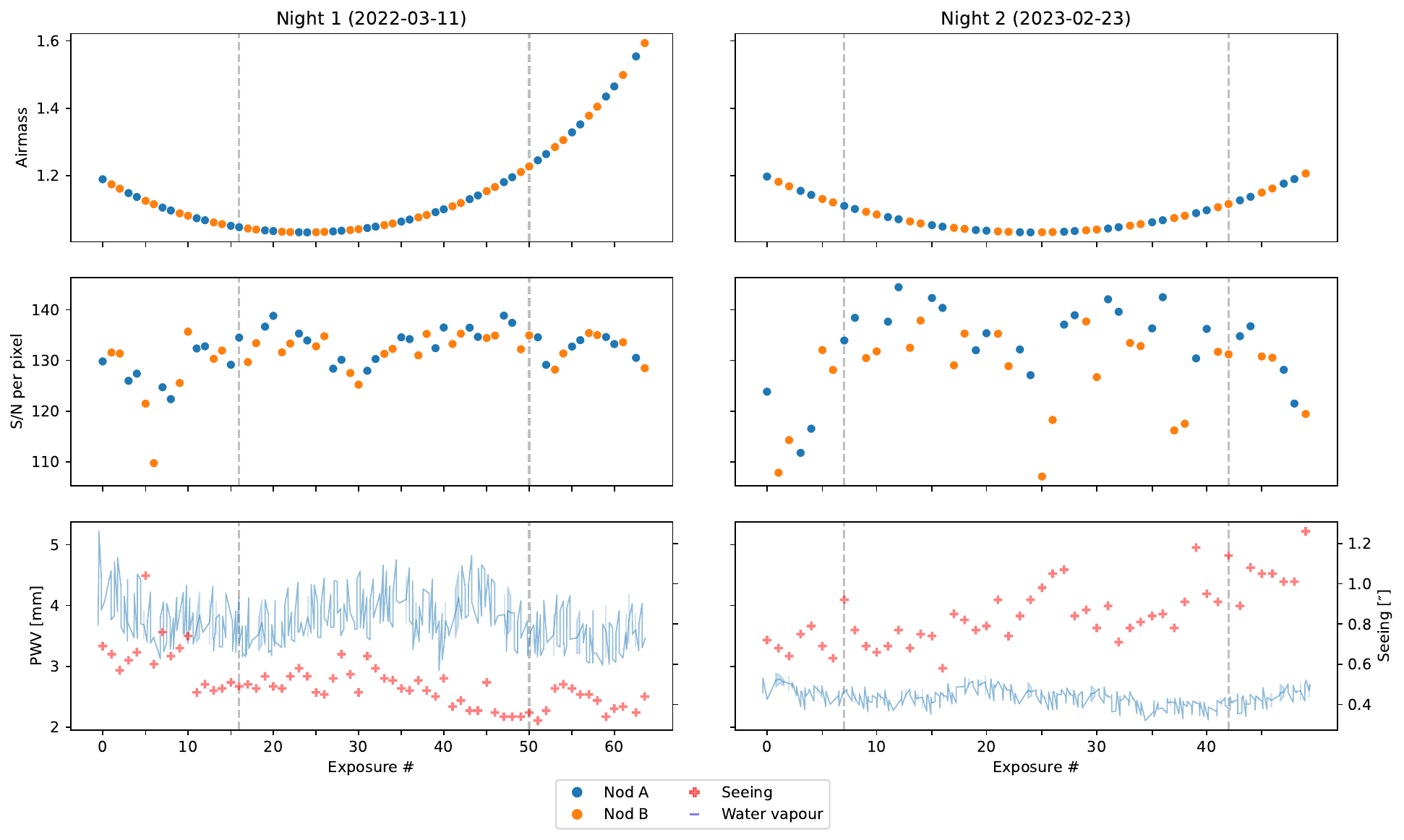}
    \vspace{-0.5cm}
    \caption{Observing conditions as a function of time for Night 1 (left) and Night 2 (right), with vertical gray lines indicating the extent of the transit event. \textit{Top:} Airmass, with coloured points blue and orange representing nodding positions A and B. \textit{Middle:} As above, but for median signal-to-noise ratio (S/N) per pixel. \textit{Bottom:} \changeo{the red crosses correspond to the seeing in arcseconds for each exposure, and the blue line to the precipitable water vapour (PWV), i.e. water vapour along our line of sight, throughout the night in millimetres}.
}
    \label{fig:airmass_snr}
\end{figure*}

Fig. \ref{fig:airmass_snr} shows the variations across N1 and N2 for airmass, S/N per exposure, and conditions (seeing and water vapour). Weather conditions were generally favourable for both nights, but conditions were overall better on N1 compared to N2. At good seeing conditions, the adaptive optics (AO) system of \criresp performs so well that it delivers a central peak that is smaller than the 0.2$\arcsec$ slit. This is an effect known as ``super-resolution'', which has occurred in observations from VLT/\criresp previously – see e.g. \citet{yan_crires_2023}, \citet{cont_exploring_2024,cont_retrieving_2025}, \citet{lesjak_retrieving_2025}, and \citet{nortmann_crires_2025}. At super-resolution, the resolving power $R$ can be as high as 150\,000 while still sampled by 2.3 detector pixels (where the nominal 0.2$\arcsec$ slit is projected on 3.5 detector pixels). The downside of super-resolution is uncertainty about where the star is located across the slit, which introduces a shift between A and B spectra obtained at different positions along the slit as part of the nodding procedure. The shift mostly manifests as RV offset (as large as 1 km/s) that is corrected for when resampling spectra on the common wavelength scale (see Sect. \ref{subsection:Method:WavelengthMorphing}).

The seeing conditions for N1 systematically stayed below 0.6\arcsec\xspace within the transit window. \change{The AO produced a stable central peak of the PSF with constant width and shape. The first 63 spectra show PSF width of 2.1 $\pm$ 0.1 km/s or $R$ = 143\,000 measured from the cross-dispersion profile constructed by the data reduction software and verified using the stellar CO lines around 2325 nm.} For N2, the seeing fluctuated above 0.6\arcsec\xspace throughout the transit, leading to noticeably variable PSF, leading to a resolving power $R$ between 120\,000 and 135\,000 or 2.5 and 2.2 km/s correspondingly. 300 m/s is certainly visible in the averaged line profile, and so we compared the results of the analysis for N1 and N2 (Section \ref{subsection:Real:CC} and Fig.~\ref{fig:CCF_per_night_frame}). No systematic differences that could be attributed to the variable spectral resolution were found.

%{\color{red}We need a plot that shows differences between R=140000 and R=120000. E.g. we can do simulated CCFs without noise at two R's or we can compare the pixel distribution functions for a pair of panels in Fig.7. Depepnding on the result this plot can go to the paper or just to the response.} 

\subsection{Data reduction}\label{subsection:Method:DataReduction}
The data were reduced using the standard ESO data reduction system for \criresp i.e. the \texttt{cr2res} pipeline (version 1.6.7)\footnote{ESO CR2RES Pipeline: \url{https://www.eso.org/sci/software/pipelines/cr2res/cr2res-pipe-recipes.html}} and recipes were executed with \texttt{EsoRex} (version 3.13.6).\footnote{ESO Recipe Execution: \url{https://www.eso.org/sci/software/cpl/esorex.html}} Raw calibrations were reduced by the recipes \texttt{cr2res\_cal\_dark}, \texttt{cr2res\_cal\_flat}, and \texttt{cr2res\_cal\_wave}, using daytime calibrations available under ESO Program ID 60.A-9051(A). Respectively, these: (i) produce the master dark used during \texttt{cr2res\_cal\_flat} and \texttt{cr2res\_cal\_wave}; (ii)  produce the master flat and adopted bad pixel mask; and (iii) determine the wavelength scale to be adopted for science exposures. Science frames were grouped into A/B nodding pairs (32 pairs on N1, and 25 pairs on N2) and reduced using the recipe \texttt{cr2res\_obs\_nodding} which performs A/B nodding subtraction to remove the sky background and influence of hot pixels, before producing 1D extracted spectra for each exposure – one A and one B. Note that we do not perform dark correction on our science frames as the A/B nodding subtraction serves the same purpose. Finally, we corrected our science spectra for the effect of the echelle blaze function using the pipeline-generated blaze file from \texttt{cr2res\_cal\_flat} and a Python script external to the pipeline.\footnote{Python software for interfacing with the \criresp data reduction pipeline, post-processing and cleaning reduced spectra, running SYSREM, performing cross-correlation, and simulating synthetic exoplanet transits per Sect. \ref{subsection:SimData:Simulations} can be found at \url{https://github.com/adrains/luciferase}. Our wavelength alignment algorithm is available as an IDL script in the folder \texttt{scripts\_reduction} on the same repository.}\label{fn:luciferase}

At this stage, these spectra are not yet science ready for our science case. The combination of super-resolution and optical effects introduced by separate A/B nodding positions produce two \textit{separate} sequences of spectra each night – one A sequence and one B sequence – that are not coherently aligned to either an absolute \textit{or} relative wavelength scale, which must be corrected before we can look for the subtle effects of a planet atmosphere on our data. We discuss how we approach this in the next section.

%%%%%%%%%%%%%%%%%%%%%%%%%%%%%%%%%%%%%%%%%%%%%%%%%%%%%%%%%%%%%%%%%%%
\subsection{Wavelength alignment}\label{subsection:Method:WavelengthMorphing}

%%%%%%%%%%%%%%%%%%%%%%%%%%%%%%%%%%%%%%%%%%%%%%%%%%%%%%%%%%%%%%%%%%%

% ==== Figure for airmass + SNR/exp ====
\begin{figure}
    \centering
    \includegraphics[width=\columnwidth, trim=2.4cm 3.cm 2.2cm 2.2cm]
    {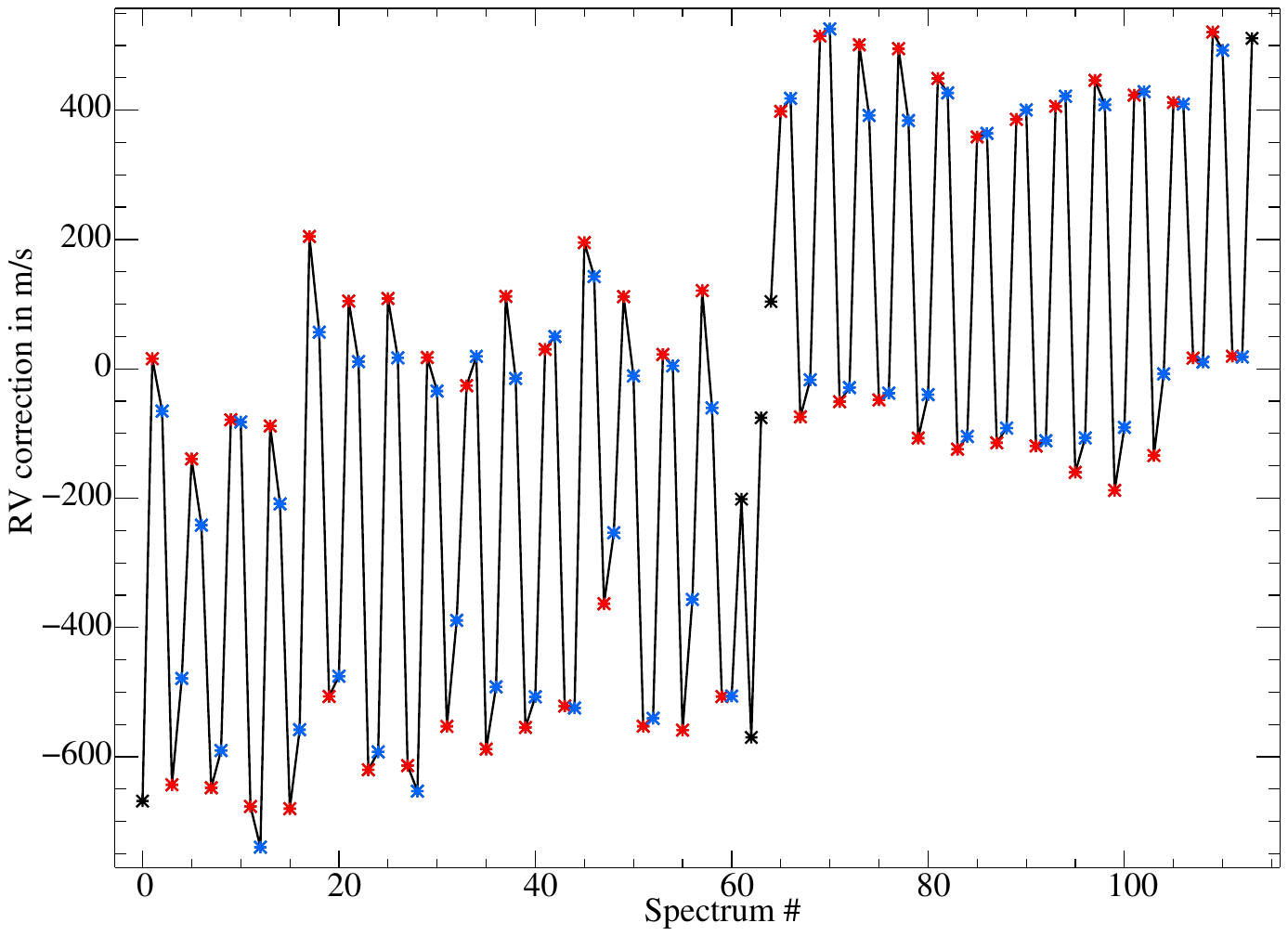}
    \caption{\changeo{Wavelength correction illustrated as RV shifts to the wavelength in the middle of a spectral order (vertical axis) determined for each spectrum. Red and blue asterisks shows AA or BB nodding pairs of consecutive exposures where no interruption in AO operation occurred and thus we expect no change in the across-the-slit position. These provides an alternative estimate of the precision. The mean differences are 80 m/s for N1 and 36 m/s for N2 to be compared to the estimates based on telluric line centers giving 91 and 53 m/s correspondingly. The large shift between the nights was measured by cross-correlating the two templates and converting it to an extra RV for each night to provide a sense of scale for the reader. In reality, most of this shift is translation (not RV) due to the reproducibility of the \criresp wavelength settings.}}
    \label{fig:wavelength_morphing}
\end{figure}

The alignment of A and B frames required in-depth investigation of the wavelength scale assigned to the spectral orders by the ESO pipeline. The calibration process is based on the calibration source illuminating the \textit{whole} slit, unlike the science target image created by the `overperforming' AO system under super-resolution conditions. The projection of the slit on the detector is tilted and curved with respect to the detector pixels, and even though the shape of the slit image changes a lot (few pixels) across the focal plane, the ESO pipeline is capable taking this into account. \changeo{What remains is the wavelength change for each science spectrum due to re-positioning the target on the slit during nodding in super-resolution conditions, which requires interruption of the AO operation.}

For K-band observations, the data product of the ESO pipeline delivers 18 spectral segments (6 spectral orders registered by 3 detectors, with gaps between detectors and no wavelength overlaps between orders). The obvious solution of cross-correlating science spectra does not work well for all spectral segments \changeo{due to the different number and origin of spectral features (stellar vs. telluric). Another problem was aligning spectra between the nights, as the change in barycentric velocity has more impact on the cross-correlation between spectral segments that have more pronounced stellar lines}. An attempt to use \texttt{molecfit} \citep{smette_molecfit_2015} to produce a robust template (that would not include stellar lines and thus will be insensitive to barycentric correction) also failed as some of the segments are essentially free of telluric features. 

\changeo{In the end, we selected a single spectral order dominated by telluric features (order 27 covering 2045.5-2089.3 nm range) and one exposure (for each night) as a template. Exposure selection was based on visual inspection for instrumental defects and cosmic ray hits missed by the data reduction system, and on the highest S/N pointed at spectrum 32 for N1 and 23 for N2 (spectrum 87 in the two-night combined sequence). In-depth discussion of the optical design of \criresp with Ernesto Oliva (priv. comm.) convinced us that the change expected due to shifts across the slit is equivalent to the change in radial velocity (RV) of the target rather than a geometrical translation of the focal plane image. With that assumption, we have written a wavelength morphing tool that transforms a spectral segment to its nominal wavelength scale given the reference template.}

\changeo{
The actual wavelength scale of a given segment is derived using a polynomial approximation to the RV applied to the nominal wavelength solution provided by the standard calibrations of \criresp for each night:}

\changeo{
\begin{equation}
w_k(x) = w(x)\cdot \sum_{m=0}^M a_m^k\cdot x^m
\end{equation}
}

\noindent \changeo{where $w(x)$ is the standard wavelength solution from calibration for pixel number $x$ and $w_k(x)$ is the morphing of that solution to phase $k$; adopted polynomial order $M$ is between 0 and 2 with index $m$; and $a_m^k$ are the unknown coefficients for phase $k$. Coefficients are found through an optimisation process that tries to maximise the cross-correlation between the reference phase and every other phase of the same transit. This means that we do not revert to an absolute reference frame and the final data may have an additional fake ``Doppler shift'' of the reference spectrum on the order of a few hundred m/s. This shift will be present as an extra offset from the expected systemic velocity in the final results.}

\changeo{Cross-correlation between the reference spectra from different transits was used to connect the two nights. This shift combines the RV change due to super-resolution and the error in echelle positioning. The latter results in a shift along the main dispersion rather than a RV shift. The reproducibility of the echelle positioning in \criresp is nominally one tenth of a pixel or about 100 m/s, but we find that in our case, the combined effect of positioning and super-resolution is around 300 m/s. This shift does not affect the analysis as the two nights are treated independently.}

%%%%%%%%%%%%%%%%%%%%%%%%%%%%%%%%%%%%%%%%%%%%%%%%%%%%%%%%%%%%%%%%%%%
\begin{figure*}[h]
    \centering
    \includegraphics[width=0.49\linewidth, trim=3.1cm 2.2cm 1.2cm 2.2cm]{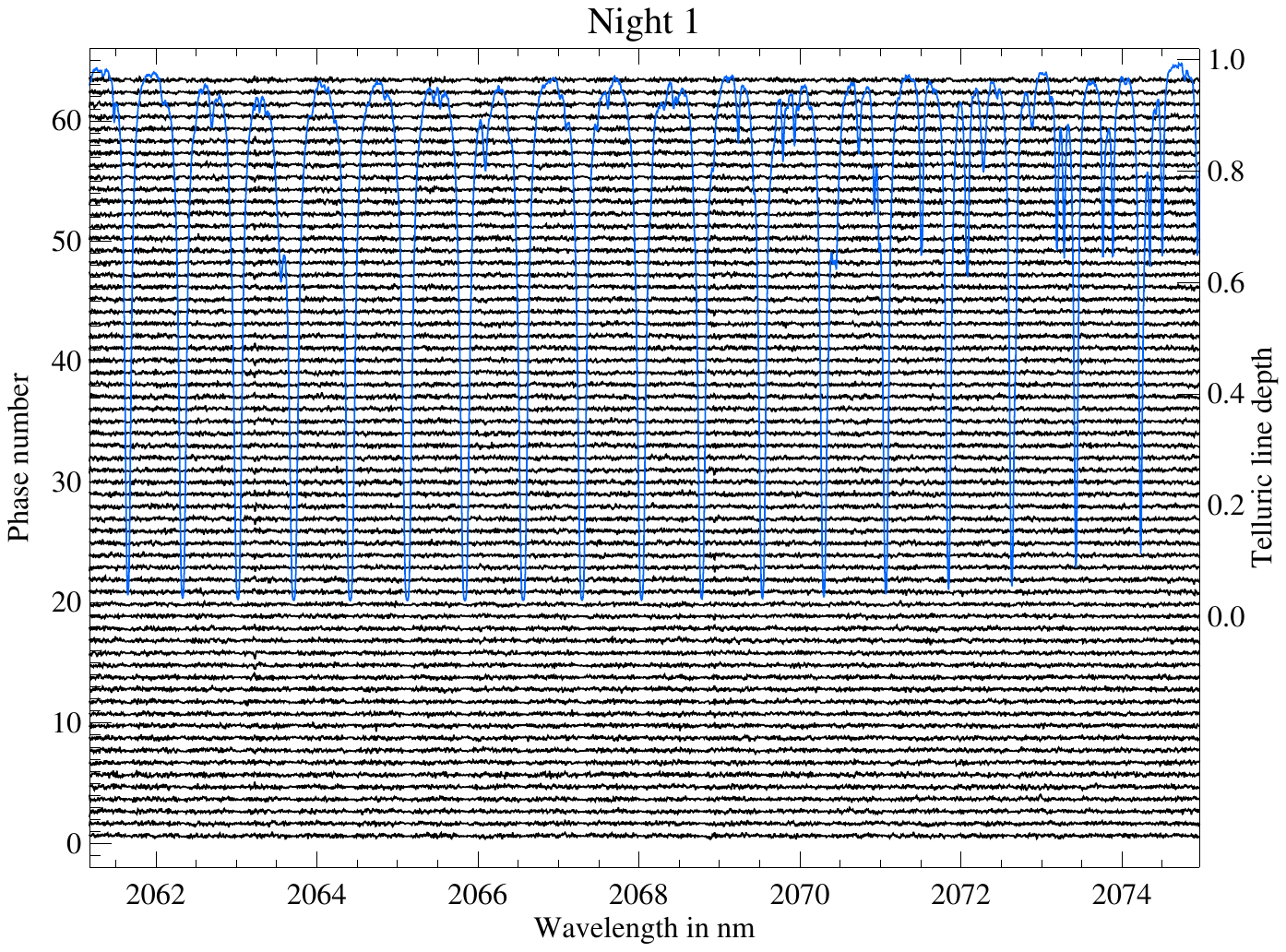}
    \includegraphics[width=0.49\linewidth, trim=3.1cm 2.2cm 1.2cm 2.2cm]{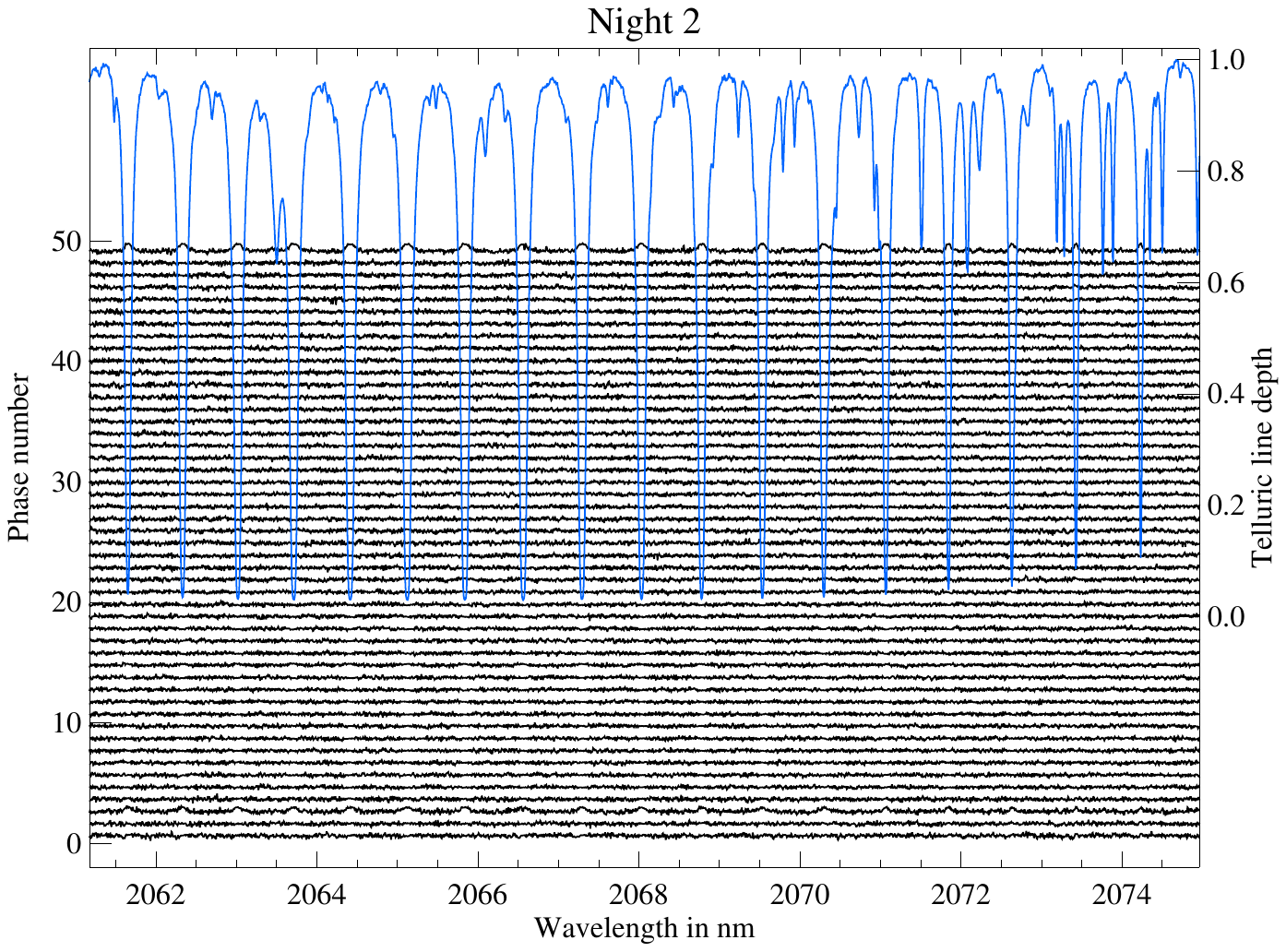}
    \caption{\changeo{Per-phase differences (black lines) between observed \wasp spectra and an inverse model of the mean nightly telluric absorption with mean PSF profile scaled to appropriate airmass. This spectral segment is totally dominated by telluric features and contains no significant stellar lines. Vertical offsets between each subsequent phase were added for visibility. A blue overplotted spectrum shows the mean telluric spectrum for each night at the lowest airmass. Note that substantial structure in the residuals is only visible for two phases on Night 2 (third from bottom, and topmost), which is indicative of the PSF changes in those phases, but also to the general stability of the PSF otherwise.}}
    \label{fig:PSF stability}
\end{figure*}

\changeo{The precision of this correction was estimated by trying 30 different phases from each night (the same spectral order) as references (900 combinations) and comparing the resulting corrections. For $M=0$, the root mean square (RMS) of the correction spread within each night was found to be 119 m/s for N1 and 99 m/s for N2. Increasing to $M=1$ reduced these numbers to 82 m/s and 67 m/s, but $M=2$ did not add any improvement. The wavelength correction derived for the reference order was applied \textit{to all spectral orders at that phase}. The corrections are illustrated in Fig.~\ref{fig:wavelength_morphing} as RV values for the central column of the middle detector. An alternative estimate of the correction precision for a single night was done by measuring the wavelength of the telluric line centres. The latter were determined by fitting a Gaussian to the line profiles. For $M=2$, we found the RMS to be 91 m/s for N1 and 53 m/s for N2. These values match nicely the correction scatter obtained for consecutive AA or BB spectra in the ABBA nodding sequences as shown in Fig.~\ref{fig:wavelength_morphing}.}

A version of the code used for this correction can be found in the repository in Footnote \hyperref[fn:luciferase]{5}, with the intention of integrating this tool into the official \criresp pipeline in the future.

\subsection{PSF stability and telluric spectrum variability}
\label{subsection:Method:psf}
Variations in the PSF or components of the telluric spectrum (e.g. H$_2$O absorption) may have strong negative consequences for detrending analysis methods applied to our \wasp data. One way to probe the severity of these effects is to study the differences between telluric lines as a function of time. The advantage of using tellurics to probe PSF stability is that these lines do not experience \change{time-varying Doppler shifts in excess of 100 m/s, unlike stellar or exoplanetary lines,} but they do change in strength and shape as a function of the airmass, AO performance, and changes in the partial pressures of various molecules in
the Earth's atmosphere (e.g. humidity). A model of telluric absorption that could be adopted for the airmass of each of our spectra and compared with the data was therefore needed, where any systematic features in the residuals will indicate changes in the PSF and/or chemical composition of the atmosphere. Ideally, this telluric spectrum should be constructed from the data itself so as to match the instrumental profile and lines present -- something more challenging to accomplish with a telluric spectrum produced via spectral synthesis methods like \texttt{molecfit}. Fortunately, as part of related project, we have developed a method to recover such a telluric spectrum from observations that we could use here.

\changeo{
The method to reconstruct telluric absorption from observations---Transmission Spectroscopy Decomposition (TSD)---is described in detail by the upcoming paper \citet{piskunov_tsd_2025}. Briefly, TSD formulates the analysis of high-resolution exoplanet transmission spectroscopy data as an inverse problem, and optimises for three spectral components (stellar, telluric, exoplanet) which can be used to best reconstruct the observed data. Each of these three components has a unique velocity frame, and by adopting a priori these velocities at each observed phase, the method can `disentangle' the components.\footnote{\changeo{This is analogous to how one might disentangle time-series double-lined eclipsing binary spectra into the three spectral components of Star A, Star B, and tellurics which together can be used to reconstruct a model of the observed data at each phase given the velocities for the binary components.}} TSD adopts the assumption that each night of data has a single constant telluric transmission spectrum affected only by airmass. While in reality this assumption is not strictly true, especially for wavelength regions affected by H$_2$O absorption, the resulting nightly telluric spectrum \textit{can} effectively be considered the mean telluric spectrum at the mean PSF for that night. By comparing this telluric spectrum to each phase of our observed \wasp data, systematic differences in the residuals would indicate a departure from this mean value, and thus highlight problems due to a changing PSF.% remaining systematic differences from the observations would indicate a problem. We applied this method to the \wasp dataset.
}

\changeo{
Fig.~\ref{fig:PSF stability} presents such a comparison for the middle detector of order 27 (2061-2075 nm). This order contains exclusively strong telluric lines, making it optimal for detecting PSF or telluric opacity variations during a transit. For N1, we see no substantial systematic deviations from the horizontal line for any of the telluric features. For N2, in two phases (third from the bottom and topmost, both out of transit phases), the residual shows inverted ``$v$'' shape in the line cores, which is indicative of additional broadening of the observed profile---consistent with worse seeing for these two phases. From this comparison, we conclude that our data set is luckily not affected with strong PSF variations, barring two phases in N2 that both occur outside the transit window.}

%%%%%%%%%%%%%%%%%%%%%%%%%%%%%%%%%%%%%%%%%%%%%%%%%%%%%%%%%%%%%%%%%%%
\subsection{Post-processing}\label{subsection:Method:cleaning}

With wavelength scales aligned and unified, the final step before analysis is to continuum normalise and clean our spectra for obvious artefacts that might interfere with the detection of an exoplanet atmosphere. To continuum normalise our data, we applied a first order polynomial correction to each of our 18 spectral segments with reference to synthetic stellar and telluric template spectra (for e.g. masking, see Sect. \ref{subsection:SimData:Simulations} for more information on our adopted MARCS and \texttt{molecfit} templates). Each night was treated separately, as was each A/B sequence. However, within a night, the slope of the correction was set constant for each spectral segment, with only the scale allowed to vary between exposures. In other words, we assumed that the \textit{shape} of the spectrum is constant across exposures, but the system throughput may change. Referencing synthetic spectra allows for better stability, especially in segments with numerous telluric absorption features. While the continuum normalisation framework is not ultimately required for the next detrending analysis, we find the results to be more robust.

For data cleaning, we looked for and sigma clipped two kinds of artefacts. The first only occur on a single pixel at a single exposure – perhaps the result of uncorrected cosmic rays – and the second are systematic differences between the A and B sequences – perhaps the result of uncorrected detector artefacts at one nodding position but not the other. For the first, we compute the standard deviation of the pixel in time for one sequence, and compared it to the median spectrum of the other sequence, clipping where this exceeds our adopted threshold. For the second, we computed the standard deviations of the entire spectral segment for the two nodding sequences, and compared these with the difference within nodding pairs. The assumption here is that systematic differences should only occur in cases of detector artefacts present in only one nodding position, in which case we must discard the entire column since {$\sim$}50\% of the data is now considered systematically aberrant. For both cases, we adopted a value of $5.0\sigma$ for sigma clipping to remove the worst of the artefacts without risking destroying subtle signals introduced by the planet, and only interpolated clipped pixels when fewer than 5 pixels  (\about10\% of the exposures on a single night) are clipped – otherwise masking the entire column. Further, we chose not to extrapolate clipped edge (in time) pixels as testing indicated this introduced spurious features observable in later analysis. The end result is spectra packaged into arrays of shape $[N_{\rm exp}, N_{\rm spec}, N_{\rm px}]$ where $N_{\rm exp}$ is the total number of exposures observed per night, $N_{\rm spec}$ is the number of spectral segments equal to $N_{\rm order} \times N_{\rm detector}$, and $N_{\rm px}$ is the number of spectral pixels per detector. 

%----------------------------------------------------
\section{Method}\label{section:Method}

This section describes how the reduced data were analysed using the cross-correlation technique. Sect. \ref{subsection:Method:SYSREM} details the removal of stellar and telluric features via SYSREM, and Sect. \ref{subsection:Method:CCF_templates} summarises the cross-correlation of SYSREM residuals with \pRT exoplanet template spectra.

%%%%%%%%%%%%%%%%%%%%%%%%%%%%%%%%%%%%%%%%%%%%%%%%%%%%%%%%%%%%%%%%%%%
\subsection{Removal of stellurics (SYSREM)}\label{subsection:Method:SYSREM}

Spectra obtained through ground-based observations of a transiting exoplanetary system consist of spectral features from three different sources: (i) the stellar spectrum from the host star; (ii) the telluric contamination from Earth's atmosphere; and finally, (iii) a very minor fraction of spectral features that have been imprinted onto the stellar light by transmission through the exoplanetary atmosphere. In this particular science case, it is only this final exoplanetary component that is of interest as it is this component that carries information about the planet's atmosphere. In order to study it, this spectrum must be isolated from the stellar and telluric contribution – or ``stellurics'' as they will be referred to collectively henceforth. 

During a single transit, stellar lines shift relative to tellurics primarily due to the rotation of Earth. This effect is of the order of a few hundred metres per second, i.e. small compared to our spectral resolution, which warrants the assumption that stelluric lines remain aligned during the relatively short observing window of a transit. From this assumption, a common method for stelluric removal at infrared wavelengths is a type of Principle Component Analysis (PCA) algorithm known as SYSREM. This algorithm was first developed by \citet{tamuz_correcting_2005} for cleaning time-series photometry and later applied to observations of exoplanets, with early examples including e.g. \citet{birkby_detection_2013,brogi_carbon_2014,schwarz_evidence_2015}. 

Conceptually, SYSREM capitalises on the fact that the Doppler shift of the exoplanet will range across the transit event from being relatively blueshifted (at the beginning of the transit) to being relatively redshifted (at the end) from the perspective of the observer by \about 350 m/s every exposure. Comparatively, the stelluric lines will not be notably shifted: the stellar lines will only shift marginally (\about 10 m/s every exposure) based on the star's relative motion, while the telluric lines will not shift at all with respect to the observer. As such, stelluric lines will be recorded on roughly the same spectral pixel throughout the observing window, whereas the exoplanetary lines will first fall on bluer pixels and then shift throughout the night across to redder pixels from exposure to exposure. Notably, this distinction between the star and planet velocity frames becomes less significant for a lower orbital velocity of the planet, even for a relatively small reduction of  radial velocity semi-amplitude \kp; in the case of a hot Jupiter ($K_p \gtrsim 120$ km/s), the large orbital velocity of the planet results in significant and spectrally resolved differences between the stellar and planetary velocity frames, but in the case of an Earth-mass planet ($K_p \sim 50$ km/s), they might be indistinguishable as this would result in a change across the transit that is smaller than the resolution element of the spectrograph. For \wasp, \kp = 105 km/s (see Table \ref{tab:WASP107_s+p_params}).

In practice, an algorithm such as SYSREM can fit and remove systematic trends – namely, the stelluric lines that are persistently present at the same spectral pixel throughout the observation. Effectively, SYSREM models the wavelength-dependent and time-dependent systematics (stellurics) before subtracting them, which should in theory leave behind only the non-systematics (planetary features) in the resulting residuals. SYSREM does this iteratively, meaning that the process of fitting a model and then removing it from the observation is repeated some number of times, at which point the spectrum should in theory be fully free of stellurics. This model is created by taking the product of two components: a spectrum $S$ that varies with wavelength $\lambda$, and a variation $A$ that varies with time $t$ to account for time-dependent fluctuations such as airmass. The model, $f$, may then be expressed as:

\begin{equation}
    f (\lambda, t) = S (\lambda) \cdot A (t),
\end{equation}

\noindent which is repeatedly fitted for each iteration to the data for each pixel, minimising the sum of the residuals squared each time. In the case of cross-correlation searches for species that are present in the host star of the observed system (e.g. CO), SYSREM may be better executed in the stellar rest frame \citep[i.e. where the telluric lines would shift marginally and the stellar lines do not; see e.g.][]{nortmann_crires_2025}. For a more detailed mathematical description of SYSREM, see Sect. 3.1 of \citet{czesla_elusive_2024}.

The benefit of SYSREM is that this procedure removes stellurics even when they densely overlap with the spectral region of the exoplanetary features, as is the case for our K-band observations, where simply discarding affected pixels would result in an unsustainably large loss of data. However, there is a challenge associated with using SYSREM in that the user does not have full insight into exactly how many iterations result in a spectrum that is neither still contaminated by stellurics (i.e. too few iterations) nor taken so far that planetary signal also begins to be removed (too many iterations). There have been investigations of how to determine the most appropriate number of iterations, such as \citet{cheverall_robustness_2023} or \citet{meech_applications_2022}, but some general limitations remain in that certain factors – such as fewer exposures, unfavourable barycentric separation, varying humidity affecting tellurics, low signal – will inevitably create obfuscation in this procedure, which impacts us adversely as will be discussed in upcoming sections. \changeo{For example, significant changes in an opacity component (e.g. \hto) of the telluric spectrum, or of the PSF width, during a transit may lead to a complete failure of SYSREM -- especially for slow-moving planets, which is why we controlled that this is not the case for our data set (Sect.\,\ref{subsection:Method:cleaning}). All of these concerns} is ultimately why, in Sect. \ref{section:RealDataAnalysis}, the results of a wider range of SYSREM iterations ($n$=8, i.e. 3–10, out of the total 0–15) are shown rather than selecting a single iteration. To account for such imperfect detrending, we weighted the residuals produced by SYSREM by inverse of the standard deviation of each spectral pixel as a function of orbital phase (exposure). This has the effect of downweighting pixels still strongly variable in time (e.g. those most affected by \ce{H2O} telluric absorption), improving the effectiveness of our cross-correlation analysis in the following section.

\subsection{Cross-correlation with synthetic templates}\label{subsection:Method:CCF_templates}

    \begin{table}\small
    \centering
        \caption{Parameters used to generate the \pRT templates.}
    \vspace{-0.4cm}
    \begin{tabularx}{\columnwidth}{Xrr} 

     \\
          \toprule
          \textbf{For all templates} \\
          \midrule
       Parameter & Units &  Value \\
     \midrule

    Wavelength range & \um & 1.8–2.7 \\
    Pressure range & bar &  1–10$^{-10}$ \\
    Eq. temperature$^W$  & $T_{\mathrm{eq}}$ [K] & 738 \\
    Int. temperature$^S$ & $T_{\mathrm{int}}$ [K] & 460 \\
    Planet radius$^W$ & $R_{\mathrm{p}}$ [\Rjup] &  0.94 \\
    Gravity$^W$ & $g$ [cm/s$^2$] & $10^{2.45} \approx 282$ \\
    Reference pressure$^D$ & $P_0$ [bar] & 0.01 \\
    Atmospheric IR opacity$^D$ & $\kappa_{\mathrm{IR}}$ [cm$^2$/g] & 0.01 \\
    Ratio of optical and IR opacity$^D$ & $\gamma$ & 0.4 \\
    \changeo{Non-metallic mass fractions}$^D$ & \changeo{H:He} & \changeo{0.72:0.28} \\
    Rayleigh species &  & H$_2$, He \\
    Continuum opacities & & H$_2$–H$_2$, H$_2$–He \\
    Convolved spec. resolution & $R$ & 140,000 \\
         \bottomrule
    \end{tabularx}

    \begin{tabularx}{\columnwidth}{Xrrrr}
    \midrule
     \textbf{Per retrieval} \\
     \midrule 
       Parameter & Units & Line list & CHI & Aur \\
       \midrule
       Mean mol. weight$^*$ & $u$ && 3.02 & 2.64 \\
    VMR per species:$^{**}$  & [dex] && \textcolor{white}{++'++}&  \\
    \ce{CH4} && HITEMP \citeyearpar{hargreaves_accurate_2020} &  -5.8 & -6.1 \\
    \ce{CO} && HITEMP \citeyearpar{rothman_hitemp_2010} &   -1.9 &  -3.0 \\
    \ce{CO2} && HITEMP \citeyearpar{rothman_hitemp_2010} &  -3.9 & -4.4 \\
    \ce{H2O} && ExoMol \citeyearpar{polyansky_exomol_2018} &  -2.1 & -2.6 \\
    \ce{H2S} && HITRAN \citeyearpar{rothman_hitran2012_2013} &  -8.5 &  -8.6 \\
    \ce{NH3} && ExoMol \citeyearpar{yurchenko_variationally_2011} &  -5.0 & -5.1 \\

    %\textit{Excluded but accounted for:} &&&&& \\
    $^{\dagger}$\textit{\ce{SO2}} &&&  -5.2 & -5.7  \\
    
     \bottomrule
      %& & & & & & \\
 
    \end{tabularx}
    
    \vspace{0.2cm}
    
    \scriptsize{\changeo{Superscripts \textit{D, S, W} correspond to values taken from \citet{dyrek_so2_2024}, \citet{sing_warm_2024}, and \citet{welbanks_high_2024} respectively.}
    
    $^*$ The mean molecular weight (MMW), denoted by $\mu$ as calculated in Eq. \ref{eq:mmw}, is here given for the atmosphere when considering the presence of all species.  
    
    $^{**}$ The values $r_V$ cited here for volume mixing ratio (VMR) are given in dex, i.e. as log$_{10}$($r_V$) for each species. These values are taken from the Extended Data Table 2 columns for \texttt{CHIMERA} (CHI) and \texttt{Aurora} (Aur) of \citet{welbanks_high_2024} as indicated, and were originally published with errors. However, we do not include these errors here as our templates are generated without uncertainties.

    $^\dagger$ The published VMR for \ce{SO2} was accounted for in our generation of templates, insofar that its weight was included in our calculation of MMW, but this species is not part of our analysis as it does not have features present in our wavelength range.}

    \label{tab:templateParams}

\vspace{-0.4cm}
\end{table}

Once observations have been cleaned of stelluric contamination, the SYSREM residuals can be searched for planetary signal; however, this signal is so extremely small that statistical methods, such as cross-correlation analysis, must be employed. The high-resolution cross-correlation spectroscopy (HRCCS) analysis uses synthetic templates computed with radiative transfer packages and an assumed model of the planetary atmosphere. Cross-correlation is then used to establish statistically significant similarity (or the lack of such) with the planetary transmission spectrum hidden in the noisy residuals of SYSREM, done over a range of radial velocity shifts for each exposure. Mathematically, this requires evaluating the normalised cross-correlation function (CCF):

% Nik's version of the CCF eqn
%\begin{equation}\label{eq:CCF_sum}
%\mathrm{CCF} (v,t) = \sum_{\lambda} \frac{x_\lambda (t) \cdot T_ {\lambda\cdot(1+v/c)}}{\sigma^2_{\lambda} (t)}
%\end{equation}

%{\color{purple}

% Adam's version of the CCF eqn
\begin{equation}\label{eq:CCF_sum}
\mathrm{CCF} (v,t) = \frac{\sum_{\lambda} x_\lambda (t) \cdot T_ {\lambda\cdot(1+v/c)}}{\sqrt{\sum_{\lambda}x^2_\lambda (t) \cdot \sum_{\lambda}T^2_ {\lambda\cdot(1+v/c)}} }
\end{equation}

%}

\noindent where $x_\lambda$ are the \textit{weighted} SYSREM residual for data taken at time $t$ and $T_\lambda$ is the template offset with radial velocity $v$. While the exact expression for CCF varies somewhat between applications, the HRCCS approach is by now well-established and has become a standard tool in the exoplanet community as it has been employed by many similar studies \citep[see e.g. the recent review by][and references therein]{snellen_exoplanet_2025}.

% , and $\sigma_\lambda(t)$ is the uncertainty of $x$

\subsubsection{Generating templates with petitRADTRANS}
    \label{subsection:Method:CCF_templates:templates}

The templates used for cross-correlation in this work were generated using the radiative transfer package \pRT \citep{molliere_petitradtrans_2019}, version 2.7.6. This package models high-resolution transmission spectra (using the \texttt{lbl} or `line-by-line' method) at $R=10^6$ that we later convolve to \changeo{$R=$ 140\,000, slightly above the average \criresp resolution across the two nights}. 

Two sets of templates were generated for our analysis, and a complete list of the parameters for all templates can be found in Table \ref{tab:templateParams}. The spectra are generated using \pRT's associated opacity data which uses a number of different line lists for the species modelled. Most parameters were adopted from the studies of \citet{dyrek_so2_2024} and \citet{welbanks_high_2024} in an attempt to facilitate comparability, favouring values from \citet{welbanks_high_2024} where the option existed as this work features more data points (including the \citet{dyrek_so2_2024} data set) across a broader wavelength range and can thus be considered more robust. Our adopted values include the assumption of the so-called ``Guillot'' temperature-pressure profile, i.e. a profile based on the often used Eq. 29 in \citet{guillot_radiative_2010}. We set the infrared atmospheric opacity to be $\kappa_{\mathrm{IR}} = 0.01$ cm$^2$ g$^{-1}$, and the ratio between optical opacity and infrared opacity to be $\gamma = 0.4$. The equilibrium temperature and intrinsic temperature were set to $T_{\mathrm{eq}} = 738$ K \citep{welbanks_high_2024} and $T_{\mathrm{int}} = 460$ K \citep{sing_warm_2024} respectively, and at the planetary radius $R_p = 0.94$ \Rjup, the pressure is $P_0 = 0.01$ bar and gravity is log$_{10}(g)$ = 2.45, which is $\approx$282 cm/s$^2$ \citep{dyrek_so2_2024, welbanks_high_2024}.

Our two sets of templates are based on the results of the two retrievals from \citet{welbanks_high_2024}, and the intention is to investigate how well each type of template fares both in contrast to one another, and also when accounting for the possible presence of a cloud deck (see Sect. \ref{subsection:Method:InclusionCloudDeck}). The two sets of retrievals from \citet{welbanks_high_2024} use \texttt{Aurora} and \texttt{CHIMERA}, two independent inference frameworks (whose outputs are found in their Extended Data Table 2), will here on be referred to as `Aur' and `CHI' respectively. The differences in the retrieved volume mixing ratios (VMR) between these two \changeo{forward models} range between tenths of dex to more than 1 dex. For example, the two values for log$_{10}$(VMR) of \ce{H2S} are \mbox{CHI = -8.5} and \mbox{Aur = -8.6} while the values for log$_{10}$(VMR) of CO are \mbox{CHI = -1.9} and \mbox{Aur = -3.0.} One purpose of this exercise is to understand how sensitive the HRCCS method is to differences in retrieved values – or, at least, to set a lower limit on the parameter differences – that can be detected with sufficient confidence, and how this combines with the impact of modelling a cloud deck presence.

We take these volume mixing ratios $r_V$ from the outcome of the \citet{welbanks_high_2024} retrievals. The mean molecular weight of $\mu$ for each retrieval is calculated from each $r_V$ of species $i$ (either metals, $Z$, or primordial, $p$, i.e. H$_2$ or He). Explicitly, $\mu$ is calculated using the relationship:

%\begin{equation}
%   X = \frac{r_V \cdot m}{\mu}
%\end{equation}

%\begin{equation}
%   X_\mathrm{total} = X_{Z+p} = 1 = \frac{\sum\limits_i ( {r_V}_i \cdot m_i) }{\mu}  
%\end{equation}

\begin{equation}\label{eq:mmw}
 \mu = \sum\limits_i ( {r_V}_i \cdot m_i) = \sum\limits_Z ( {r_V}_Z \cdot m_Z) + \sum\limits_p( {r_V}_p \cdot m_p).
\end{equation}

This equation is solved noting that $\sum {r_V}_p=1-\sum {r_V}_Z$, and assuming H:He mass fractions ($X$) of 0.72:0.28 gives $m_p = 2 \times 0.72 m_\mathrm{H} + 0.28 m_\mathrm{He}$. Values for ${r_V}_Z$ are known from the retrievals and $m_i$ are basic atomic and molecular data from the NIST database.\footnote{National Institute of Standards and Technology (NIST) database: \url{https://webbook.nist.gov/chemistry/name-ser/}} This mean molecular weight $\mu$ is assumed to be constant throughout the atmospheric layer that we are observing, thereby neglecting potential stratification due to photochemistry at lower altitudes or evaporation at high altitudes, as sufficient mixing is expected at the altitudes we are observing, even though these processes may vary significantly and are currently poorly understood \citepeg{miguel_exploring_2014,lavvas_aerosol_2017,soni_signature_2024}.

All templates are generated for the wavelength range covering the \criresp K-band plus a margin of approx. 0.2 \um in either direction (i.e. generated across $1.8-2.7$ \um), and convolved to match the resolution of \criresp. Templates were generated for individual species, as well as a ``global'' template that includes contributions from all species. We included all species from the retrievals of \citet{welbanks_high_2024} except for \ce{SO2} as this species is not expected to have features in the K-band. In all templates, however, \ce{SO2} is still accounted for in our calculations of the mean molecular weight.

% ==== Figure for spectrum w and wo cloud ====
\begin{figure*}
    \centering
    \vspace{-0.45cm}
    \makebox[\textwidth]{\includegraphics[width=1.0\textwidth]{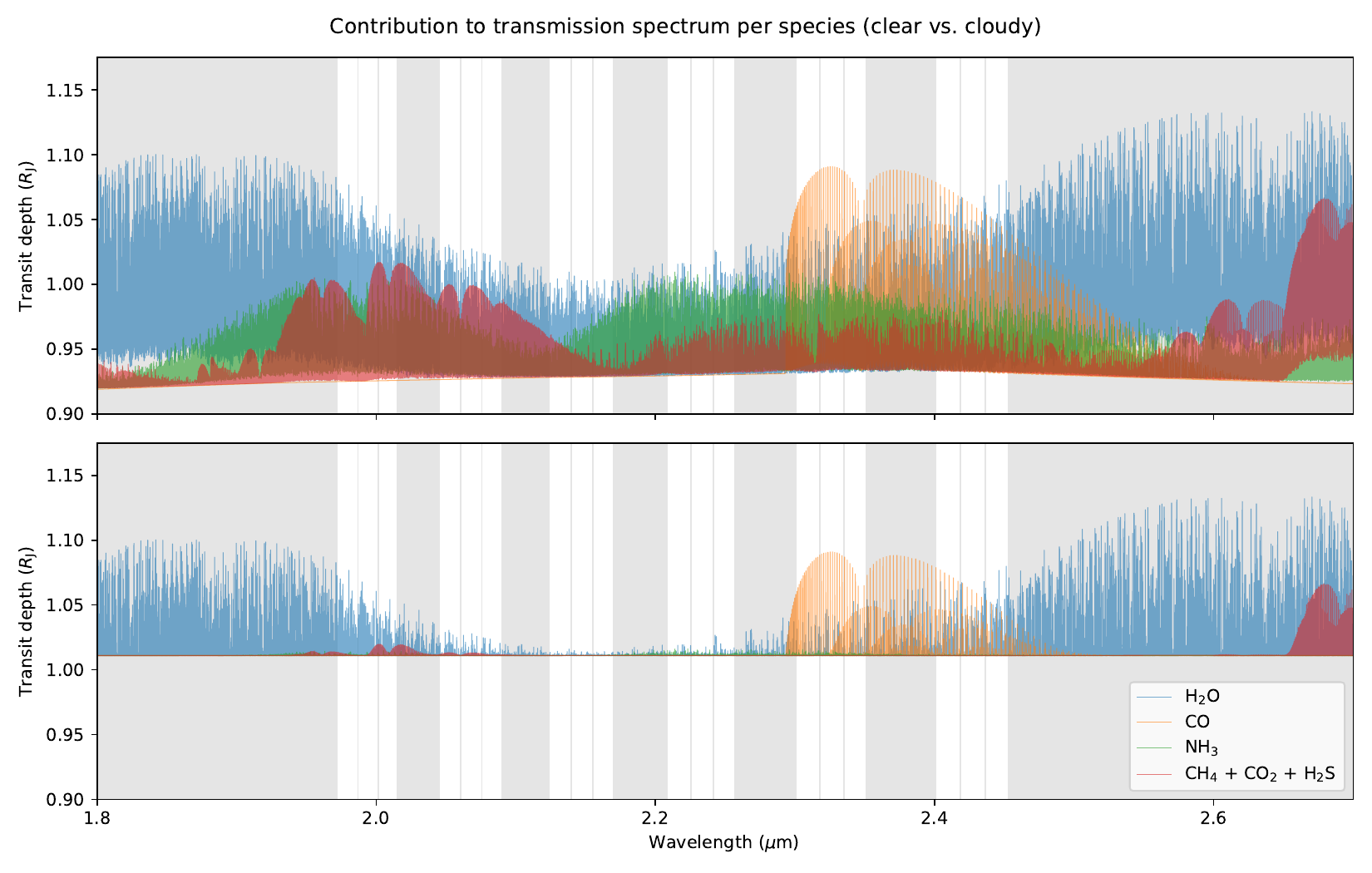}}
    \vspace{-0.8cm}\caption{Plot showing the simulated global (i.e. including all species) transmission spectrum of \wasp, coloured by species contribution, based on parameters from the results of the \citet{welbanks_high_2024} CHIMERA \changeo{forward model} and as generated by \pRT at a resolving power of $R=10^6$. The plot demonstrates the notable change of the spectrum's shape when either excluding (top plot, showing a  ``clear'' atmosphere) or including (bottom plot, a ``cloudy'' atmosphere) the presence of a grey cloud deck at $P=10^{-5}$ bar. The grey areas denote wavelength ranges that fall outside of the spectral orders of the VLT/\criresp K-band of $\sim$2.0–2.5 \um (1972–2452 nm).}
    \label{fig:clearVScloudy_pRT}
\end{figure*}

\subsubsection{Inclusion of cloud deck in templates}\label{subsection:Method:InclusionCloudDeck}

The presence of clouds, hazes, and/or other condensates (here all included in the broader term of ``aerosols'') in an exoplanetary atmosphere significantly impacts its transmission spectrum, and are thus very important to accurately include in one's analysis. Generally, aerosols scatter light and therefore quench the spectral features i.e. decrease the amplitudes of present species' spectral lines. In the best case, this will only discreetly manifest as increased scattering slopes or reduced line depths \citep{pinhas_signatures_2017}, but in less favourable cases, this can result in spectra that are at worst completely featureless. This is because the effect of aerosols on a transmission spectrum can be hugely varied as the total impact will be different depending on several different variables: planetary chemistry, atmospheric scale height, thermal and wind gradients across the planet's limb, wavelength regime being studied, and many more. This large parameter space is partially why the field of exoplanetary cloud studies is so active today, as there are always several effects working in tandem for any given target that create complex results that can be challenging to interpret. For a more detailed discussion on the impact of aerosols on transmission spectra, see Sect. 3.1 of \citet{gao_aerosols_2021}.

While this impact is therefore well-known, in high resolution (ground-based) spectroscopy studies of hot exoplanets, \changeo{there are arguments that are often made for not including} clouds in the generation of templates for cross-correlation. Generally speaking, the reasoning for doing so would be that any potential clouds, if at all present, would be (i) below the atmospheric layer being probed, and (ii) sufficiently flat across our (comparatively narrow) wavelength range that their only impact should be an effective overall dampening of our line strength, meaning their inclusion or exclusion will not significantly impact the results of our cross-correlation detections \citep[unless doing a simultaneous analysis over very broad wavelength regions; see][]{pino_diagnosing_2018}. 

However, in the case of \wasp, the matter of exactly how its clouds may substantiate is an active research question where recent studies have shed significant light on the topic but unknowns still remain. \citet{dyrek_so2_2024} describe a high-altitude cloud deck at $P = 10^{-5}$ bar (see their Extended Data Fig. 4), while \citet{murphy_evidence_2024} and \citet{murphy_panchromatic_2025} demonstrate that \wasp's cloud formation differ across its morning limb and evening limb. Coupling this with the knowledge of its ongoing photochemistry from \citet{dyrek_so2_2024}, \citet{sing_warm_2024}, and \citet{welbanks_high_2024}, and the varied shortcomings in our ability to model atmospheres (as acknowledged by all of the above), it is clear that a full description of clouds in \wasp is still a work in progress with many unrestrained parameters. This is also supported by \citet{changeat_cloud_2025} who demonstrated the importance of cloud modelling in retrievals for JWST data across a range of targets including \wasp.

Therefore, we only generate templates that assume two extreme border cases to straddle the upper and lower boundaries of reality: a grey (i.e. uniform across all wavelengths) cloud deck at $P = 10^{-5}$ bar versus no cloud deck. The decision to not model more complex or iterative manifestations of clouds was effectively due to the aforementioned reasons of why clouds are often neglected in ground-based analysis: if cloud decks are present at pressures higher than $P\approx 10^{-3}$ bar (i.e. at lower altitudes), we find no discernible differences appear in our \pRT generated spectra compared to the clear spectra (the standard deviation $s$ of respective transit depths between the two being $s < 0.005$), and only minor differences ($s< 0.01$) appear at $P\approx 10^{-4}$ bar. Only at $P=10^{-5}$ bar do differences begin to manifest relatively noticeably ($s = 0.01$), and as such, lower altitude cloud decks are in our case effectively equivalent to the clear case. Thus, the on/off treatment of clouds was considered acceptable for this work; in subsequent sections, we demonstrate why this question may need to be revisited in the future.  

The assumed chemical composition of \wasp means that the spectral features of the transmission spectrum are not uniformly affected by clouds across the wavelength regime of the K-band. This is illustrated by Fig. \ref{fig:clearVScloudy_pRT}, which shows the transmission spectrum of \wasp based on the results of the CHI-\changeo{model} from \citet{welbanks_high_2024}. In the K-band, significant quenching can be seen in the spectral lines found between $\sim$2.1–2.3 \um, for which the lines are only barely visible above the cloud deck, compared to the lines at $\sim$2.0–2.1 \um and $\sim$2.3–2.5 \um that are relatively speaking less quenched. Crucially, not only does this mean that a significant amount of signal is lost compared to the cloud-free case, but this means that the cloud deck is not merely reducing the line strength of the spectrum's lines across the whole spectral window as can be assumed in cross-correlation studies – instead, it is effectively altering the shape of the spectrum as a whole, which will have a significant effect on our HRCCS results (see Sect. \ref{subsection:SimData:Simulations}). Fig. \ref{fig:clearVScloudy_pRT} also demonstrates the relatively small and/or hidden contribution of \ce{CH4} + \ce{CO2} + \ce{H2S} compared to \ce{H2O}, CO, and \ce{NH3}, which is why no attempt is made to detect these species individually in Sect. \ref{section:RealDataAnalysis}. 

In spite of previously referenced findings that suggest differences between morning and evening limbs of \wasp, we use one template at a time to construct the CCFs for the whole transit i.e. do not analyse the limbs differently. This was largely motivated by the claim by \citet{murphy_panchromatic_2025} in the caption of their Fig. 1, which states that ``the combination of [their] morning and evening spectra are consistent with the panchromatic limb-combined spectrum [...] presented in Welbanks et al. (2024)''. Supported by tests from simulations, we consider this to be a second order effect as only a small number of exposures – nine per transit – would carry the imprint of the `morning' and `evening' spectrum. Furthermore, the discontinuity arising from a sudden change of templates is non-physical, and implementing a smooth transition between the templates would require further considerations, making a simple switch unrealistic.

\subsubsection{Construction of \kpvsys diagrams}
    \label{subsection:Method:CCF_templates:kpvsys}

Once the templates are generated, these can be used to perform cross-correlation with the output of the stelluric removal i.e. the residuals containing the extracted planetary spectrum. As denoted by Eq. \ref{eq:CCF_sum}, the cross-correlation analysis is performed individually on each exposure of total $N$ exposures before these are summed. As the amount by which the exoplanet has been Doppler shifted varies over time $t$, the amount of radial velocity shift $v$ at which the CCF is maximised will vary between each exposure. By plotting the CCF of each exposure in a heat map (where each exposures are plotted on the vertical axis, i.e. each exposure is denoted by a single row), the maximal cross-correlation value in S/N is indicated through brightness at the value of $v$ where this CCF maximisation happens. Due to the movement of the planet, the first exposures (pre-transit) and last exposures (post-transit) should contain no planetary signal whereas the exposures taken during transit should contain planetary signal along the slanted line of CCF peaks, following the changing radial velocity of the planet. In our analysis, CCF plots are generated separately for exposures in each night (N1 or N2) and nodding position (A or B) to create four groups of CCF plots of N1-A, N1-B, N2-A, N2-B. These plots are shown later in Sect. \ref{section:RealDataAnalysis}, confirming that no obvious artefacts or anomalous features are present.

While the information necessary to infer a detection is technically available at this point, \changeo{no obvious planetary trail is visible, and the CCF for any individual phase remains too noisy to detect the planet.} % maximal cross-correlation values are still very noisy as can be seen by the lack of noticeable planetary trails in the CCF plots. 
\changeo{In order to boost this, we can collapse our set of $N_{\rm phase} \times N_v$ CCFs in the phase dimension---i.e. collapse the data in the time dimension---by aligning the CCFs in the exoplanet rest frame and combining. At each phase there are three velocity components to consider: the time varying barycentric velocity (known), the constant systemic velocity of WASP-107 relative to the Solar System barycentre ($v_{\rm sys}$, known), and the time varying projected exoplanet orbital velocity \change{(not necessarily well-known)}. To get around not perfectly knowing the exoplanet velocity, we repeatedly collapse our CCFs for some set of trial exoplanet RV semi-amplitude (\kp) values, and expect our signal to be maximised at the real \kp of the planet.}
%In order to boost this, we benefit from the fact that the CCF should in theory be maximised when shifted by the target's true radial velocity semi-amplitude \kp (which is known for \wasp to be \kp $= 105.2 \pm 2.5$ km/s; see Table \ref{tab:WASP107_s+p_params}) at the target's velocity \vsys = \vsys$_\mathrm{sys}$ \textcolor{red}{\textbf{SORT THIS OUT}} relative to the observer. 
Shifting CCFs according to trial radial velocities for a range of \kp values (extending on both sides of the target's actual \kp, i.e. from \kp = 0–400 km/s, and \vsys = $\pm$ 200 km/s) and co-adding all the exposures produces a so-called \kpvsys map that shows the CCF surface in \kp and \vsys coordinates. A well-defined peak in the \kpvsys maps indicates that the template has matches in the SYSREM residuals, and if this peak occurs near the true \kp and \vsys values,\footnote{\changeo{Note that the choice of $x$-axis zero point is somewhat arbitrary, with the two main conventions being for $v = 0$ km/s to be the Solar System barycentre or, as we adopt here, $v=v_{\rm sys}$ i.e. the rest frame of WASP-107.}} we have a detection. Significance of detection is estimated by calculating the standard deviation (computed from the full map excluding the central region around \vsys = 0 km/s by $\pm$ 30 km/s, i.e. the outer regions only), and comparing the maximum S/N of the map to the standard deviation, where a larger detection significance is implied by a larger deviation. 

%----------------------------------------------------
%\section{Analysis}\label{section:Analysis}

\section{Analysis of simulated data}\label{section:SimDataAnalysis}

As the field of high-resolution cross-correlation spectroscopy of exoplanet atmospheres has been active for nearly two decades by now, a large sample of exoplanets have been successfully characterised in that time as citations throughout this paper show. For understandable reasons, the strongest detections have come from targets that are (i) close-in to their host star, meaning frequent opportunities for observations, large velocity excursions, and usually with circular orbits and thus tidal locking; (ii) with extended atmospheres, and sufficiently hot that their atmospheric species produce strong, atomic spectral lines with minimal likelihood of cloud formation due to dissociation. This has resulted in an over-representation of hot Jupiters and ultra-hot Jupiters in the catalogue of well-studied exoplanetary atmospheres, as can be shown by statistics by the IAC ExoAtmospheres database (see Footnote \hyperref[fn:IAC]{2}) and as summarised in Table 1 of \citet{cont_exploring_2024}.

In the case of \wasp, there are a number of factors that complicate our analysis in terms of how strong the anticipated cross-correlation signal should be. The target is not hot (below 800 K), which partly explains why its atmosphere is dominated by molecular species at \changeo{these} temperatures. Due to the fact that molecular species produce orders of magnitude more spectral lines than atomic species – especially in the infrared wavelengths – their features are intrinsically more challenging to identify as their line lists are often less accurate and complete in comparison with atoms. For further discussions on the impact of molecular line lists on exoplanet studies, see e.g. \citet{hoeijmakers_search_2015} for titanium oxide; \citet{hedges_effect_2016} for water; \citet{bowesman_high-resolution_2021} for aluminium oxide; \citet{tannock_146-248_2022} for methane; and \citet{rengel_about_2022} for an overview of atomic and molecular databases used in the (exo)planetary community. Furthermore, as described in Sect. \ref{subsection:Method:InclusionCloudDeck}, the presence of a cloud deck may be detrimental in our analysis; and as discussed, the detailed properties of the \wasp clouds are largely unknown. 

To understand these challenges, we developed a transit simulator to study and predict how some of these effects might substantiate and impact our interpretation of the real data. Investigating how our target may appear through the use of simulations is a crucial guiding tool in this new and under-explored regime of transmission spectroscopy of cool planets, as we do not expect to see the canonical \kpvsys maps that are more closely associated with past studies of hotter exoplanets.
%%%%%%%%%%%%%%%%%%%%%%%%%%%%%%%%%%%%%%%%%%%%%%%%%%%%%%%%%%%%%%%%%%%%%%%%%%%
% ==== Figure for matrix comparing clouds yes/no in sims + temp ====
% ==== Figure for arg periastron variation (all real e) ====
\begin{figure*}
    \centering
    %\vspace{-0.45cm}
    \makebox[\textwidth]{\includegraphics[width=1.0\textwidth]{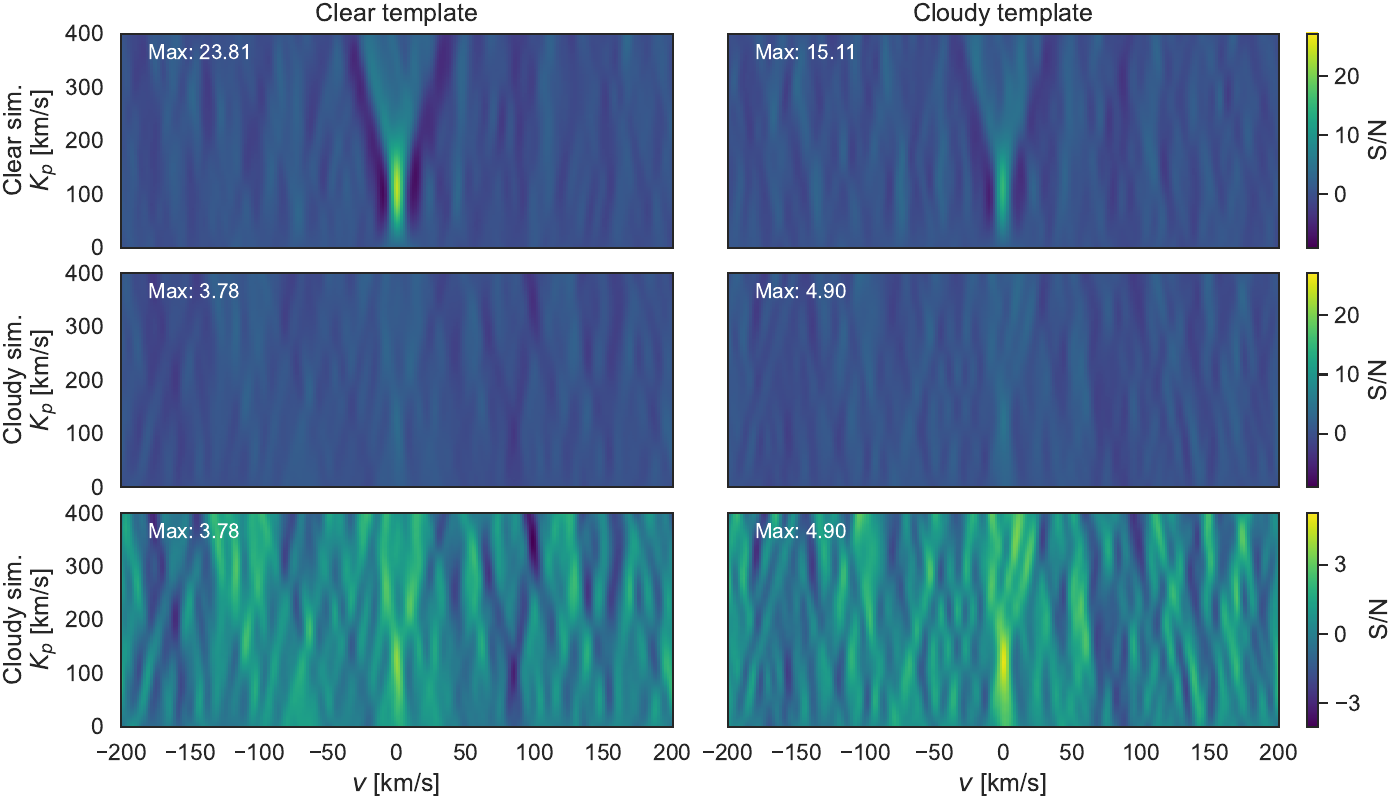}}
    \vspace{-0.5cm}\caption{Simulated \kpvsys plots to test the impact of including or excluding cloud decks in four different test cases. \textit{Top:} Simulations of a ``clear'' \wasp that has no cloud deck present, cross-correlated with a ``clear'' template that excludes clouds (left) and with a ``cloudy'' template that includes clouds (right). \textit{Middle:} Same test as the top row, now simulating a ``cloudy'' \wasp with a cloud deck present at $P=10^{-5}$ bar, shown on the same colour scale as the test above for comparison. \textit{Bottom:} Same test as the middle row, now shown on its own colour scale for reference.}
    \label{fig:test1_matrixYesNoClouds}
\end{figure*}

%%%%%%%%%%%%%%%%%%%%%%%%%%%%%%%%%%%%%%%%%%%%%%%%%%%%%%%%%%%%%%%%%%%%%%%%%%%
\subsection{Simulated observations}\label{subsection:SimData:Simulations}

Our simulated observations require six main ingredients: (1) a simulated stellar spectrum for WASP-107; (2) a model of Earth's atmospheric transmission incorporating key telluric species; (3) a synthetic transmission spectrum for \wasp; (4) instrumental throughput and a wavelength scale for \criresp; (5) system parameters describing the star, planet, and their orbits; (6) velocities (star, planet, barycentric), airmasses, and slit losses computed at each simulated epoch. 

To simplify this process and increase the explanatory utility of the simulations, we opt to simulate the same two transits as we actually observed down to the timing and number of exposures as described in Table \ref{tab:Observations}. Our star, planet, and orbital parameters are taken from Table \ref{tab:WASP107_s+p_params}, and the velocities of each rest frame (star, planet, barycentric) are computed for each epoch using these system parameters. Our planet transmission spectra are modelled as per Table \ref{tab:templateParams} \changeo{(CHI-values)}, and our telluric transmission spectra are two \texttt{molecfit} models \changeo{each computed and then} fitted to the master spectrum from each night of our observed \criresp data. Finally, our stellar spectrum is based on a custom 1D LTE\footnote{Local Thermodynamic Equilibrium} spherically-symmetric model atmosphere computed using the latest version of the MARCS code \citep{gustafsson_grid_2008}. The stellar spectrum is generated at $R{\sim}100\,000$ adopting parameters from \citet{piaulet_wasp-107bs_2021} and using MARCS opacity tables, the solar abundance ratios of \citet{grevesse_solar_2007}, and VALD database  line data \citep{ryabchikova_major_2015}. This stellar intensity spectrum was computed for 49 radial stellar disk positions (i.e. $\mu$ angles), which enables us to interpolate specific intensities directly without needing to invoke limb-darkening approximations. A full description of our simulation methodology is available in the methodology paper \changeo{\citep{piskunov_tsd_2025}}.%$T_{\rm eff}=4\,420\,$K, $\log g=4.61\,$dex, and ${\rm[M/H]} = 0.05\,$dex, 

While we do our best to simulate our real \criresp observations, there are several approximations made which prevent a 1:1 simulation of reality, but we do not perceive these limit the interpretability and utility of our simulations. Firstly, beyond adopted instrumental transfer functions for \criresp as made available by ESO, we make no attempt to model the optics of the spectrograph, the pipeline reduction of data from raw 2D to extracted 1D spectra, nor effects like correlated noise, hot pixels, nodding, cosmic rays, or super-resolution. Due to the low $v\sin i$ of WASP-107 \citep[$0.45\,$km/s determined by][]{rubenzahl_tess-keck_2021} as compared to the motion of the planet and velocity resolution of \criresp, we neglect stellar rotation – and thus the Rossiter-McLaughlin effect \citep{rossiter_detection_1924,mclaughlin_results_1924} – when modelling. Despite evidence indicating that \wasp has observable 3D structure like limb-variations that presumably result in complex line profiles visible at high-resolution, we treat the planet atmosphere as 1D and constant with time. We model slit losses and thus varying `seeing' conditions, but we do not model time-varying tellurics (except for the changing airmass), e.g. as the result of changing humidity conditions throughout a transit. Finally, we simulate \wasp transits at S/N \about 130 to match our real data assuming normal distribution of noise.

\subsubsection{Note on fluctuation in detection significance}\label{subsection:SimData:DetSigFluc}

For the simulations presented in the following sections, the detection significance can be seen to vary at the $\sim$1\sig level. This is expected behaviour, and is the result of random noise and slit losses applied to each spectral pixel and exposure respectively. These random excursions from the average have already been proven to exist in a previous simulation study aimed at finding the best possible observing strategy – for further discussion of this effect, see Sect. 4.3 of \citet{boldt-christmas_optimising_2024}.

\subsubsection{Test 1: inclusion versus exclusion of cloud deck}\label{subsection:SimData:test1_clouds}

One of our first tests was to measure the impact of including or excluding a cloud deck in our cross-correlation (CC) templates, considering a hypothetical scenario where we do not know if our target in reality has a cloud deck present (cloudy) or not (clear). In these simulations, we demonstrate four cases:

\begin{enumerate}
\item Using \textit{clear} CC template with \textit{clear} atmosphere. 
\item Using \textit{clear} CC template with \textit{cloudy} atmosphere. 
\item Using \textit{cloudy} CC template with \textit{clear} atmosphere. 
\item Using \textit{cloudy} CC template with \textit{cloudy} atmosphere.
\end{enumerate}

In all four cases, no other parameters are changed between each simulation. In order to recreate the most common HRCCS scenario of searching for a singular species at a time, both the cloudy and clear CC templates include spectra from \ce{H2O} only. The resulting \kpvsys maps (combined N1+N2) are shown in Fig. \ref{fig:test1_matrixYesNoClouds}, and demonstrate two important effects. Firstly, there are relative differences in detection significance between the cases where the planet's actual physical characteristics (i.e. in this case, whether a cloud deck truly is present or not) matches or does not match the template's characteristics. In both the case of a clear planet + clear template and cloudy planet + cloudy template, the detection significance is higher than that of a clear planet + cloudy template and cloudy planet + clear template respectively, i.e. higher when the template matches the atmosphere. It may not be surprising that one is rewarded if using a template that is truer to reality, but it may be surprising that the magnitude of this reward is so significant, with the ``mismatched'' cases achieving only approximately \changeo{65\% and 75\%} of the total possible detection significances respectively. Secondly, there is a very stark difference in detection significance between the cases of cloudy and clear planets, with the clear planet + clear template case resulting in a maximum S/N that is \changeo{nearly five} times higher than the cloudy planet + cloudy template case; even the clear but ``mismatched" case of clear planet + cloudy template fares better as its maximum S/N is over three times higher (even when recalling the expected fluctuations described in Sect. \ref{subsection:SimData:DetSigFluc}). \changeo{This is in line with previous work that lead to similar conclusions \citep{pino_diagnosing_2018,allart_wasp-127b_2020,lafarga_hot_2023} for hotter targets.}

This \changeo{effect} underlines a clear obstacle in studying exoplanets with cloudy atmospheres, which becomes increasingly relevant as we move towards characterising cooler (and thus cloudier) exoplanets. In the specific case of \wasp, the presence of a cloud deck clearly reduces the detection significance from a strong detection to one barely above the noise floor even under these idealised circumstances, making the choice of ``correct'' template even more important. 

\subsubsection{Test 2: varying the argument of periastron}\label{subsection:SimData:test2_argPeri}

\begin{figure*}
    \centering
    %\vspace{-0.45cm}
    \makebox[\textwidth]{\includegraphics[width=1.0\textwidth]{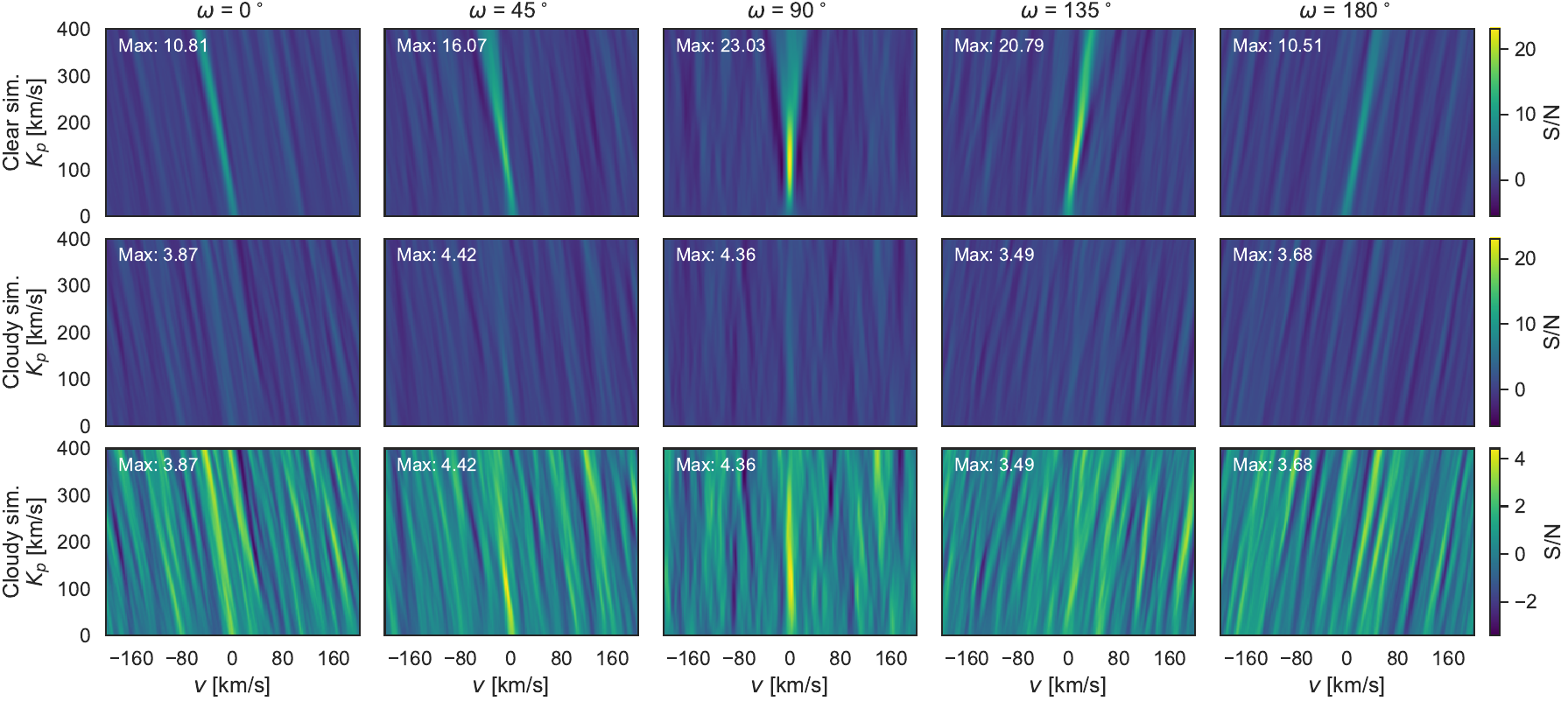}}
   %\vspace{-0.8cm}
    \caption{Simulated \kpvsys plots to test varying the argument of periastron $\omega$ in ten different test cases. \textit{Top:} Simulations of a ``clear'' \wasp that has no cloud deck present, cross-correlated with a ``clear'' template that excludes clouds, for five different values of $\omega$. \textit{Middle:} Same test as the top row, now simulating a ``cloudy'' \wasp cross-correlated with a ``cloudy'' template, where both the simulation and template include a cloud deck at $P=10^{-5}$ \changeo{bar} and is shown on the same colour scale as the test above for comparison. \textit{Bottom:} Same test as the middle row, now shown on its own colour scale for reference.}
    \label{fig:test2_w_real_e}
\end{figure*}

The role of orbital configurations on exoplanet observations has been discussed extensively in the literature, which is understandable considering its potentially great impact on many systems. As described in the beginning of this section, the field of cross-correlation spectroscopy has been primarily raised on case studies of hotter, close-in planets that tend to have circular orbits – but as the field continues to attempt to characterise increasingly smaller and cooler exoplanets and wider orbits, the likelihood of a given target having a more complex orbit increases. As such, many feasibility and methodology studies have been recently produced in the context of a wide range of parameters – to name a few, \citet{prinoth_high-resolution_2024} for high eccentricity and the argument of periastron; \citet{cheverall_feasibility_2024} for low velocity planets; and \citet{hong_velocity_2025} for orbital periods.

For the next test, we study the impact of the argument of periastron, also known as the argument of periapsis, which is one of the orbital elements required to fully describe an \textit{elliptical} orbit. The argument of periastron, $\omega$, is the angle between the ascending node (i.e. the point where the orbital plane crosses the reference plane, in the rising direction) and the orbit's closest point to the star (measured in the plane of the orbit). $\omega$ defines the orientation of the semimajor axis in the plane of the orbit relative to the observer – the ``twist'' in the spherical coordinate direction – and therefore determines what segment of the orbit is actually captured during the primary eclipse and how the radial velocity of the planet changes during transit.

Here, we want to test the impact of arguments of periastron values given the eccentricity suggested for \wasp, which is $e=0.06 \pm 0.04$ \citep{piaulet_wasp-107bs_2021}, in the context of the previous test on cloud inclusion. In this simulation, we model \wasp using its known orbital parameters (including eccentricity) across a range of $\omega$ values, but construct our \kpvsys map using the equation for a circular orbit in order to mimic the often true reality of not perfectly knowing orbital parameters. This is also relevant considering the well-known bias in orbital eccentricity determinations where the $e$ of a purely circular orbit may come out as slightly positive due to random noise, meaning non-zero $e$ values $<$3\sig away from zero (such as in our case) may be less trustworthy \citep{lucy_spectroscopic_1971,shen_eccentricity_2008,zakamska_observational_2011}.

Fig. \ref{fig:test2_w_real_e} demonstrates the effect of varying $\omega$ across a full half-revolution of 0–180$^\circ$ for both clear simulations and cloudy simulations. Both cases are cross-correlated with their ``matching'' respective templates of \ce{H2O}, i.e. the clear simulations are cross-correlated with a clear template and the cloudy simulations with a cloudy template. All other parameters between them remain the same, and are otherwise the same as in the first test.

For both the clear and cloudy case, $\omega=90^\circ$ provides the best possible detection significance in the \kpvsys map. This is to be expected to be the case for a slightly eccentric orbit such as ours, as demonstrated by \citet{prinoth_high-resolution_2024} as this will align the observation with capturing the primary eclipse at a point during which the planet's radial velocity is maximally changing. The other angles, however, have a more varied impact on the clear and cloudy cases. For the clear simulation, it is notable that the detection significance is effectively halved at $\omega = 0^\circ$ and $\omega = 180^\circ$ compared to $\omega = 90^\circ$, and that the impact on detection significance is not immediately clear with only a single realisation for $\omega = 45, 135 ^\circ$ due to the reasons detailed regarding noise realisations in Sect. \ref{subsection:SimData:DetSigFluc}. Meanwhile, for the cloudy simulation, while $\omega = 90^\circ$ fares better than other angles, the variation between them is significantly less notable; at this point, the signal is already diminished to the point that the effect of cloud-quenching is more significant than the effect of $\omega$.

\begin{figure*}
    \centering
    %\vspace{-0.45cm}
    \makebox[\textwidth]{\includegraphics[width=1.0\textwidth]{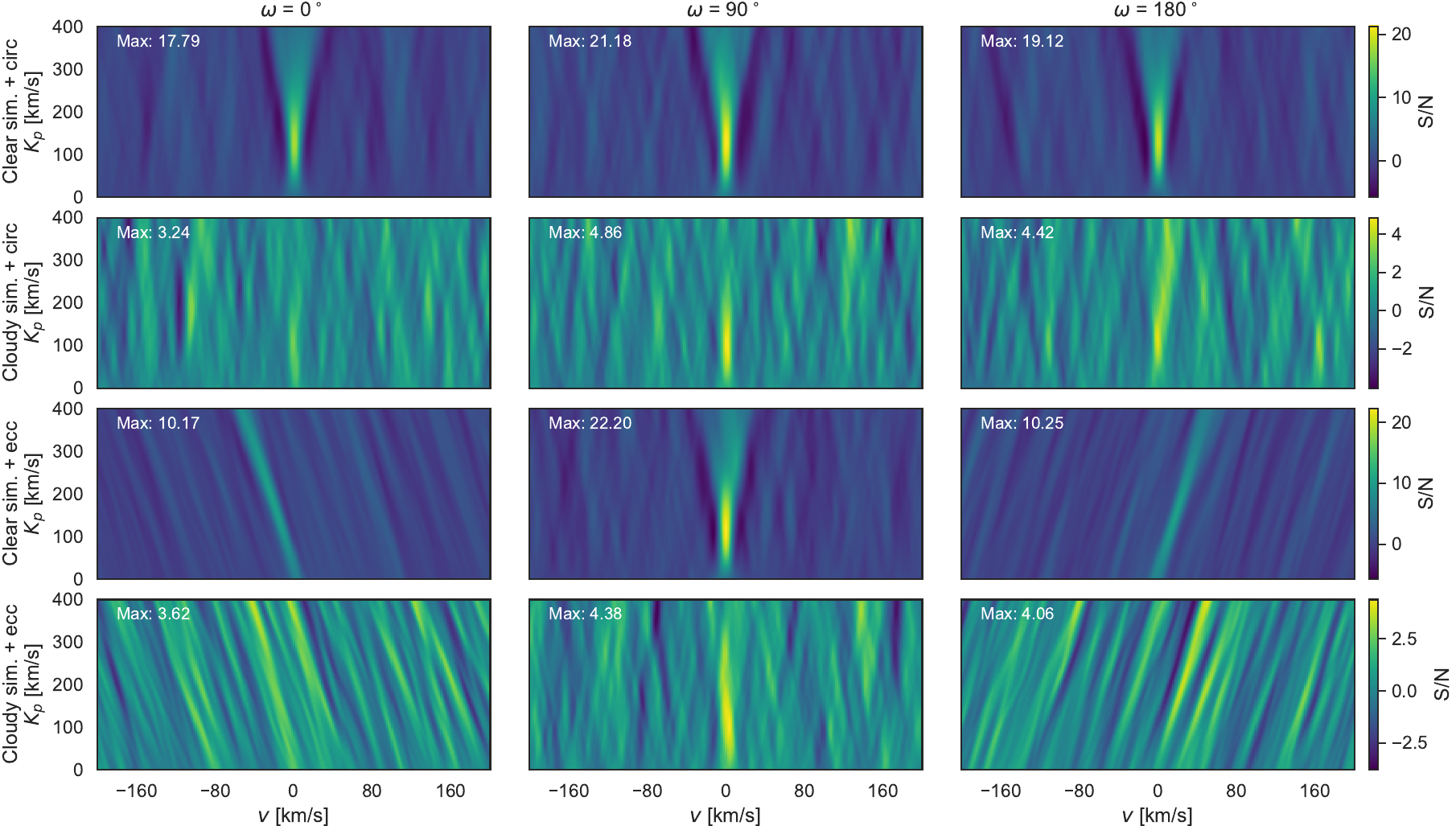}}
    \vspace{-0.5cm}\caption{Simulated \kpvsys plots to test varying the argument of periastron $\omega$ and orbital eccentricity in six different test cases. \textit{First row:} Simulations of a ``clear'' WASP-107 b that has no cloud deck present, cross-correlated with a ``clear'' template that excludes clouds, for three different values of $\omega$, all for a WASP-107 b on a fully circular orbit. \textit{Second row:} Same test as the top row, now simulating a ``cloudy'' WASP-107 b cross-correlated with a ``cloudy'' template, where both the simulation and template include a cloud deck at $P=10^{-5}$ bar. \textit{Third row:} Same test as the first row, but now for a WASP-107 b on its true (somewhat) eccentric orbit of $e=0.06$. \textit{Fourth row:} Same test as the second row, but now for a WASP-107 b on its true orbit of $e=0.06$.}
    \label{fig:test3_eccorbit}
\end{figure*}

One will also note the varying tilt of the signal in the \kpvsys map, which reflects the asymmetry of the planet's radial velocity change throughout the transit. This is a known phenomenon that creates a `stripe' structure on the \kpvsys plot, where stripes tend to be vertical for $\omega = 90^\circ$ and slanted for other values. This effect was previously described in \citet{basilicata_gaps_2024}, who saw this in GIANO-B transmission spectra of the warm Neptune-like planet HAT-P-11 b (Kepler-3 b), which is even more eccentric at $e\approx 0.26$ and shows a pattern similar to our plots. 

\changeo{\subsubsection{Test 3: varying orbital eccentricity}\label{subsection:SimData:test2_argPeri}}

 \changeo{In the case of WASP-107 b's only slightly eccentric orbit where $e=0.06$, it is fairly intuitive that $\omega=90^\circ$ should give a maximised detection significance. To compare: for a less eccentric (i.e. circular) orbit, the planet's \change{orbital velocity $v_\mathrm{orb}$ is constant at all points across the orbit, and so there will be no particular angle that will create a higher $\Delta v_\mathrm{orb}$ across the transit; for a more eccentric orbit, the angle at which $\Delta v_\mathrm{orb}$ is maximised} during the transit window would be further away from $\omega = 90^\circ$ (with exact extent varying on eccentricity $e$). In order to demonstrate this intuition, our third test varies the argument of periastron $\omega$ once more, but in this case, also for the case of a fully circular orbit, both with and without clouds. The results are shown in Fig. \ref{fig:test3_eccorbit}.}

In these simulations, it is demonstrated that (i) the tilt arising in the \kpvsys plots as mentioned in Sect. 5.1.2. and in \citet{basilicata_gaps_2024} can indeed be attributed to the eccentricity, as it is absent in the simulations for the circular orbit; (ii) within the noise fluctuations expected for a single realisation (see again Sect. 5.1.1.), the detection significance remains broadly the same for all $\omega$ angles in the case of the clear, circular orbit, but the detection significance drops by almost half for the clear, eccentric orbit case. In the cloudy cases, mirroring the findings of the previous tests, the variation between the circular and eccentric cases are almost negligible as the signal strengths are here severely quenched.

\vspace{0.4cm}
\section{Analysis of real data}\label{section:RealDataAnalysis}

 % ==== Figure for CCF ====
\begin{figure*}
    \centering
    %\vspace{-0.45cm}
    \makebox[\textwidth]{\includegraphics[width=1.0\textwidth]{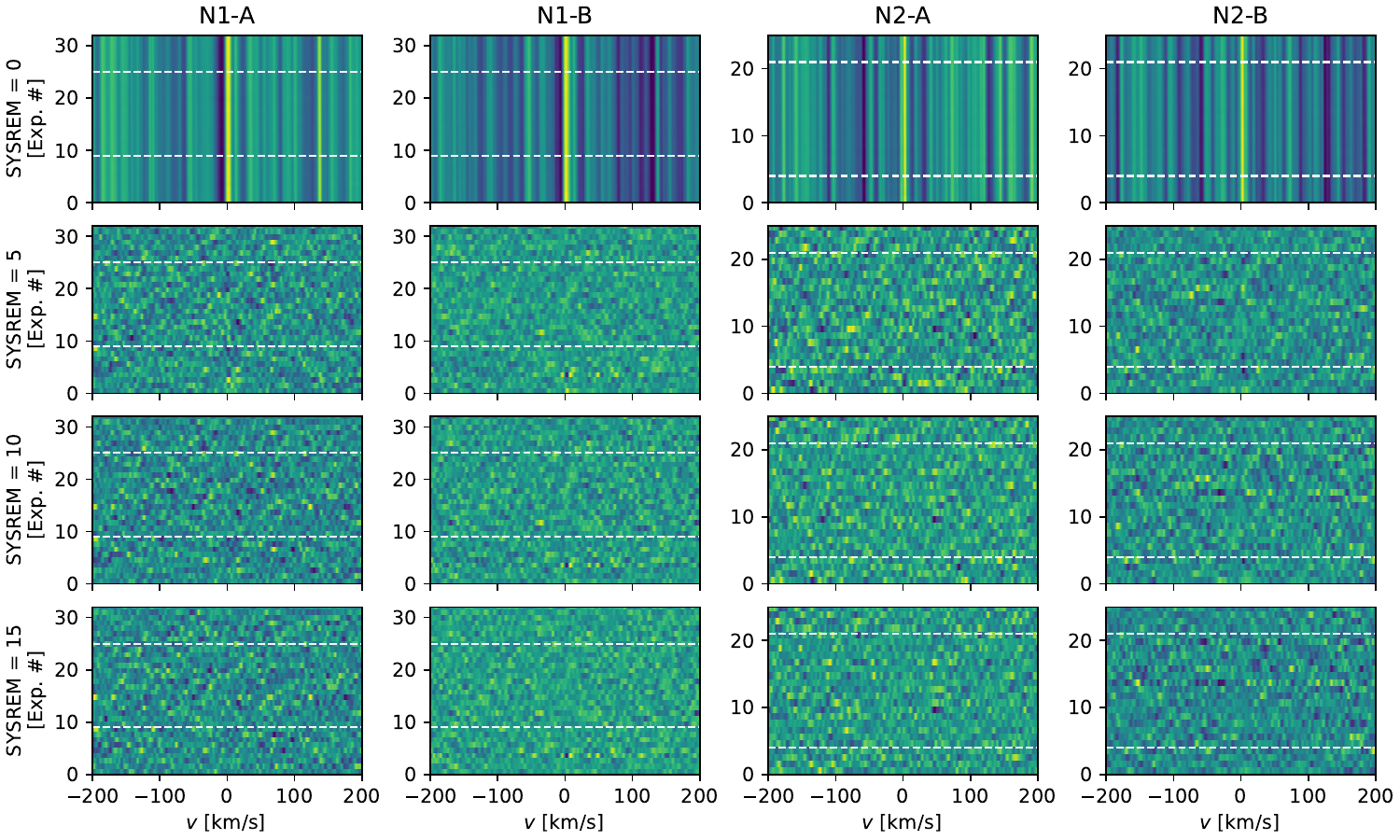}}
    \vspace{-0.5cm}
    \caption{Cross-correlation function (CCF) maps of the A and B frames for Night 1 (N1) and Night 2 (N2) from our VLT/\criresp data. For each plot, the white dotted lines denote the start and end of the transit. The data are shown for SYSREM iterations 0, 5, 10, and 15 in order to demonstrate that all non-zero iterations are clear of any obvious contamination or artefacts. The maps show cross-correlation with the ``global'' template from the CHI-\changeo{model} (see Table \ref{tab:templateParams}) that includes a cloud deck at $P=10^{-5}$ bar.}
    \label{fig:CCF_per_night_frame}
\end{figure*}

\subsection{Cross-correlation analysis of real data}\label{subsection:Real:CC}

Once the two nights of data from VLT/\criresp have been pipeline reduced, wavelength corrected  and detrended, we begin the cross-correlation analysis using the technique and templates described in Sect. \ref{subsection:Method:CCF_templates}. Following this, there are $4 \times 4$ templates to use in our cross-correlation analysis: three templates contain the spectral contribution of the single species (with continuum) \ce{CO}, \ce{H2O}, and \ce{NH3} while a fourth template contains all six species and is referred to as the ``global'' template. The three species included in the global template that do not have their own templates (\ce{CO4}, \ce{CO2}, and \ce{H2S}) were determined to have vanishingly small contributions at our wavelength range – see Fig. \ref{fig:clearVScloudy_pRT} – and thus no analysis was attempted for them individually. Each of the four templates were generated in four models: based on values of the two \citet{welbanks_high_2024} \changeo{model} frameworks \texttt{Aurora} and \texttt{CHIMERA}; and including or not a grey cloud deck at $P_\mathrm{bar} = 10^{-5}$ bar. The templates without clouds are referred to as \texttt{nC} for ``no cloud'' and \texttt{yC} those with as ``yes cloud'', creating the four labels \texttt{Aur+yC}, \texttt{Aur+nC}, \texttt{CHI+yC}, and \texttt{CHI+nC}. All cross-correlations were done in the instrument (telluric) rest frame, except for CO, which was shifted to the stellar rest frame considering the presence of CO in the stellar spectrum of \wasp. 

%\textcolor{red}{(do I need a ref to justify this?)}

% ==== Figure ccf across sysrem - part 1 ====
\begin{figure*}
    \centering
    %\vspace{-0.45cm}
    \makebox[\textwidth]{\includegraphics[width=1.03\textwidth]{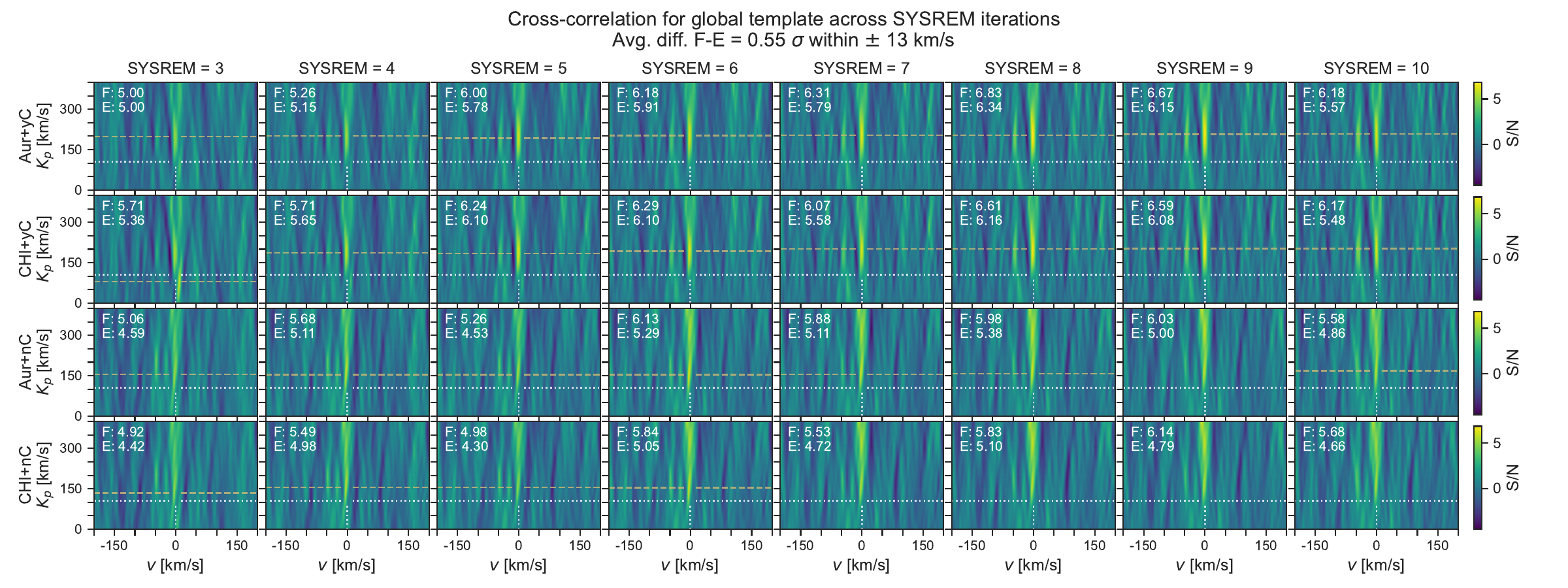}}
   %\vspace{-0.8cm}
    \makebox[\textwidth]{\includegraphics[width=1.03\textwidth]{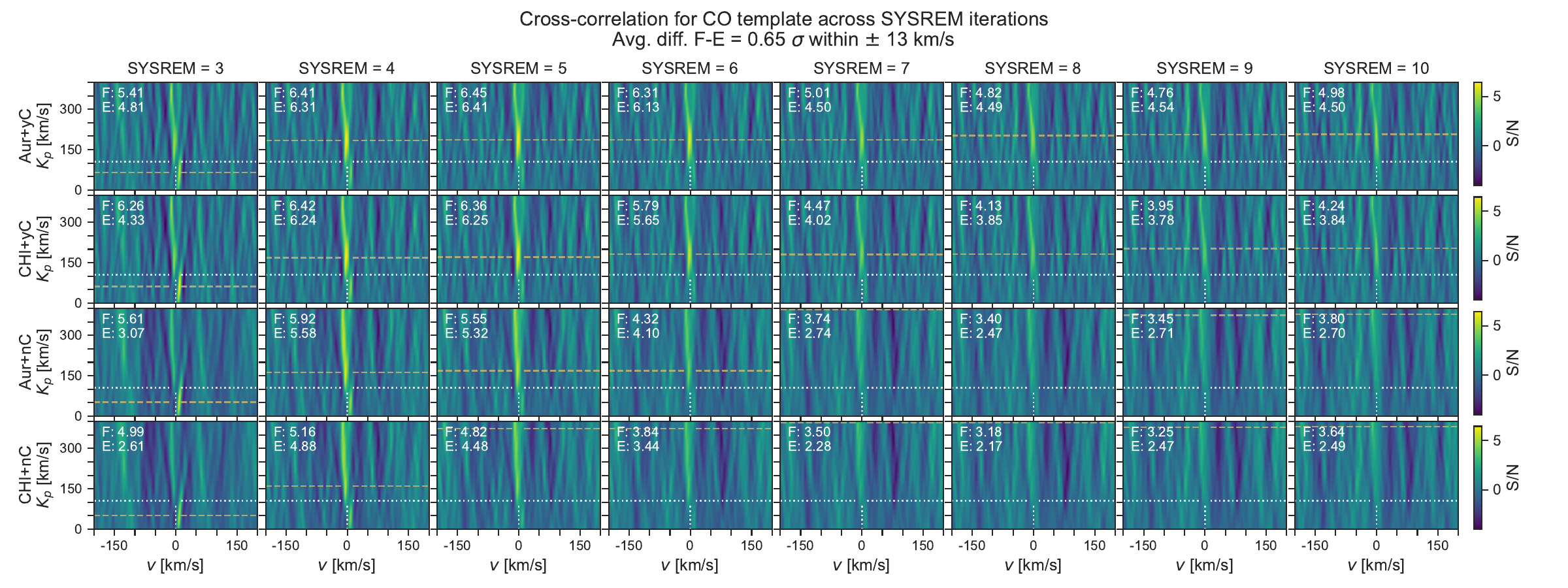}}
   %\vspace{-0.8cm}
    \caption{\kpvsys plots for SYSREM iterations 3-10, cross-correlating with the global template \textit{(top)} and with the CO template \textit{(bottom)}. In each plot, the white dotted line denotes the expected location of the peak\change{, and the yellow dashed line denotes the actual location of the peak}. In the top left, the F-value denotes the maximum value found across the ``full'' map, and the E-value denotes the maximum value found within the ``expected'' location, i.e. within 13 km/s of \kp = 105 km/s and \vsys = 0 km/s.}
    \label{fig:kpvsys_global_CO}
\end{figure*}

% ==== Figure ccf across sysrem - part 2 ====
\begin{figure*}
    \centering
    %\vspace{-0.45cm}
    \makebox[\textwidth]{\includegraphics[width=1.03\textwidth]{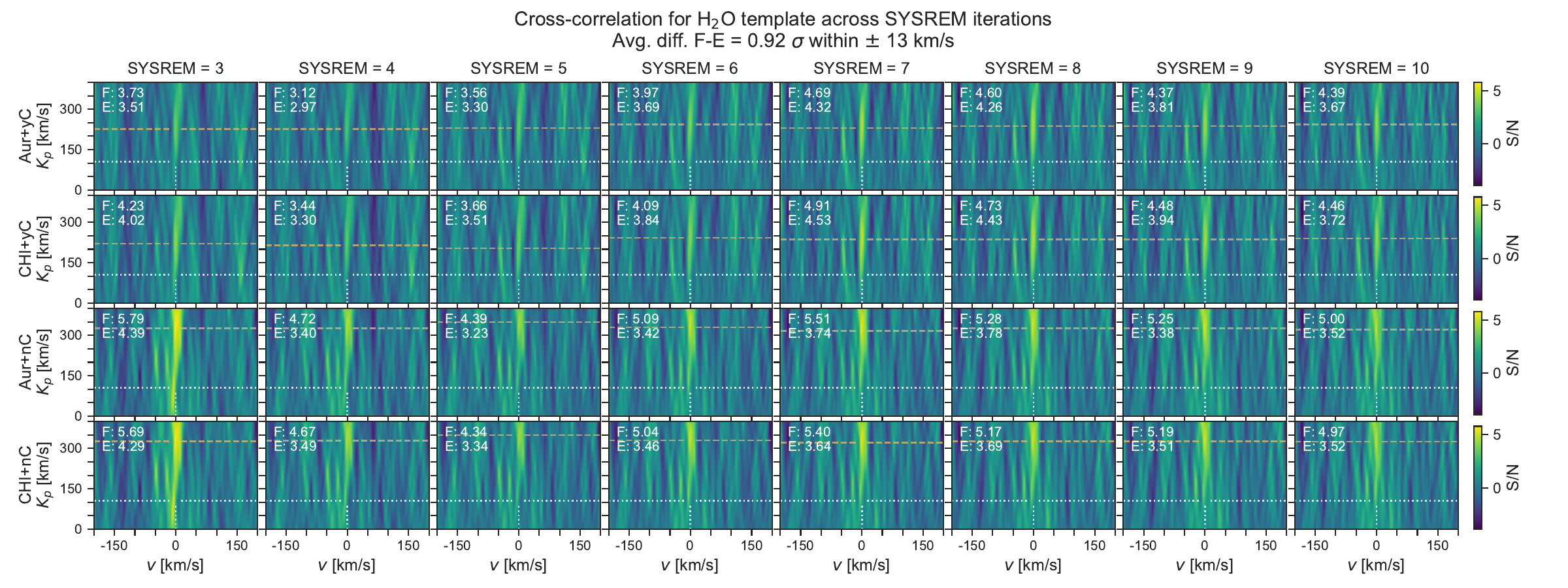}}
   %\vspace{-0.8cm}
    \makebox[\textwidth]{\includegraphics[width=1.03\textwidth]{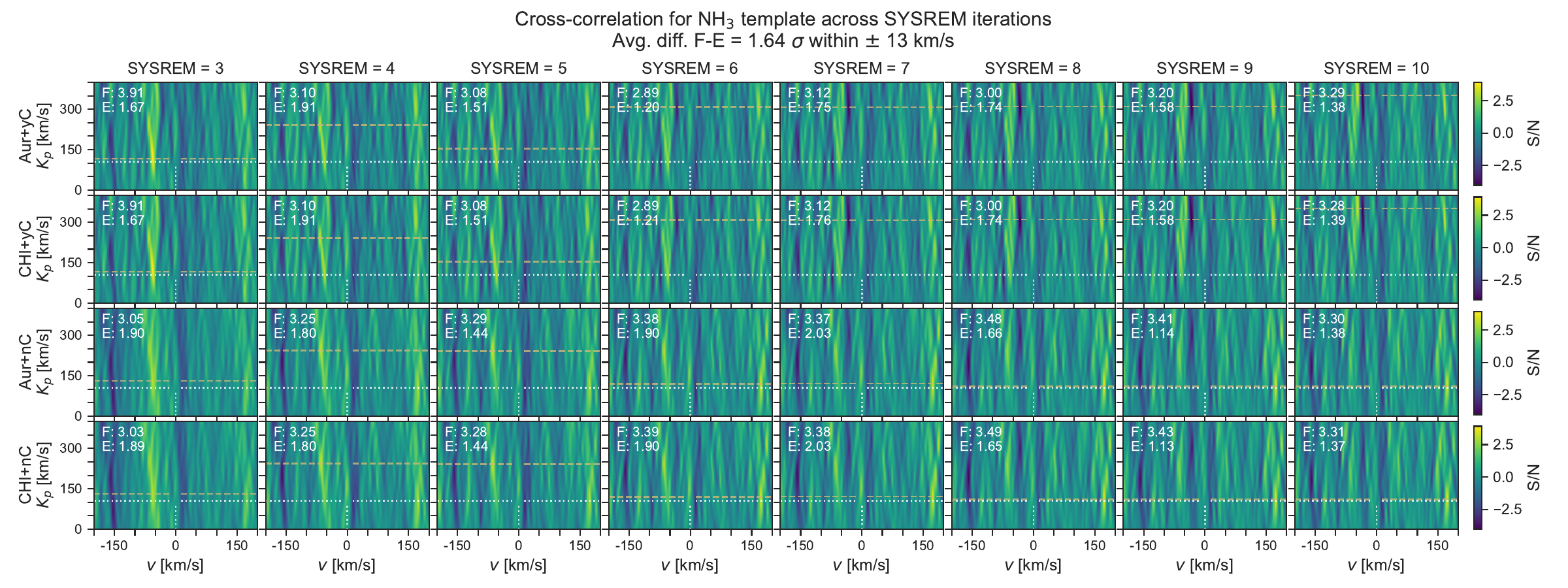}}
   %\vspace{-0.8cm}
    \caption{\kpvsys plots for SYSREM iterations 3-10, cross-correlating with the \ce{H2O} template \textit{(top)} and with the \ce{NH3} template \textit{(bottom)}. In each plot, the white dotted line denotes the expected location of the peak\change{, and the yellow dashed line denotes the actual location of the peak}. In the top left, the F-value denotes the maximum value found across the ``full'' map, and the E-value denotes the maximum value found within the ``expected'' location, i.e. within 13 km/s of \kp = 105 km/s and \vsys = 0 km/s.}
    \label{fig:kpvsys_h2o_nh3}
\end{figure*}

Before creating the \kpvsys plots, we inspected the CCF maps that span the offset of \vsys = $\pm$ 200 km/s for N1-A, N1-B, N2-A, and N2-B as described in Sect. \ref{subsection:Method:CCF_templates:kpvsys}. The data have insufficient S/N to visually show a planetary trace, and so these plots are merely inspected to ensure that there are no obvious artefacts or clear contamination in the data. The CCF plots with global template \texttt{CHI+yC} for telluric frame SYSREM iterations $n$ = 0, 5, 10, 15 are shown in Fig. \ref{fig:CCF_per_night_frame}, where the white dotted lines represent the beginning and end of the transits.

Each template was cross-correlated across a \kpvsys space in the range of \vsys = $\pm$ 200 km/s centred on the stellar RV, then shifted and co-added in steps of 1 km/s across \kp = 0–400 km/s. Thus, the expected position of the \kpvsys peak should be at \kp = 105 km/s (Table \ref{tab:WASP107_s+p_params}) and \vsys = 0 km/s as we \changeo{adopt} the rest frame of the \changeo{exoplanetary system}. For each of the 16 templates, a \kpvsys plot was generated after each iteration of SYSREM, which we run for 15 iterations. Considering the first few SYSREM iterations clearly contain residual stelluric signal, and considering that detection significances drop off towards $\sim$12 iterations, we only plot SYSREM iterations 3–10 for each template. The motivation behind plotting this wide range of SYSREM iterations rather than selecting a single iteration is to demonstrate the consistency of the plots' broader features across multiple iterations. The plots for the global template and the CO template are found in Fig. \ref{fig:kpvsys_global_CO}, and the plots for the \ce{H2O} template and \ce{NH3} template are found in Fig. \ref{fig:kpvsys_h2o_nh3}. 

In these plots, two detection significances are shown in the top left corner of each iteration\change{:} one value for the maximum peak of S/N across the full \kpvsys map, and one value for the  maximum peak of S/N closer to the expected value \change{i.e. the maximum value within a narrow column centred on \vsys = 0 km/s}. These maxima are denoted by ``F'' for ``full'' and ``E'' for ``expected''. The F-maximum is calculated from the entire \kp space, as described in Sect. \ref{subsection:Method:CCF_templates:kpvsys}, i.e. by subtracting the background map (the \kpvsys values outside the central column around $\pm$30 km/s from \vsys = 0 km/s) from the full map, all divided by the standard deviation of the non-central region. For all species except \ce{NH3}, the F-maximum is generally between 3–6\sig, where the F-maximum of \ce{NH3} is generally $<$3\sig, and so \ce{NH3} can thus already at this point be considered a non-detection. However, the F-maximum for all species consistently appears at an offset from the expected \kpvsys maximum position for all templates, and for all SYSREM iterations in the direction of larger \kp.

\change{The E-maximum is then} calculated in the same way as the F-maximum, \change{but this value is instead the maximum from within a box of $N$ = 13 km/s from the expected value in the \kpvsys map. This value of $N$ was chosen by first} calculating the average difference (across all SYSREM iterations and all parameter sets for that species) \change{between the maximum found within boxes of different $N$-values and the previous F-maximum. By} requiring the average difference between \change{these values} to be $<$1\sig, i.e. that both the F-maximum and the E-maximum coordinates are both within where a 1\sig contour would be, \change{we calculate the lowest value of $N$ that fulfils this requirement}. Excluding the non-detection \ce{NH3} maps, we find \change{this $N$-value} to be 13 km/s, and so the E-maximum is then the highest S/N found within this 26$\times$26 km/s box, i.e. within 13 km/s from the expected \kpvsys value. Across all SYSREM iterations and for all parameter-sets in Fig. \ref{fig:kpvsys_global_CO} and Fig. \ref{fig:kpvsys_h2o_nh3}, the F-maximum and E-maximum are shown in the top left of each \kpvsys plot.

The E-maximum is $\sim$6\sig for the global template (maximum = 6.34\sig, SYSREM = 8) and CO templates (maximum = 6.41\sig, \changeo{SYSREM = 5}), and $\sim$4\sig for the \ce{H2O} template (maximum = 4.53\sig, \changeo{SYSREM = 7}). In all three cases, the average E-maximum is higher for the two parameter sets that do include clouds than those that do not, with the average difference in E-maxima for \texttt{yC} and \texttt{nC} respectively being: 0.90\sig for the global template; 1.64\sig for CO; and 0.19\sig for \ce{H2O}.

The differences between the \changeo{retrievals from the two forward models} (Aur versus CHI) have a much smaller impact on the detection significance than the inclusion of clouds (yC versus nC). To separate the effects, we construct four \kpvsys difference maps after each SYSREM iteration from 3 to 10, and compare their average values ($\Delta$Map) and the differences of F-maxima ($\Delta$Max). The purpose of the difference maps is to isolate the effect of the clouds from the effect of \changeo{forward models}, and so we compute the following differences: (Aur+yC) – (Aur+nC) and (CHI+yC) – (CHI+nC) for the clouds and (Aur+yC) – (CHI+yC) and (Aur+nC) – (CHI+nC) for \changeo{models}. We then take the absolute value of the average, $\mu$, for differences at each SYSREM iterations 3 to 10, and calculate the standard deviation, $s$, in that set of differences. A table summarising these values can be found in Table \ref{tab:AvgStdDiffs}.

In this table, it can be seen that the differences of $\Delta$Map have $\mu \sim 0.02$ and $s \sim 0.22$ for \changeo{forward models} (i.e. templates of the same cloud inclusion) but $\mu \sim 0.06$ and $s \sim 0.87$ for differences in cloud inclusions (i.e. templates of the same \changeo{forward models}). For $\Delta$Max, the trend is the same, with $\mu \sim 0.23$ and $s \sim 0.50$ for differences in \changeo{forward models} but $\mu \sim 0.72$ and $s \sim 0.73$ for differences in cloud inclusions. Across the board, this implies that the differences between the inclusion or exclusion of a cloud deck are more impactful than a different volume mixing ratio of a species, even when the ratio is varied by fairly notable quantities (for example, 1.1 dex in the case of CO). This is particularly notable considering that the two retrievals do account for their respective MMW, which differ by 0.38 $u$ (see Table \ref{tab:templateParams}), and HRCCS being seemingly more sensitive to clouds than to MMW may indicate that high-resolution observations could be a way to resolve the metallicity-cloud degeneracy. 

\begin{table}[h!]
\label{tab:AvgStdDiffs}
\centering
\renewcommand{\arraystretch}{1.2}
\resizebox{\columnwidth}{!}{%
\begin{tabular}{llcccccc}
& & \multicolumn{2}{c}{\textbf{Global}} & \multicolumn{2}{c}{\textbf{CO}} & \multicolumn{2}{c}{\textbf{\ce{H2O}}} \\
\cmidrule(lr){3-4} \cmidrule(lr){5-6} \cmidrule(lr){7-8}
& & $\mu$ & $s$ & $\mu$ & $s$ & $\mu$ & $s$ \\
\midrule

\multirow{4}{*}{\textbf{$\Delta$ Map}} 
& (Aur+yC) – (CHI+yC) & 0.01 & 0.40 & 0.03 & 0.24 & 0.03 & 0.18 \\
& (Aur+nC) – (CHI+nC) & 0.01 & 0.22 & 0.01 & 0.18 & 0.00 & 0.09 \\
& (Aur+yC) – (Aur+nC) & 0.05 & 1.04 & 0.05 & 0.70 & 0.11 & 0.96 \\
& (CHI+yC) – (CHI+nC) & 0.04 & 0.98 & 0.04 & 0.65 & 0.09 & 0.90 \\

\midrule

\multirow{4}{*}{\textbf{$\Delta$ Max}} 
& (Aur+yC) – (CHI+yC) & 0.61 & 1.55 & 0.28 & 0.42 & 0.11 & 0.15 \\
& (Aur+nC) – (CHI+nC) & 0.10 & 0.48 & 0.25 & 0.28 & 0.00 & 0.09 \\
& (Aur+yC) – (Aur+nC) & 0.32 & 0.65 & 1.02 & 0.57 & 0.63 & 0.59 \\
& (CHI+yC) – (CHI+nC) & 0.84 & 1.61 & 0.99 & 0.52 & 0.52 & 0.41 \\
\bottomrule
\end{tabular}%
}
\caption{Summary of differences between \kpvsys\ results from the four parameter-set templates for each species. $\Delta$Map is the subtraction of full \kpvsys\ maps, and $\Delta$Max is the subtraction of F-maxima. For both  $\Delta$Map and $\Delta$Max, the first two rows correspond to differences where the cloud inclusion is the same, and the last two rows correspond to differences where the \changeo{forward models} is the same. Values shown are the modulus average ($\mu$) and standard deviation ($s$) across SYSREM iterations 3–10. All values are rounded to 2 d.p.}
\end{table}

\subsection{Investigation of \kpvsys maximum location deviation}\label{subsection:Real:InvestKpvsysOffset}

For all templates, the location of the F-maximum along \kp varied across all SYSREM iterations and parameter sets from as low as $\sim$150 km/s (i.e. $\sim$45 km/s from the expected location of 105 km/s) to as high as $\sim$400 km/s (the edge of \kpvsys map). In the \vsys direction, no significant variation was found, with all species-parameter combinations in all SYSREM iterations generally deviating less than 5\,km/s (most often only by \about1\,km/s). 

The F-maximum location is relatively consistent across all templates, indicating that its offset in \kp from the expected location comes from the data and our analysis. In the global template, the maximum \changeo{mostly} appears at $\sim$150 km/s in the two \texttt{nC} templates\changeo{, except some SYSREM iterations that have F-maxima at the edge of the map,} but \changeo{consistently} at $\sim$200 km/s in the two \texttt{yC} templates. In \ce{CO} and \ce{H2O}, it is instead the clear models that produce a larger deviation, placing the F-maximum closer to $\sim$300 km/s in all four \texttt{nC} templates \changeo{(again with some SYSREM iterations showing the F-maxima at the map edges for CO)} but \changeo{consistently} at $\sim$200 km/s in the two \texttt{yC} templates (ranging from approx. $\sim$180 km/s to $\sim$250 km/s, similar to the global \texttt{yC}-templates). 

% ==== sim w and wo tellurics vs real ====
\begin{figure*}
    \centering
    %\vspace{-0.45cm}
    \makebox[\textwidth]{\includegraphics[width=1.03\textwidth]{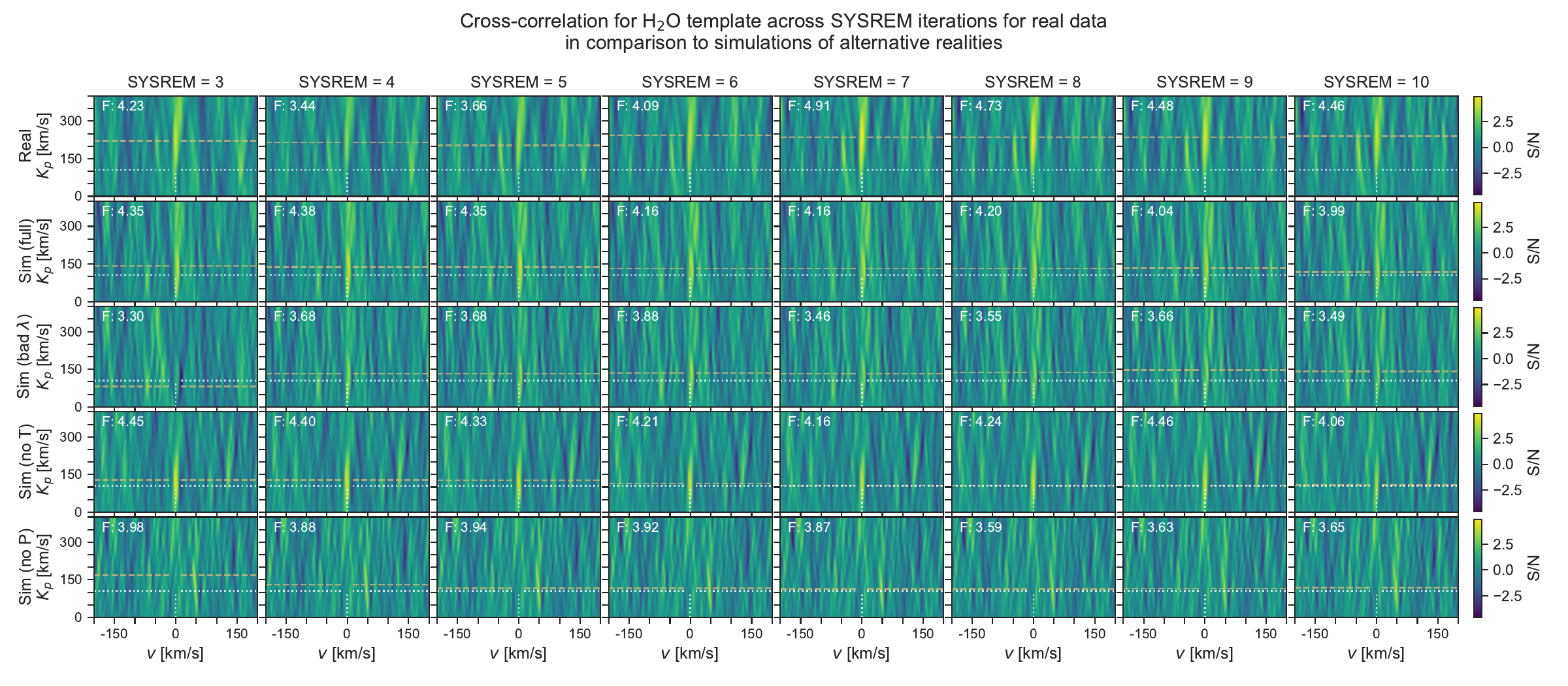}}
    \vspace{-0.45cm}
    \caption{Real and simulated \kpvsys plots for SYSREM iterations 3-10, cross-correlating with the \ce{H2O} template including clouds.  In each plot, the white dotted line denotes the expected location of the peak, and the yellow dashed line denotes the actual location of the peak. \changeo{\textit{First row:} the real data \kpvsys map as shown in Fig. \ref{fig:kpvsys_h2o_nh3}. \textit{Second row:} a simulated observation including all expected components (a ``full'' simulation). \textit{Third row:} a simulated observation with an intentionally perturbed wavelength solution (``bad $\lambda$''). \textit{Fourth row:} a simulated observation excluding telluric contamination (``no T''). \textit{Fifth row:} a simulated observation excluding a planetary atmosphere (``no P'').}  }\label{fig:sim_noTell_yesTell}
\end{figure*}

Some of these commonalities can be seen in Fig. \ref{fig:kpvsys_global_CO} and Fig. \ref{fig:kpvsys_h2o_nh3}. Here, a characteristic ``chimney smoke'' pattern is visible, where the more centralised peak is accompanied by higher S/N at effectively all \kp values above it, and appears to be broadening towards higher \kp values. This pattern varies from being more or less pronounced for different species and at different SYSREM iterations, but this region is most distinct in the \texttt{nC} templates of \ce{H2O} (which also show a larger difference in F-E than any other combination).

To explore whether this effect may be an artefact in the data (possibly from a specific detector, a specific nodding position, etc.), we experimented with excluding different combinations of data from Night 1, data from Night 2, data from A-frames, and data from B-frames. We also tried excluding specific spectral orders in several different permutations, but none of these exclusion tests produced any notable changes in the F-maxima location (only lowering the detection significances). 

Instead, we revisited the simulations that we generated in Sect. \ref{section:SimDataAnalysis}, as we had now determined them to be in relatively good agreement with the real data results. A plot showing \changeo{our new sets of \kpvsys plots can be found in Fig. \ref{fig:sim_noTell_yesTell}. We first took our simulated observation that is as close as possible to our real \wasp observation; this is referred to as the ``full'' simulation (as it includes all three stellar, telluric, and planetary components).} 

\changeo{Next, considering the challenges we faced in our wavelength alignment (Sect. \ref{subsection:Method:WavelengthMorphing}), we took the same simulation but perturbed its wavelength scale by Doppler shifting and interpolating fluxes and uncertainties by RVs drawn from a normal distribution ($\sigma_\mathrm{RV}$ = 100 m/s) before creating new \kpvsys plots from this new data set (labelled ``bad $\lambda$''). By doing the interpolation and Doppler shift after the simulation had been created, noise should be constant between this perturbed simulation and the original ``full'' simulation.} 

\changeo{We then created two new simulated observations: one that did not include any telluric lines but maintained all other parameters, and one that did not include planetary signal (bare rock) but maintained all other parameters. The plots from these simulations are labelled ``no T'' and ``no P'' for no tellurics and no planetary atmosphere respectively. Together, these two final simulations should test if any of the many manifestations, off-sets, and other sources of spurious signal detailed in the paper -- arising from telluric contamination or elsewhere -- could combine to create a signal that appears to be seemingly genuine but is in fact not.}

\begin{figure*}
    \makebox[\textwidth]{\includegraphics[width=1.03\textwidth]{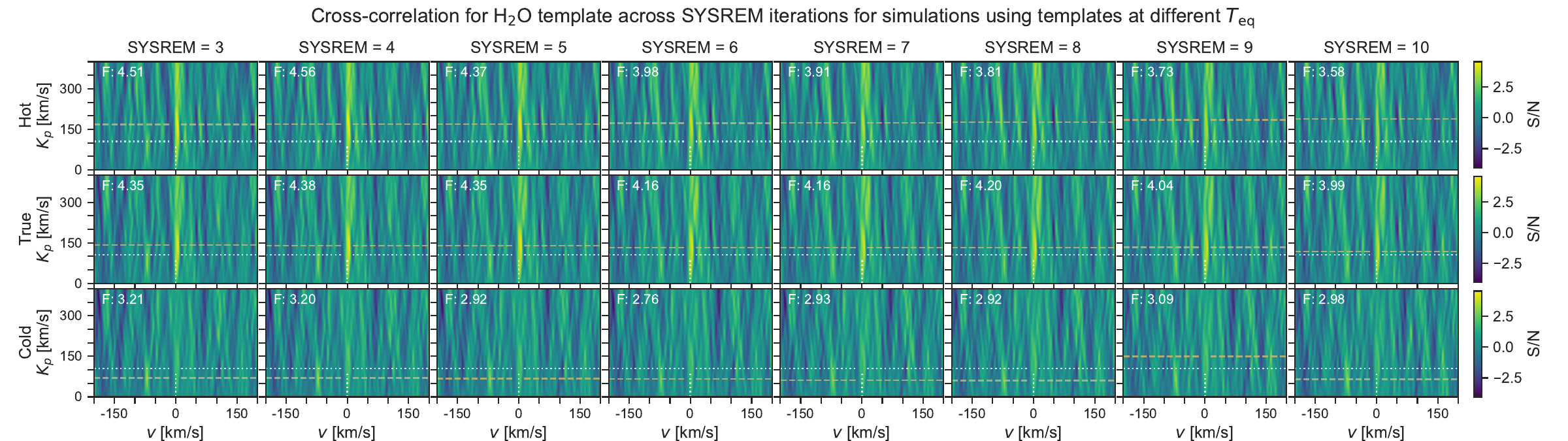}}
    \vspace{-0.45cm}
    \caption{Simulated \kpvsys plots for SYSREM iterations 3-10, cross-correlating with the \ce{H2O} template including clouds. In each plot, the white dotted line denotes the expected location of the peak, and the yellow dashed line denotes the actual location of the peak. All simulated observations include tellurics. \textit{Top row:} cross-correlation with a template with equilibrium temperature $T_\mathrm{eq} = 1\,200$ K (``hot''). \textit{Middle row:} cross-correlation with a template with equilibrium temperature $T_\mathrm{eq} = 738$ K (``true''). \textit{Bottom row:} cross-correlation with a template with equilibrium temperature $T_\mathrm{eq} = 300$ K (``cold'').}
    \label{fig:sim_hotcold}
\end{figure*}

\changeo{Finally, we contrasted all these simulations to what was shown in the real data (``real'').} All \changeo{data sets in Fig. \ref{fig:sim_noTell_yesTell}, both real and simulated,} are cross-correlated with the \texttt{CHI+yC} template for \ce{H2O}; and in the simulations, the simulated atmosphere is the \texttt{CHI+yC} global template.

In \changeo{Fig. \ref{fig:sim_noTell_yesTell}}, it becomes clear that the ``chimney smoke'' pattern is present in the \changeo{full simulation including telluric lines, and that the simulation without tellurics shows no such feature.} In the \changeo{full} simulation, it also becomes obvious that the effect is not actually an extension of the signal itself, but rather its own region of increased signal that happens to be located relatively near the true detection maximum in the expected location. In the real data, the small gap between the ``smoke'' and the true detection becomes less pronounced or vanishes, possibly due to noise properties of the data. One also notices that the F-maximum (denoted by the yellow dashed lines) is closer to the expected location (the white dotted lines) in the \changeo{no T}-simulation than in the \changeo{full} simulation, but that the shift in \kp is smaller for both simulations compared to the real data.

\changeo{Our bad $\lambda$-simulation shows that a poor wavelength alignment will indeed reduce our detection significance by at least $\sim$0.5--1\sig  without changing its location. Interestingly enough, the chimney smoke effect also appears reduced here. This could be interpreted as the chimney smoke being in some way a manifestation of real signal, but it is not clear how this would combine with the finding above that removing telluric lines also reduces the presence of chimney smoke.}

\changeo{In our no P-simulation, we find that no signal of significant S/N -- or chimney smoke effect -- can be detected anywhere near the expected location, indicating that our signal found along \vsys = 0 km/s is indeed genuine. The F-maximum of this map reaches towards, but is never at or above, 4.0\sig\xspace -- but importantly, that signal is clearly from a manifestation that is significantly removed from the expected location in \vsys-space. In our \kpvsys map, the no P-case \textit{never} results in any signal above 2.0\sig within $\pm$ 40 km/s of \vsys = 0 km/s for any of the SYSREM iterations shown. In contrast, for all our simulations containing planetary features (and for our real data), the signal \textit{always} appears within approx. $\pm$ 5 km/s of \vsys (and most often at 0--1 km/s). This significantly supports our interpretation that the signal in our real data is in fact genuine.} 

\changeo{Out of concern for the spurious > 3\sig signal appearing in our no P-simulation, we then created three more such simulations of a bare rock, i.e. only modelling a transiting sphere of radius $R_\mathrm{p}$ with no wavelength-dependent absorption. We find that with each new simulation, and thus each new noise realisation, the location of this artefact (and often-times multiple artefacts) changes location and strength at random, appearing minimally at 2.6\sig  and maximally at 3.9\sig  (average and median both 3.5) across the different bare rock simulations for SYSREM = 3 to SYSREM = 10. In terms of displacement in the \vsys-space, artefacts appear between $\sim \pm\,30-120$\,km/s from \vsys = 0, with their respective strengths varying across SYSREM iterations. This is to be expected considering that, as explored in detail in \citet{boldt-christmas_optimising_2024} and as mentioned in Sect. \ref{subsection:SimData:DetSigFluc}, there is evidence of the precedent that different noise realisations can have major impacts. Nonetheless, the persistence of the planet signal at (roughly) the expected location in all other simulations, aligned with what we see in the real data, gives us confidence that the potential for spurious signals there do not give us less confidence in our planet detection.}

Following \changeo{these findings}, we considered that \changeo{the displaced chimney smoke} effect could perhaps be related to the relative line strengths of the molecular bands in the \ce{H2O} template. This should differ for different values of equilibrium temperature by changing the level populations of water molecules and thus the relative strength of different groups of lines. We therefore created two more sets of \kpvsys maps from the \changeo{full} simulation, this time using new generations of the \texttt{CHI+yC} template that now changed $T_\mathrm{eq}$ from 738 K to $T_\mathrm{eq}$ = 1\,200 K and $T_\mathrm{eq}$ = 300 K (i.e. approx. $\pm$ 450 K from the initial $T_\mathrm{eq}$). The result of cross-correlating the same simulated observation with these templates of different $T_\mathrm{eq}$ values (labelled ``hot'' for 1\,200 K, ``true'' for 738 K, and ``cold'' for 300 K) are shown in Fig. \ref{fig:sim_hotcold}. In this plot, it can be seen that an incorrect template clearly affects both detection significance and \kp of the F-maximum, where the expected \kpvsys maximum and the F-maximum are once again denoted by white dotted lines and yellow dashed lines respectively. Here, the effect is somewhat more pronounced and produces an F-maximum at higher \kp for the hotter template, and is less pronounced and at lower \kp for the cold template. The F-maxima for the hot template peak at earlier SYSREM iterations, and for the cold template at later ones, but both have lower significance than with the true template.

%----------------------------------------------------
\section{Results and discussion}
    \label{section:ResultsDiscussion}

In this section, we summarise our results and provide some further context, \changeo{hypotheses}, and proposed interpretations of what has been demonstrated in this study. The regime of relatively cool gas giants is clearly understudied compared to hot and ultra-hot Jupiters (as discussed in Sect. \ref{section:SimDataAnalysis}), and this is even more true for HRCCS analyses. In this study, a dedicated data reduction effort was required for handling super-resolution (Sect. \ref{subsection:Method:WavelengthMorphing}), and the use of simulations informed our interpretation (Sect. \ref{subsection:SimData:Simulations} and \ref{subsection:Real:InvestKpvsysOffset}), with the combination of both ultimately leading to our \change{detection of cross-correlation signal}. Only with multiple previous studies having repeatedly confirmed the presence of species like \ce{H2O}, essentially guaranteeing its existence somewhere in the data, and with reassurances from our simulations that the detection should not be particularly high even under ideal circumstances, was there an incentive to explore this data set more thoroughly. As such, one of the key results of this paper is a firm recommendation that authors of other similar, future studies try to reproduce this environment of strong guidance from simulations and previous literature if attempting to study targets whose signal strengths are anticipated to be relatively low for whatever reason (e.g. cooler, cloudier, smaller, or some combination).

Through this process, \change{we arrive at the reporting of} two individual species, CO and \ce{H2O} at detection significances of $\sigma \gtrsim$ 4.5, in the atmosphere of \wasp using VLT/\criresp data \change{near the expected \kpvsys location}. This is in line with previous space-based studies that have also found these species to be present, and with these detections, we confirm that transmission HRCCS is indeed able to detect molecular species in warm planetary atmosphere regimes of $T_\mathrm{eq}\,<\,800$\,K. While we are not able to identify any other individual species, we also confirm that HRCCS using templates including all six species based on the parameters of \citet{welbanks_high_2024} provides a \change{signal of $\sigma \sim$ 6, with the maxima found near the expected \kpvsys location but at an off-set in the \kp direction}. Our results also indicate a non-detection of \ce{NH3} in spite of this species being confirmed by previous studies; however, this is not particularly surprising considering its low amount of spectroscopic signal at our wavelength range. 

It is important to underline the context in which these detections are claimed. Under other circumstances, e.g. if \wasp had been a less-studied target for which no previous detections had been made, our detections would have lacked the necessary context to favour one interpretation over others. However, thanks to the wealth of existing space-based literature to build upon, these species have already been detected in this planet multiple times and so their physical existence is known a priori. Combining this with our simulations demonstrating that our detection significances are in line with expectations, we do consider these detections to be genuine, even if caution is advised against showing the same degree of confidence for lesser studied targets.

With the previous space-based literature providing many atmospheric parameters, we shifted our focus towards investigating the sensitivity of our detections to the properties of the template. In particular, we compared the impact of the two \changeo{forward models} from \citet{welbanks_high_2024} and the inclusion of clouds. We discovered that clouds play a much more significant role in HRCCS, while different \changeo{forward models} have only a marginal effect. This points to the importance of combining broadband low-resolution spectra from space with high-resolution ground-based observations to get the full picture, echoing the sentiments in Sect. \ref{section:Introduction} on the benefits of the two types of data.

Proving the success of HRCCS for cooler targets is crucial to the future of the field, considering that high-resolution spectroscopy is our best method for analysing atmospheric circulation such as winds and jet streams. In light of the great interest within the community regarding the dynamics of atmospheres of planets such as \wasp, honing our ability to analyse exoplanetary atmospheric spectra at high spectral resolution for cooler targets is the key to future success in this research area. This study provides important insight to some of the obstacles and challenges that come with this novel and thus relatively unexplored regime, so the methodology and lessons learned from this work will hopefully be of use to such future studies.

The obstacle that we are unable to describe fully in this paper is the deviation of the F-maximum from the expected location (Sect. \ref{subsection:Real:InvestKpvsysOffset}). Our interpretation is not that this is a physical offset where the planetary signal peaks at a significantly different \kp, but rather that this is an area of additional higher S/N above the expected location in the \kp direction of our \kpvsys plots. Future work will be required to fully understand and explain its origin, but considering its partial appearance in our simulations, it can be speculated that it must arise due to a combination of both intrinsic and extrinsic reasons. Below, we discuss some primary suspects based on findings and tests so far:

(i) \textit{Poor choice of templates}: The simulations used to test the origin of the ``chimney smoke'' artefact show that using an incorrect atmospheric model may create trails in the \kpvsys plots very similar to what we get from the real data (see Fig. \ref{fig:sim_hotcold}). This could be an indicator of a problem arising from our choice of planetary atmospheric model used to generate our templates,  from which there are many parameters that may be impactful beyond the equilibrium temperature tested in Sect. \ref{subsection:Real:InvestKpvsysOffset}. This question certainly requires further investigation before we embark on more HRCCS analysis of cooler gas giants. This is especially true considering our very simplistic treatment of models, and the clear impact of template selection seen in the simulated tests in Sect. \ref{subsection:SimData:Simulations}, and the fact that the clouds of \wasp are still being researched. Considering our finding that the inclusion of clouds is arguably the parameter that has the most striking impact on our detection significance, it can be expected that more accurate cloud models may improve detection significance further, although the extent to which this is true will presumably vary for different targets and wavelength ranges.

(ii) \textit{Residual stellurics}: Our simulations demonstrated that the \kp artefact could also be caused by the presence of tellurics (Fig. \ref{fig:sim_noTell_yesTell}). It is reasonable to assume that SYSREM residuals remain correlated with tellurics, causing a similar effect in the real data analysis. This is further supported by the fact that the artefact is most pronounced in the cross-correlation with the \ce{H2O} \texttt{nC} templates (which are a close match to the water vapour lines in the Earth atmosphere) and by \citet{gandhi_seeing_2020} who also finds that \ce{H2O} detections with HRCCS should be diminished due to telluric absorption obscuring the signal for cloudy cases in particular. In reality, the artefact may be caused by the combination of erroneous template (based on overly simplistic atmospheric model) and SYSREM residuals. Stelluric contamination and the challenges of its removal is a well-known obstacle at infrared wavelengths \citep[see e.g.][and discussions within]{maguire_assessing_2024}, and so it is not surprising that this may have been done imperfectly – or at least imperfectly to a degree that perhaps would have been acceptable or negligible in a hot Jupiter study, but not for this target where the error budget is significantly depleted as our detection is closer to the noise floor.

(iii) \textit{SYSREM performance across spectral orders}:  Across our wavelength range, telluric contamination changes dramatically in strength and origin, where the middle spectral orders have relatively little tellurics. \ce{H2O} is present everywhere, but changes its strength dramatically, while \ce{CO2} and \ce{CH4} also contribute significantly to some spectral intervals. Considering these differences across wavelength, different spectral orders will almost certainly require different numbers of SYSREM iterations before they are fully detrended – but at each iteration, SYSREM will not have fully removed tellurics in certain spectral intervals, while it will start ``eating'' into planetary signal in the others. In the future, a robust estimate of convergence for SYSREM should be developed and be ran on a subset of wavelength points that are expected to behave similarly (i.e. points mostly affected by only one species), as the amount of telluric absorption may change throughout a transit due to changes in humidity and temperature.

(iv) \textit{Wavelength solution}: The wavelength solution for \criresp is not perfect, and even the fine tuning described in Sect. \ref{subsection:Method:WavelengthMorphing} may still have some (small) contribution to the artefact that we see in our results \changeo{-- although our test exploring this in Fig. \ref{fig:sim_noTell_yesTell} indicates that a poor wavelength solution should only impact our detection significance, not necessarily create offsets}. Future work should investigate the impact of this on our analysis, both in the context of SYSREM performance and possible impact on evaluating CCF.

(v) \textit{Physical effects}: It is worth noting that there are also physical reasons to expect a deviation of the maximum S/N in a \kpvsys plot. In \citet{wardenier_modelling_2023}, the location of maxima in \kpvsys plots for hot Jupiters are explored, and they find that several atmospheric effects such as morning-to-evening limb variation and cloud decks can create offsets of up to $\sim$20 km/s due to the combination of several effects arising from the 3D nature of the exoplanet atmosphere. In this work, they underline the fact that a \kp offset of a species merely reflects the \textit{rate of change} of its Doppler shift in the planetary rest frame, and as such, $\Delta$\kp can be quite large compared to $\Delta$\vsys. The relatively small \kp of a smaller planet like \wasp (compared to hot Jupiters) also creates a challenge as the lower the velocity of the planet, the harder it is to distinguish the planetary signal from stelluric lines, compounding the concern raised in point (iii) above. As such, it can be rationalised that at least a partial offset in the peak signal is to be expected, but further work will be required to differentiate this from other possible effects \changeo{-- especially considering that these effects should not be significant enough to explain an offset of our magnitude on their own}.

For all of the above, there is a thread of commonality in that these are all aspects that will realistically exist in all high resolution transmission spectroscopy studies to some extent – but at relatively low S/N where each of these effects may only impact the result by $\lesssim$2 \sig, meaning they should be present but not decisive for many studies. The difference is that for this target and observation, only a relatively low maximum S/N can be anticipated even under ideal circumstances as demonstrated by our simulations. As such, for targets such as \wasp, the error budget is too limited for these effects to be affordably overlooked, and so they must be accounted for – at least qualitatively. This finding is also applicable to observations of hotter (or otherwise more favourable) targets in the search for species of fainter signals, where the same factors become relevant. The impact of a vast range of uncertainties in HRCCS is thoroughly explored in \citet{savel_peering_2025}, and the interplay between HRCCS detections and laboratory data (line lists) is examined in a recent review by \citet{yurchenko_data_2025}, with many previous studies already acknowledging the many challenges associated with smaller, cooler, and/or cloudy planets in particular \citepeg{molliere_detecting_2019,hood_prospects_2020,finnerty_keck_2023,dubey_quantified_2025}. In practice, a number of recent attempts at transmission HRCCS for smaller planets resulting in non-detections also shed light on the challenges involved, such as \citet{grasser_peering_2024}, \citet{dash_constraints_2024}, and \citet{parker_limits_2025}.

It has also been found that detection significances in HRCCS are broadly speaking punished in the use of templates that do not match the reality of the target atmosphere. However, not all parameters are punished (or rewarded) equally, and this work finds that certain parameters such as volume mixing ratios and thus mean molecular weight are less impactful on the cross-correlation detection than e.g. correctly accounting for cloud inclusion and temperature $T_\mathrm{eq}$. This should be considered in work on the metallicity-cloud degeneracy, as this finding implies that cross-correlating with the two types of template for a target where this degeneracy is unresolved may help us determine which is a better fit and therefore reality. These factors appear to be somewhat species-dependent, reflecting the diversity of possible spectral bands, and will thus also vary across exact wavelength ranges. Considering that $T_\mathrm{eq}$ is largely calculated from the radius of the host star, and considering that stellar radius is difficult to constrain for cooler stars such as M-dwarfs \citepeg{shields_habitability_2016,parsons_scatter_2018,cassisi_effective_2019}, this consideration is particularly important for planets around cooler stars. In transmission especially, this needs to be accounted for carefully against the concern of spot-crossings, which is inherently more of a concern in cooler stars and has already been found to be a concern for this target by \citet{murphy_panchromatic_2025}. 

%----------------------------------------------------
\section{Conclusions}
    \label{section:Conclusions}

    In this study, we characterise the atmosphere of the warm Neptune-like exoplanet \wasp using two transit observations from VLT/\criresp in the K-band. We use cross-correlation to confirm the detection of two individual species, \ce{CO} at \about 6\sig and \ce{H2O} at \about 4.5\sig, in \wasp within a reasonable distance (13 km/s) of its expected location in the \kpvsys detection map. We confirm that the global transmission spectrum as presented by \citet{welbanks_high_2024} can also be detected at \about 6\sig within the same distance from the expected location. We also search for \ce{NH3} but do not detect it ($\sim$1–2\sig). By the use of simulations, we demonstrate that these findings are in line with the expected detection significances for these species, for this target, and for these observations. These detections represent the first molecular detections made using HRCCS with transmission spectra for a target of $T_\mathrm{eq}$$<$\,800 K, marking the community moving towards the characterisation of smaller, cooler exoplanets. 

Through both simulations and analysis of real data, we demonstrate that our HRCCS analysis is sensitive to the inclusion of a cloud deck in our cross-correlation templates, and to the equilibrium temperature of the template, yet not particularly sensitive to the exact volume mixing ratio of different species from space-based retrievals. This is true even for the maximum difference in volume mixing ratio between the two \changeo{forward models} we tested, which is 1.1 dex for CO. We also find that our maximum \kpvsys peak deviates from its expected \kp location, which we speculate to arise from a combination of possible reasons such as insufficient removal of stellar and telluric signal, and the model template equilibrium temperature. Our interpretation is that this is not a true offset but rather an artefact of our HRCCS analysis. Understanding its exact nature will need further work.

As the combination of new instrumentation and maturing methodology brings research further into the realm of characterising smaller and cooler exoplanets with high-resolution spectroscopy, the parameter space of possible exoplanetary atmospheres expands – and with it, the parameter space of uncertainties. This diversification marks exciting new scientific frontiers, but associated with this reduced error budget are many sources of noise and uncertainty that graduate from being negligible to impactful. Only by mapping out these pitfalls may we avoid them, and only then will we continue our progression towards characterising even cooler and smaller exoplanet targets with high-resolution transmission spectroscopy.

%----------------------------------------------------
\begin{acknowledgements}
    We thank Bengt Edvardsson for computing our custom model atmosphere of the star WASP-107. L.B.-Ch., A.D.R., and N.P. acknowledge support by the Knut and Alice Wallenberg Foundation (grant 2018.0192). \changeo{A.D.R acknowledges support from  ANID/Fondo ALMA 2024/N°31240064.} F.L. acknowledges the support by the Deutsche Forschungsgemeinschaft (DFG, German Research Foundation) – Project number 31466515. D.C. is supported by the LMU-Munich Fraunhofer-Schwarzschild Fellowship and by the Deutsche Forschungsgemeinschaft (DFG, German Research Foundation) under Germany's Excellence Strategy -- EXC 2094 -- 390783311. O.K. acknowledges support by the Swedish Research Council (grant agreement no. 2023-03667) and the Swedish National Space Agency. U.H. acknowledges support from the Swedish National Space Agency (SNSA/Rymdstyrelsen). M.R. acknowledges the support by the DFG priority program SPP 1992 ``Exploring the Diversity of Extrasolar Planets'' (DFG PR 36 24602/41). D.S. acknowledges financial support from the project PID2021-126365NB-C21(MCI/AEI/FEDER, UE) and from the Severo Ochoa grant CEX2021-001131-S funded by MCIN/AEI/ 10.13039/501100011033. \changeo{E.N. acknowledges the support from the Deutsches Zentrum für Luft- und
Raumfahrt (DLR, German Aerospace Center) - project number 50OP2502.} Based on observations collected at the European Organisation for Astronomical Research in the Southern Hemisphere under ESO programme 108.C-0267(D) and 110.C-4127(D). CRIRES+ is an ESO upgrade project carried out by Thüringer Landessternwarte Tautenburg, Georg-August Universität Göttingen, and Uppsala University. The project is funded by the Federal Ministry of Education and Research (Germany) through Grants 05A11MG3, 05A14MG4, 05A17MG2 and the Knut and Alice Wallenberg Foundation. This work was co-funded by the European Union (ERC-CoG, EVAPORATOR, Grant agreement No. 101170037). Views and opinions expressed are however those of the author(s) only and do not necessarily reflect those of the European Union or the European Research Council. Neither the European Union nor the granting authority can be held responsible for them. This research has made use of the NASA Exoplanet Archive, which is operated by the California Institute of Technology, under contract with the National Aeronautics and Space Administration under the Exoplanet Exploration Program. This work has also made use of \texttt{EsoRex} \citep{eso_cpl_development_team_esorex_2015} and the following Python packages: \texttt{Astropy} \citep{astropy_collaboration_astropy_2013}, \texttt{iPython} \citep{perez_ipython_2007}, \texttt{Matplotlib} \citep{hunter_matplotlib_2007}, \texttt{NumPy} \citep{harris_array_2020}, \texttt{Pandas} \citep{mckinney_data_2010,team_pandas-devpandas_2023}, \texttt{PyAstronomy} \citep{czesla_pya_2019}, \texttt{SciPy} \citep{virtanen_scipy_2020}, and \texttt{Seaborn} \citep{waskom_mwaskomseaborn_2017}. \changeo{Finally, we wish to thank the anonymous referee of this paper, whose questions and suggestions has significantly improved this manuscript}. 
\end{acknowledgements}

\bibliography{2024CRIRESpaperLBC.bib} % your references Yourfile.bib

@article{molliere_petitradtrans_2019,
	title = {{petitRADTRANS} - {A} {Python} radiative transfer package for exoplanet characterization and retrieval},
	volume = {627},
	copyright = {© ESO 2019},
	issn = {0004-6361, 1432-0746},
	url = {https://www.aanda.org/articles/aa/abs/2019/07/aa35470-19/aa35470-19.html},
	doi = {10.1051/0004-6361/201935470},
	language = {en},
	urldate = {2022-09-01},
	journal = {A\&A},
	author = {Molliere, P. and Wardenier, J. P. and Boekel, R. van and Henning, Th and Molaverdikhani, K. and Snellen, I. a. G.},
	month = jul,
	year = {2019},
	eid = {A67},
}

@article{piaulet_wasp-107bs_2021,
	title = {{WASP}-107b's {Density} {Is} {Even} {Lower}: {A} {Case} {Study} for the {Physics} of {Planetary} {Gas} {Envelope} {Accretion} and {Orbital} {Migration}},
	volume = {161},
	issn = {1538-3881},
	shorttitle = {{WASP}-107b's {Density} {Is} {Even} {Lower}},
	url = {https://doi.org/10.3847/1538-3881/abcd3c},
	doi = {10.3847/1538-3881/abcd3c},
	language = {en},
	number = {2},
	urldate = {2022-09-01},
	journal = {AJ},
	author = {Piaulet, Caroline and Benneke, Björn and Rubenzahl, Ryan A. and Howard, Andrew W. and Lee, Eve J. and Thorngren, Daniel and Angus, Ruth and Peterson, Merrin and Schlieder, Joshua E. and Werner, Michael and Kreidberg, Laura and Jaouni, Tareq and Crossfield, Ian J. M. and Ciardi, David R. and Petigura, Erik A. and Livingston, John and Dressing, Courtney D. and Fulton, Benjamin J. and Beichman, Charles and Christiansen, Jessie L. and Gorjian, Varoujan and Hardegree-Ullman, Kevin K. and Krick, Jessica and Sinukoff, Evan},
	month = jan,
	year = {2021},
	eid = {70},
}

@article{smette_molecfit_2015,
	title = {Molecfit: {A} general tool for telluric absorption correction - {I}. {Method} and application to {ESO} instruments},
	volume = {576},
	copyright = {© ESO, 2015},
	issn = {0004-6361, 1432-0746},
	shorttitle = {Molecfit},
	url = {https://www.aanda.org/articles/aa/abs/2015/04/aa23932-14/aa23932-14.html},
	doi = {10.1051/0004-6361/201423932},
	language = {en},
	urldate = {2022-09-01},
	journal = {A\&A},
	author = {Smette, A. and Sana, H. and Noll, S. and Horst, H. and Kausch, W. and Kimeswenger, S. and Barden, M. and Szyszka, C. and Jones, A. M. and Gallenne, A. and Vinther, J. and Ballester, P. and Taylor, J.},
	month = apr,
	year = {2015},
	eid = {A77},
}

@article{tamuz_correcting_2005,
	title = {Correcting systematic effects in a large set of photometric light curves},
	volume = {356},
	issn = {0035-8711},
	url = {https://doi.org/10.1111/j.1365-2966.2004.08585.x},
	doi = {10.1111/j.1365-2966.2004.08585.x},
	number = {4},
	urldate = {2022-09-01},
	journal = {MNRAS},
	author = {Tamuz, O. and Mazeh, T. and Zucker, S.},
	month = feb,
	year = {2005},
	eid = {1466--1470},
}

@INPROCEEDINGS{guillot_giant_2022,
       author = {{Guillot}, T. and {Fletcher}, L.~N. and {Helled}, R. and {Ikoma}, M. and {Line}, M.~R. and {Paramentier}, V.},
        title = "{Giant Planets from the Inside-Out}",
    booktitle = {Protostars and Planets VII},
         year = 2023,
       series = {ASP Conf. Ser.},
       volume = {534},
        month = jul,
        pages = {947},
          doi = {10.48550/arXiv.2205.04100},
archivePrefix = {arXiv},
       eprint = {2205.04100},
 primaryClass = {astro-ph.EP},
       adsurl = {https://ui.adsabs.harvard.edu/abs/2023ASPC..534..947G}
}

@article{gao_aerosols_2021,
	title = {Aerosols in {Exoplanet} {Atmospheres}},
	volume = {126},
	issn = {0148-0227},
	url = {https://ui.adsabs.harvard.edu/abs/2021JGRE..12606655G},
	doi = {10.1029/2020JE006655},
	urldate = {2022-09-19},
	journal = {J. Geophys. Res. (Planets)},
	author = {Gao, Peter and Wakeford, Hannah R. and Moran, Sarah E. and Parmentier, Vivien},
	month = apr,
	year = {2021},
	eid= {e06655},
}

@article{birkby_detection_2013,
	title = {Detection of water absorption in the day side atmosphere of {HD} 189733 b  using ground-based high-resolution spectroscopy at 3.2μm.},
	volume = {436},
	issn = {0035-8711},
	url = {https://ui.adsabs.harvard.edu/abs/2013MNRAS.436L..35B},
	doi = {10.1093/mnrasl/slt107},
	urldate = {2022-09-21},
	journal = {MNRAS},
	author = {Birkby, J. L. and de Kok, R. J. and Brogi, M. and de Mooij, E. J. W. and Schwarz, H. and Albrecht, S. and Snellen, I. A. G.},
	month = nov,
	year = {2013},
	eid= {L35--L39},
}

@article{ehrenreich_nightside_2020,
	title = {Nightside condensation of iron in an ultrahot giant exoplanet},
	volume = {580},
	issn = {1476-4687},
	doi = {10.1038/s41586-020-2107-1},
	language = {eng},
	number = {7805},
	journal = {Nature},
	author = {Ehrenreich, David and Lovis, Christophe and Allart, Romain and Zapatero Osorio, María Rosa and Pepe, Francesco and Cristiani, Stefano and Rebolo, Rafael and Santos, Nuno C. and Borsa, Francesco and Demangeon, Olivier and Dumusque, Xavier and González Hernández, Jonay I. and Casasayas-Barris, Núria and Ségransan, Damien and Sousa, Sérgio and Abreu, Manuel and Adibekyan, Vardan and Affolter, Michael and Allende Prieto, Carlos and Alibert, Yann and Aliverti, Matteo and Alves, David and Amate, Manuel and Avila, Gerardo and Baldini, Veronica and Bandy, Timothy and Benz, Willy and Bianco, Andrea and Bolmont, Émeline and Bouchy, François and Bourrier, Vincent and Broeg, Christopher and Cabral, Alexandre and Calderone, Giorgio and Pallé, Enric and Cegla, H. M. and Cirami, Roberto and Coelho, João M. P. and Conconi, Paolo and Coretti, Igor and Cumani, Claudio and Cupani, Guido and Dekker, Hans and Delabre, Bernard and Deiries, Sebastian and D'Odorico, Valentina and Di Marcantonio, Paolo and Figueira, Pedro and Fragoso, Ana and Genolet, Ludovic and Genoni, Matteo and Génova Santos, Ricardo and Hara, Nathan and Hughes, Ian and Iwert, Olaf and Kerber, Florian and Knudstrup, Jens and Landoni, Marco and Lavie, Baptiste and Lizon, Jean-Louis and Lendl, Monika and Lo Curto, Gaspare and Maire, Charles and Manescau, Antonio and Martins, C. J. a. P. and Mégevand, Denis and Mehner, Andrea and Micela, Giusi and Modigliani, Andrea and Molaro, Paolo and Monteiro, Manuel and Monteiro, Mario and Moschetti, Manuele and Müller, Eric and Nunes, Nelson and Oggioni, Luca and Oliveira, António and Pariani, Giorgio and Pasquini, Luca and Poretti, Ennio and Rasilla, José Luis and Redaelli, Edoardo and Riva, Marco and Santana Tschudi, Samuel and Santin, Paolo and Santos, Pedro and Segovia Milla, Alex and Seidel, Julia V. and Sosnowska, Danuta and Sozzetti, Alessandro and Spanò, Paolo and Suárez Mascareño, Alejandro and Tabernero, Hugo and Tenegi, Fabio and Udry, Stéphane and Zanutta, Alessio and Zerbi, Filippo},
	month = apr,
	year = {2020},
	pmid = {32161364},
	pmcid = {PMC7212060},
	eid= {597--601},
}

@article{allart_high-resolution_2019,
	title = {High-resolution confirmation of an extended helium atmosphere around {WASP}-107b},
	volume = {623},
	copyright = {© ESO 2019},
	issn = {0004-6361, 1432-0746},
	url = {https://www.aanda.org/articles/aa/abs/2019/03/aa34917-18/aa34917-18.html},
	doi = {10.1051/0004-6361/201834917},
	language = {en},
	urldate = {2022-11-02},
	journal = {A\&A},
	author = {Allart, R. and Bourrier, V. and Lovis, C. and Ehrenreich, D. and Aceituno, J. and Guijarro, A. and Pepe, F. and Sing, D. K. and Spake, J. J. and Wyttenbach, A.},
	month = mar,
	year = {2019},
	eid= {A58},
}

@article{kirk_confirmation_2020,
	title = {Confirmation of {WASP}-107b’s {Extended} {Helium} {Atmosphere} with {Keck} {II}/{NIRSPEC}},
	volume = {159},
	issn = {1538-3881},
	url = {https://dx.doi.org/10.3847/1538-3881/ab6e66},
	doi = {10.3847/1538-3881/ab6e66},
	language = {en},
	number = {3},
	urldate = {2022-11-02},
	journal = {AJ},
	author = {Kirk, James and Alam, Munazza K. and Lopez-Morales, Mercedes and Zeng, Li},
	month = feb,
	year = {2020},
	eid= {115},
}

@article{kreidberg_water_2018,
	title = {Water, {High}-altitude {Condensates}, and {Possible} {Methane} {Depletion} in the {Atmosphere} of the {Warm} {Super}-{Neptune} {WASP}-107b},
	volume = {858},
	issn = {2041-8205},
	url = {https://dx.doi.org/10.3847/2041-8213/aabfce},
	doi = {10.3847/2041-8213/aabfce},
	language = {en},
	number = {1},
	urldate = {2022-11-02},
	journal = {ApJL},
	author = {Kreidberg, Laura and Line, Michael R. and Thorngren, Daniel and Morley, Caroline V. and Stevenson, Kevin B.},
	month = may,
	year = {2018},
	eid= {L6},
}

@article{wang_metastable_2021,
	title = {Metastable {Helium} {Absorptions} with {3D} {Hydrodynamics} and {Self}-consistent {Photochemistry}. {II}. {WASP}-107b, {Stellar} {Wind}, {Radiation} {Pressure}, and {Shear} {Instability}},
	volume = {914},
	issn = {0004-637X},
	url = {https://dx.doi.org/10.3847/1538-4357/abf1ed},
	doi = {10.3847/1538-4357/abf1ed},
	language = {en},
	number = {2},
	urldate = {2022-11-05},
	journal = {ApJ},
	author = {Wang, Lile and Dai, Fei},
	month = jun,
	year = {2021},

	eid= {99},
}

@article{guillot_radiative_2010,
	title = {On the radiative equilibrium of irradiated planetary atmospheres},
	volume = {520},
	copyright = {© ESO, 2010},
	issn = {0004-6361, 1432-0746},
	url = {https://www.aanda.org/articles/aa/abs/2010/12/aa13396-09/aa13396-09.html},
	doi = {10.1051/0004-6361/200913396},
	language = {en},
	urldate = {2022-11-08},
	journal = {A\&A},
	author = {Guillot, T.},
	month = sep,
	year = {2010},
	eid= {A27},
}

@article{dorn_crires_2023,
	title = {{CRIRES}+ on sky at the {ESO} {Very} {Large} {Telescope} - {Observing} the {Universe} at infrared wavelengths and high spectral resolution},
	volume = {671},
	copyright = {© The Authors 2023},
	issn = {0004-6361, 1432-0746},
	url = {https://www.aanda.org/articles/aa/abs/2023/03/aa45217-22/aa45217-22.html},
	doi = {10.1051/0004-6361/202245217},
	language = {en},
	urldate = {2023-03-22},
	journal = {A\&A},
	author = {Dorn, R. J. and Bristow, P. and Smoker, J. V. and Rodler, F. and Lavail, A. and Accardo, M. and Ancker, M. van den and Baade, D. and Baruffolo, A. and Courtney-Barrer, B. and Blanco, L. and Brucalassi, A. and Cumani, C. and Follert, R. and Haimerl, A. and Hatzes, A. and Haug, M. and Heiter, U. and Hinterschuster, R. and Hubin, N. and Ives, D. J. and Jung, Y. and Jones, M. and Kaeufl, H.-U. and Kirchbauer, J.-P. and Klein, B. and Kochukhov, O. and Korhonen, H. H. and Köhler, J. and Lizon, J.-L. and Moins, C. and Molina-Conde, I. and Marquart, T. and Neeser, M. and Oliva, E. and Pallanca, L. and Pasquini, L. and Paufique, J. and Piskunov, N. and Reiners, A. and Schneller, D. and Schmutzer, R. and Seemann, U. and Slumstrup, D. and Smette, A. and Stegmeier, J. and Stempels, E. and Tordo, S. and Valenti, E. and Valenzuela, J. J. and Vernet, J. and Vinther, J. and Wehrhahn, A.},
	month = mar,
	year = {2023},
	
	eid= {A24},
}

@article{cheverall_robustness_2023,
	title = {Robustness measures for molecular detections using high-resolution transmission spectroscopy of exoplanets},
	volume = {522},
	issn = {0035-8711},
	url = {https://doi.org/10.1093/mnras/stad648},
	doi = {10.1093/mnras/stad648},
	number = {1},
	urldate = {2023-05-23},
	journal = {MNRAS},
	author = {Cheverall, Connor J and Madhusudhan, Nikku and Holmberg, Måns},
	month = jun,
	year = {2023},
	eid= {661--677},
}

@article{seidel_wind_2020,
	title = {Wind of change: retrieving exoplanet atmospheric winds from high-resolution spectroscopy},
	volume = {633},
	copyright = {© ESO 2020},
	issn = {0004-6361, 1432-0746},
	shorttitle = {Wind of change},
	url = {https://www.aanda.org/articles/aa/abs/2020/01/aa36892-19/aa36892-19.html},
	doi = {10.1051/0004-6361/201936892},
	language = {en},
	urldate = {2023-05-24},
	journal = {A\&A},
	author = {Seidel, J. V. and Ehrenreich, D. and Pino, L. and Bourrier, V. and Lavie, B. and Allart, R. and Wyttenbach, A. and Lovis, C.},
	month = jan,
	year = {2020},
	
	eid= {A86},
}

@article{rothman_hitemp_2010,
	series = {{XVIth} {Symposium} on {High} {Resolution} {Molecular} {Spectroscopy} ({HighRus}-2009)},
	title = {{HITEMP}, the high-temperature molecular spectroscopic database},
	volume = {111},
	issn = {0022-4073},
	url = {https://www.sciencedirect.com/science/article/pii/S002240731000169X},
	doi = {10.1016/j.jqsrt.2010.05.001},
	language = {en},
	number = {15},
	urldate = {2023-05-24},
	journal = {JQSRT},
	author = {Rothman, L. S. and Gordon, I. E. and Barber, R. J. and Dothe, H. and Gamache, R. R. and Goldman, A. and Perevalov, V. I. and Tashkun, S. A. and Tennyson, J.},
	month = oct,
	year = {2010},
	eid= {2139--2150},
}

@article{jwst_transiting_exoplanet_community_early_release_science_team_identification_2023,
	title = {Identification of carbon dioxide in an exoplanet atmosphere},
	volume = {614},
	issn = {0028-0836},
	url = {https://ui.adsabs.harvard.edu/abs/2023Natur.614..649J},
	doi = {10.1038/s41586-022-05269-w},
	urldate = {2023-05-30},
	journal = {Nature},
	author = {{JWST Transiting Exoplanet Community Early Release Science Team} and Ahrer, Eva-Maria and Alderson, Lili and Batalha, Natalie M. and Batalha, Natasha E. and Bean, Jacob L. and Beatty, Thomas G. and Bell, Taylor J. and Benneke, Björn and Berta-Thompson, Zachory K. and Carter, Aarynn L. and Crossfield, Ian J. M. and Espinoza, Néstor and Feinstein, Adina D. and Fortney, Jonathan J. and Gibson, Neale P. and Goyal, Jayesh M. and Kempton, Eliza M. -R. and Kirk, James and Kreidberg, Laura and López-Morales, Mercedes and Line, Michael R. and Lothringer, Joshua D. and Moran, Sarah E. and Mukherjee, Sagnick and Ohno, Kazumasa and Parmentier, Vivien and Piaulet, Caroline and Rustamkulov, Zafar and Schlawin, Everett and Sing, David K. and Stevenson, Kevin B. and Wakeford, Hannah R. and Allen, Natalie H. and Birkmann, Stephan M. and Brande, Jonathan and Crouzet, Nicolas and Cubillos, Patricio E. and Damiano, Mario and Désert, Jean-Michel and Gao, Peter and Harrington, Joseph and Hu, Renyu and Kendrew, Sarah and Knutson, Heather A. and Lagage, Pierre-Olivier and Leconte, Jérémy and Lendl, Monika and MacDonald, Ryan J. and May, E. M. and Miguel, Yamila and Molaverdikhani, Karan and Moses, Julianne I. and Murray, Catriona Anne and Nehring, Molly and Nikolov, Nikolay K. and Petit dit de la Roche, D. J. M. and Radica, Michael and Roy, Pierre-Alexis and Stassun, Keivan G. and Taylor, Jake and Waalkes, William C. and Wachiraphan, Patcharapol and Welbanks, Luis and Wheatley, Peter J. and Aggarwal, Keshav and Alam, Munazza K. and Banerjee, Agnibha and Barstow, Joanna K. and Blecic, Jasmina and Casewell, S. L. and Changeat, Quentin and Chubb, K. L. and Colón, Knicole D. and Coulombe, Louis-Philippe and Daylan, Tansu and de Val-Borro, Miguel and Decin, Leen and Dos Santos, Leonardo A. and Flagg, Laura and France, Kevin and Fu, Guangwei and García Muñoz, A. and Gizis, John E. and Glidden, Ana and Grant, David and Heng, Kevin and Henning, Thomas and Hong, Yu-Cian and Inglis, Julie and Iro, Nicolas and Kataria, Tiffany and Komacek, Thaddeus D. and Krick, Jessica E. and Lee, Elspeth K. H. and Lewis, Nikole K. and Lillo-Box, Jorge and Lustig-Yaeger, Jacob and Mancini, Luigi and Mandell, Avi M. and Mansfield, Megan and Marley, Mark S. and Mikal-Evans, Thomas and Morello, Giuseppe and Nixon, Matthew C. and Ortiz Ceballos, Kevin and Piette, Anjali A. A. and Powell, Diana and Rackham, Benjamin V. and Ramos-Rosado, Lakeisha and Rauscher, Emily and Redfield, Seth and Rogers, Laura K. and Roman, Michael T. and Roudier, Gael M. and Scarsdale, Nicholas and Shkolnik, Evgenya L. and Southworth, John and Spake, Jessica J. and Steinrueck, Maria E. and Tan, Xianyu and Teske, Johanna K. and Tremblin, Pascal and Tsai, Shang-Min and Tucker, Gregory S. and Turner, Jake D. and Valenti, Jeff A. and Venot, Olivia and Waldmann, Ingo P. and Wallack, Nicole L. and Zhang, Xi and Zieba, Sebastian},
	month = feb,
	year = {2023},
	eid= {649--652},
}

@article{yan_crires_2023,
	title = {{CRIRES}+ detection of {CO} emissions lines and temperature inversions on the dayside of {WASP}-18b and {WASP}-76b},
	volume = {672},
	issn = {0004-6361},
	url = {https://ui.adsabs.harvard.edu/abs/2023A&A...672A.107Y},
	doi = {10.1051/0004-6361/202245371},
	urldate = {2023-06-05},
	journal = {A\&A},
	author = {Yan, F. and Nortmann, L. and Reiners, A. and Piskunov, N. and Hatzes, A. and Seemann, U. and Shulyak, D. and Lavail, A. and Rains, A. D. and Cont, D. and Rengel, M. and Lesjak, F. and Nagel, E. and Kochukhov, O. and Czesla, S. and Boldt-Christmas, L. and Heiter, U. and Smoker, J. V. and Rodler, F. and Bristow, P. and Dorn, R. J. and Jung, Y. and Marquart, T. and Stempels, E.},
	month = apr,
	year = {2023},
	eid= {A107},
}

@article{cutri_vizier_2003,
	title = {{VizieR} {Online} {Data} {Catalog}: {2MASS} {All}-{Sky} {Catalog} of {Point} {Sources} ({Cutri}+ 2003)},
	shorttitle = {{VizieR} {Online} {Data} {Catalog}},
	url = {https://ui.adsabs.harvard.edu/abs/2003yCat.2246....0C},
	urldate = {2023-06-29},
	journal = {VizieR Online Data Catalog},
	author = {Cutri, R. M. and Skrutskie, M. F. and van Dyk, S. and Beichman, C. A. and Carpenter, J. M. and Chester, T. and Cambresy, L. and Evans, T. and Fowler, J. and Gizis, J. and Howard, E. and Huchra, J. and Jarrett, T. and Kopan, E. L. and Kirkpatrick, J. D. and Light, R. M. and Marsh, K. A. and McCallon, H. and Schneider, S. and Stiening, R. and Sykes, M. and Weinberg, M. and Wheaton, W. A. and Wheelock, S. and Zacarias, N.},
	month = jun,
	year = {2003},
	eid= {II/246},
}

@article{astropy_collaboration_astropy_2013,
	title = {Astropy: {A} community {Python} package for astronomy},
	volume = {558},
	issn = {0004-6361},
	shorttitle = {Astropy},
	url = {https://ui.adsabs.harvard.edu/abs/2013A&A...558A..33A},
	doi = {10.1051/0004-6361/201322068},
	urldate = {2023-07-06},
	journal = {A\&A},
	author = {{Astropy Collaboration} and Robitaille, Thomas P. and Tollerud, Erik J. and Greenfield, Perry and Droettboom, Michael and Bray, Erik and Aldcroft, Tom and Davis, Matt and Ginsburg, Adam and Price-Whelan, Adrian M. and Kerzendorf, Wolfgang E. and Conley, Alexander and Crighton, Neil and Barbary, Kyle and Muna, Demitri and Ferguson, Henry and Grollier, Frédéric and Parikh, Madhura M. and Nair, Prasanth H. and Unther, Hans M. and Deil, Christoph and Woillez, Julien and Conseil, Simon and Kramer, Roban and Turner, James E. H. and Singer, Leo and Fox, Ryan and Weaver, Benjamin A. and Zabalza, Victor and Edwards, Zachary I. and Azalee Bostroem, K. and Burke, D. J. and Casey, Andrew R. and Crawford, Steven M. and Dencheva, Nadia and Ely, Justin and Jenness, Tim and Labrie, Kathleen and Lim, Pey Lian and Pierfederici, Francesco and Pontzen, Andrew and Ptak, Andy and Refsdal, Brian and Servillat, Mathieu and Streicher, Ole},
	month = oct,
	year = {2013},
	eid= {A33},
}

@article{perez_ipython_2007,
	title = {{IPython}: {A} {System} for {Interactive} {Scientific} {Computing}},
	volume = {9},
	issn = {1558-366X},
	shorttitle = {{IPython}},
	doi = {10.1109/MCSE.2007.53},
	number = {3},
	journal = {Comput. Sci. Eng.},
	author = {Perez, Fernando and Granger, Brian E.},
	month = may,
	year = {2007},
	eid= {21--29},
}

@article{hunter_matplotlib_2007,
	title = {Matplotlib: {A} {2D} {Graphics} {Environment}},
	volume = {9},
	issn = {1558-366X},
	shorttitle = {Matplotlib},
	doi = {10.1109/MCSE.2007.55},
	number = {3},
	journal = {Comput. Sci. Eng.},
	author = {Hunter, John D.},
	month = may,
	year = {2007},
	eid= {90--95},
}

@article{harris_array_2020,
	title = {Array programming with {NumPy}},
	volume = {585},
	copyright = {2020 The Author(s)},
	issn = {1476-4687},
	url = {https://www.nature.com/articles/s41586-020-2649-2},
	doi = {10.1038/s41586-020-2649-2},
	language = {en},
	number = {7825},
	urldate = {2023-07-06},
	journal = {Nature},
	author = {Harris, Charles R. and Millman, K. Jarrod and van der Walt, Stéfan J. and Gommers, Ralf and Virtanen, Pauli and Cournapeau, David and Wieser, Eric and Taylor, Julian and Berg, Sebastian and Smith, Nathaniel J. and Kern, Robert and Picus, Matti and Hoyer, Stephan and van Kerkwijk, Marten H. and Brett, Matthew and Haldane, Allan and del Río, Jaime Fernández and Wiebe, Mark and Peterson, Pearu and Gérard-Marchant, Pierre and Sheppard, Kevin and Reddy, Tyler and Weckesser, Warren and Abbasi, Hameer and Gohlke, Christoph and Oliphant, Travis E.},
	month = sep,
	year = {2020},
	eid= {357--362},
}

@article{waskom_mwaskomseaborn_2017,
	title = {Mwaskom/{Seaborn}: {V0}.8.1 ({September} 2017)},
	shorttitle = {Mwaskom/{Seaborn}},
	url = {https://ui.adsabs.harvard.edu/abs/2017zndo....883859W},
	doi = {10.5281/zenodo.883859},
	urldate = {2023-07-06},
	journal = {{Seaborn} (10.5281/zenodo.883859)},
	author = {Waskom, Michael and Botvinnik, Olga and O'Kane, Drew and Hobson, Paul and Lukauskas, Saulius and Gemperline, David C and Augspurger, Tom and Halchenko, Yaroslav and Cole, John B. and Warmenhoven, Jordi and de Ruiter, Julian and Pye, Cameron and Hoyer, Stephan and Vanderplas, Jake and Villalba, Santi and Kunter, Gero and Quintero, Eric and Bachant, Pete and Martin, Marcel and Meyer, Kyle and Miles, Alistair and Ram, Yoav and Yarkoni, Tal and Williams, Mike Lee and Evans, Constantine and Fitzgerald, Clark and {Brian} and Fonnesbeck, Chris and Lee, Antony and Qalieh, Adel},
	month = sep,
	year = {2017},
}

@inproceedings{mckinney_data_2010,
        address = {{56--61}},
	booktitle = {Proc. SciPy Conf.},
	url = {https://conference.scipy.org/proceedings/scipy2010/mckinney.html},
	doi = {10.25080/Majora-92bf1922-00a},
	language = {en},
	urldate = {2023-07-06},
	author = {McKinney, Wes},
	year = {2010},
}

@misc{team_pandas-devpandas_2023,
	title = {{Pandas}  (10.5281/zenodo.8092754)},
	shorttitle = {pandas-dev/pandas},
	url = {https://zenodo.org/record/8092754},
	urldate = {2023-07-06},
	author = {{The pandas development team}},
	month = jun,
	year = {2023},
	doi = {10.5281/zenodo.8092754},
}

@article{virtanen_scipy_2020,
	title = {{SciPy} 1.0: fundamental algorithms for scientific computing in {Python}},
	volume = {17},
	copyright = {2020 The Author(s)},
	issn = {1548-7105},
	shorttitle = {{SciPy} 1.0},
	url = {https://www.nature.com/articles/s41592-019-0686-2},
	doi = {10.1038/s41592-019-0686-2},
	language = {en},
	number = {3},
	urldate = {2023-07-06},
	journal = {Nature Methods},
	author = {Virtanen, Pauli and Gommers, Ralf and Oliphant, Travis E. and Haberland, Matt and Reddy, Tyler and Cournapeau, David and Burovski, Evgeni and Peterson, Pearu and Weckesser, Warren and Bright, Jonathan and van der Walt, Stéfan J. and Brett, Matthew and Wilson, Joshua and Millman, K. Jarrod and Mayorov, Nikolay and Nelson, Andrew R. J. and Jones, Eric and Kern, Robert and Larson, Eric and Carey, C. J. and Polat, İlhan and Feng, Yu and Moore, Eric W. and VanderPlas, Jake and Laxalde, Denis and Perktold, Josef and Cimrman, Robert and Henriksen, Ian and Quintero, E. A. and Harris, Charles R. and Archibald, Anne M. and Ribeiro, Antônio H. and Pedregosa, Fabian and van Mulbregt, Paul},
	month = mar,
	year = {2020},
	eid= {261--272},
}

@software{czesla_pya_2019,
       author = {{Czesla}, Stefan and {Schr{\"o}ter}, Sebastian and {Schneider}, Christian P. and {Huber}, Klaus F. and {Pfeifer}, Fabian and {Andreasen}, Daniel T. and {Zechmeister}, Mathias},
        title = "{PyA: Python astronomy-related packages}",
 howpublished = {Astrophysics Source Code Library, record ascl:1906.010},
         year = 2019,
        month = jun,
          eid = {ascl:1906.010},
archivePrefix = {ascl},
       eprint = {1906.010},
       adsurl = {https://ui.adsabs.harvard.edu/abs/2019ascl.soft06010C}
}

@article{prinoth_titanium_2022,
	title = {Titanium oxide and chemical inhomogeneity in the atmosphere of the exoplanet {WASP}-189 b},
	volume = {6},
	copyright = {2022 The Author(s), under exclusive licence to Springer Nature Limited},
	issn = {2397-3366},
	url = {https://www.nature.com/articles/s41550-021-01581-z},
	doi = {10.1038/s41550-021-01581-z},
	language = {en},
	number = {4},
	urldate = {2023-09-14},
	journal = {Nat. Astron.},
	author = {Prinoth, Bibiana and Hoeijmakers, H. Jens and Kitzmann, Daniel and Sandvik, Elin and Seidel, Julia V. and Lendl, Monika and Borsato, Nicholas W. and Thorsbro, Brian and Anderson, David R. and Barrado, David and Kravchenko, Kateryna and Allart, Romain and Bourrier, Vincent and Cegla, Heather M. and Ehrenreich, David and Fisher, Chloe and Lovis, Christophe and Guzmán-Mesa, Andrea and Grimm, Simon and Hooton, Matthew and Morris, Brett M. and Oreshenko, Maria and Pino, Lorenzo and Heng, Kevin},
	month = apr,
	year = {2022},
	eid= {449--457},
}

@article{lesjak_retrieval_2023,
	title = {Retrieval of the dayside atmosphere of {WASP}-43b with {CRIRES}+},
	volume = {678},
	issn = {0004-6361},
	url = {https://ui.adsabs.harvard.edu/abs/2023A&A...678A..23L},
	doi = {10.1051/0004-6361/202347151},
	urldate = {2023-10-17},
	journal = {A\&A},
	author = {Lesjak, F. and Nortmann, L. and Yan, F. and Cont, D. and Reiners, A. and Piskunov, N. and Hatzes, A. and Boldt-Christmas, L. and Czesla, S. and Heiter, U. and Kochukhov, O. and Lavail, A. and Nagel, E. and Rains, A. D. and Rengel, M. and Rodler, F. and Seemann, U. and Shulyak, D.},
	month = oct,
	year = {2023},
	eid= {A23},
}

@article{mocnik_starspots_2017,
	title = {Starspots on {WASP}-107 and pulsations of {WASP}-118},
	volume = {469},
	issn = {0035-8711},
	url = {https://ui.adsabs.harvard.edu/abs/2017MNRAS.469.1622M},
	doi = {10.1093/mnras/stx972},
	urldate = {2023-12-08},
	journal = {MNRAS},
	author = {Močnik, T. and Hellier, C. and Anderson, D. R. and Clark, B. J. M. and Southworth, J.},
	month = aug,
	year = {2017},
	eid= {1622--1629},
}

@article{dressing_characterizing_2019,
	title = {Characterizing {K2} {Candidate} {Planetary} {Systems} {Orbiting} {Low}-mass {Stars}. {IV}. {Updated} {Properties} for 86 {Cool} {Dwarfs} {Observed} during {Campaigns} 1–17},
	volume = {158},
	issn = {1538-3881},
	url = {https://dx.doi.org/10.3847/1538-3881/ab2895},
	doi = {10.3847/1538-3881/ab2895},
	language = {en},
	number = {2},
	urldate = {2023-12-08},
	journal = {AJ},
	author = {Dressing, Courtney D. and Hardegree-Ullman, Kevin and Schlieder, Joshua E. and Newton, Elisabeth R. and Vanderburg, Andrew and Feinstein, Adina D. and Duvvuri, Girish M. and Arnold, Lauren and Bristow, Makennah and Thackeray, Beverly and Abrahams, Ellianna Schwab and Ciardi, David R. and Crossfield, Ian J. M. and Yu, Liang and Martinez, Arturo O. and Christiansen, Jessie L. and Crepp, Justin R. and Isaacson, Howard},
	month = jul,
	year = {2019},

	eid= {87},
}

@article{nortmann_ground-based_2018,
	title = {Ground-based detection of an extended helium atmosphere in the {Saturn}-mass exoplanet {WASP}-69b},
	volume = {362},
	issn = {0036-8075},
	url = {https://ui.adsabs.harvard.edu/abs/2018Sci...362.1388N},
	doi = {10.1126/science.aat5348},
	urldate = {2023-12-20},
	journal = {Science},
	author = {Nortmann, Lisa and Pallé, Enric and Salz, Michael and Sanz-Forcada, Jorge and Nagel, Evangelos and Alonso-Floriano, F. Javier and Czesla, Stefan and Yan, Fei and Chen, Guo and Snellen, Ignas A. G. and Zechmeister, Mathias and Schmitt, Jürgen H. M. M. and López-Puertas, Manuel and Casasayas-Barris, Núria and Bauer, Florian F. and Amado, Pedro J. and Caballero, José A. and Dreizler, Stefan and Henning, Thomas and Lampón, Manuel and Montes, David and Molaverdikhani, Karan and Quirrenbach, Andreas and Reiners, Ansgar and Ribas, Ignasi and Sánchez-López, Alejandro and Schneider, P. Christian and Zapatero Osorio, María R.},
	month = dec,
	year = {2018},
	eid= {1388--1391},
}

@article{linssen_constraining_2022,
	title = {Constraining planetary mass-loss rates by simulating {Parker} wind profiles with {Cloudy}},
	volume = {667},
	issn = {0004-6361},
	url = {https://ui.adsabs.harvard.edu/abs/2022A&A...667A..54L},
	doi = {10.1051/0004-6361/202243830},
	urldate = {2024-01-08},
	journal = {A\&A},
	author = {Linssen, D. C. and Oklopčić, A. and MacLeod, M.},
	month = nov,
	year = {2022},
	eid= {A54},
}

@article{millholland_tidal_2020,
	title = {Tidal {Inflation} {Reconciles} {Low}-density {Sub}-{Saturns} with {Core} {Accretion}},
	volume = {897},
	issn = {0004-637X},
	url = {https://ui.adsabs.harvard.edu/abs/2020ApJ...897....7M},
	doi = {10.3847/1538-4357/ab959c},
	urldate = {2024-01-08},
	journal = {ApJ},
	author = {Millholland, Sarah and Petigura, Erik and Batygin, Konstantin},
	month = jul,
	year = {2020},
	eid= {7},
}

@article{schlawin_clear_2018,
	title = {Clear and {Cloudy} {Exoplanet} {Forecasts} for {JWST}: {Maps}, {Retrieved} {Composition}, and {Constraints} on {Formation} with {MIRI} and {NIRCam}},
	volume = {156},
	issn = {0004-6256},
	shorttitle = {Clear and {Cloudy} {Exoplanet} {Forecasts} for {JWST}},
	url = {https://ui.adsabs.harvard.edu/abs/2018AJ....156...40S},
	doi = {10.3847/1538-3881/aac774},
	urldate = {2024-01-08},
	journal = {AJ},
	author = {Schlawin, Everett and Greene, Thomas P. and Line, Michael and Fortney, Jonathan J. and Rieke, Marcia},
	month = jul,
	year = {2018},
	eid= {40},
}

@article{dyrek_so2_2024,
	title = {{SO2}, silicate clouds, but no {CH4} detected in a warm {Neptune}},
	volume = {625},
	copyright = {2023 The Author(s), under exclusive licence to Springer Nature Limited},
	issn = {1476-4687},
	url = {https://www.nature.com/articles/s41586-023-06849-0},
	doi = {10.1038/s41586-023-06849-0},
	language = {en},
	number = {7993},
	urldate = {2024-01-29},
	journal = {Nature},
	author = {Dyrek, Achrène and Min, Michiel and Decin, Leen and Bouwman, Jeroen and Crouzet, Nicolas and Mollière, Paul and Lagage, Pierre-Olivier and Konings, Thomas and Tremblin, Pascal and Güdel, Manuel and Pye, John and Waters, Rens and Henning, Thomas and Vandenbussche, Bart and Ardevol Martinez, Francisco and Argyriou, Ioannis and Ducrot, Elsa and Heinke, Linus and van Looveren, Gwenael and Absil, Olivier and Barrado, David and Baudoz, Pierre and Boccaletti, Anthony and Cossou, Christophe and Coulais, Alain and Edwards, Billy and Gastaud, René and Glasse, Alistair and Glauser, Adrian and Greene, Thomas P. and Kendrew, Sarah and Krause, Oliver and Lahuis, Fred and Mueller, Michael and Olofsson, Goran and Patapis, Polychronis and Rouan, Daniel and Royer, Pierre and Scheithauer, Silvia and Waldmann, Ingo and Whiteford, Niall and Colina, Luis and van Dishoeck, Ewine F. and Östlin, Göran and Ray, Tom P. and Wright, Gillian},
	month = jan,
	year = {2024},
	eid= {51--54},
}

@article{spake_helium_2018,
	title = {Helium in the eroding atmosphere of an exoplanet},
	volume = {557},
	issn = {0028-0836, 1476-4687},
	url = {https://www.nature.com/articles/s41586-018-0067-5},
	doi = {10.1038/s41586-018-0067-5},
	language = {en},
	number = {7703},
	urldate = {2024-02-05},
	journal = {Nature},
	author = {Spake, J. J. and Sing, D. K. and Evans, T. M. and Oklopčić, A. and Bourrier, V. and Kreidberg, L. and Rackham, B. V. and Irwin, J. and Ehrenreich, D. and Wyttenbach, A. and Wakeford, H. R. and Zhou, Y. and Chubb, K. L. and Nikolov, N. and Goyal, J. M. and Henry, G. W. and Williamson, M. H. and Blumenthal, S. and Anderson, D. R. and Hellier, C. and Charbonneau, D. and Udry, S. and Madhusudhan, N.},
	month = may,
	year = {2018},
	eid= {68--70},
}

@article{rubenzahl_tess-keck_2021,
	title = {The {TESS}-{Keck} {Survey}. {IV}. {A} {Retrograde}, {Polar} {Orbit} for the {Ultra}-low-density, {Hot} {Super}-{Neptune} {WASP}-107b},
	volume = {161},
	issn = {0004-6256},
	url = {https://ui.adsabs.harvard.edu/abs/2021AJ....161..119R},
	doi = {10.3847/1538-3881/abd177},
	urldate = {2024-02-08},
	journal = {AJ},
	author = {Rubenzahl, Ryan A. and Dai, Fei and Howard, Andrew W. and Chontos, Ashley and Giacalone, Steven and Lubin, Jack and Rosenthal, Lee J. and Isaacson, Howard and Batalha, Natalie M. and Crossfield, Ian J. M. and Dressing, Courtney and Fulton, Benjamin and Huber, Daniel and Kane, Stephen R. and Petigura, Erik A. and Robertson, Paul and Roy, Arpita and Weiss, Lauren M. and Beard, Corey and Hill, Michelle L. and Mayo, Andrew and Mocnik, Teo and Murphy, Joseph M. Akana and Scarsdale, Nicholas},
	month = mar,
	year = {2021},
	eid= {119},
}

@article{khodachenko_simulation_2021,
	title = {Simulation of 10 830 Å absorption with a {3D} hydrodynamic model reveals the solar {He} abundance in upper atmosphere of {WASP}-107b},
	volume = {503},
	issn = {1745-3925},
	url = {https://doi.org/10.1093/mnrasl/slab015},
	doi = {10.1093/mnrasl/slab015},
	number = {1},
	urldate = {2024-02-08},
	journal = {MNRAS},
	author = {Khodachenko, M L and Shaikhislamov, I F and Fossati, L and Lammer, H and Rumenskikh, M S and Berezutsky, A G and Miroshnichenko, I B and Efimof, M A},
	month = may,
	year = {2021},
	eid= {L23--L27},
}

@article{kesseli_search_2020,
	title = {A {Search} for {FeH} in {Hot}-{Jupiter} {Atmospheres} with {High}-dispersion {Spectroscopy}},
	volume = {160},
	issn = {1538-3881},
	url = {https://dx.doi.org/10.3847/1538-3881/abb59c},
	doi = {10.3847/1538-3881/abb59c},
	language = {en},
	number = {5},
	urldate = {2024-02-08},
	journal = {AJ},
	author = {Kesseli, Aurora Y. and Snellen, I. A. G. and Alonso-Floriano, F. J. and Mollière, P. and Serindag, D. B.},
	month = oct,
	year = {2020},

	eid= {228},
}

@article{spake_posttransit_2021,
	title = {The {Posttransit} {Tail} of {WASP}-107b {Observed} at 10830 Å},
	volume = {162},
	issn = {1538-3881},
	url = {https://dx.doi.org/10.3847/1538-3881/ac178a},
	doi = {10.3847/1538-3881/ac178a},
	language = {en},
	number = {6},
	urldate = {2024-02-23},
	journal = {AJ},
	author = {Spake, J. J. and Oklopcic, A. and Hillenbrand, L. A.},
	month = dec,
	year = {2021},
	eid= {284},
}

@article{dai_oblique_2017,
	title = {The {Oblique} {Orbit} of {WASP}-107b from {K2} {Photometry}},
	volume = {153},
	issn = {0004-6256},
	url = {https://ui.adsabs.harvard.edu/abs/2017AJ....153..205D},
	doi = {10.3847/1538-3881/aa65d1},
	urldate = {2024-02-26},
	journal = {AJ},
	author = {Dai, Fei and Winn, Joshua N.},
	month = may,
	year = {2017},
	eid= {205},
}

@ARTICLE{guilluy_gaps_2024,
       author = {{Guilluy}, G. and {D'Arpa}, M.~C. and {Bonomo}, A.~S. and {Spinelli}, R. and {Biassoni}, F. and {Fossati}, L. and {Maggio}, A. and {Giacobbe}, P. and {Lanza}, A.~F. and {Sozzetti}, A. and {Borsa}, F. and {Rainer}, M. and {Micela}, G. and {Affer}, L. and {Andreuzzi}, G. and {Bignamini}, A. and {Boschin}, W. and {Carleo}, I. and {Cecconi}, M. and {Desidera}, S. and {Fardella}, V. and {Ghedina}, A. and {Mantovan}, G. and {Mancini}, L. and {Nascimbeni}, V. and {Knapic}, C. and {Pedani}, M. and {Petralia}, A. and {Pino}, L. and {Scandariato}, G. and {Sicilia}, D. and {Stangret}, M. and {Zingales}, T.},
        title = "{The GAPS Programme at TNG. LIV. A He I survey of close-in giant planets hosted by M-K dwarf stars with GIANO-B}",
      journal = {\aap},
         year = 2024,
        month = jun,
       volume = {686},
          eid = {A83},
          doi = {10.1051/0004-6361/202348997},
       adsurl = {https://ui.adsabs.harvard.edu/abs/2024A&A...686A..83G}
}

@article{kokori_exoclock_2023,
	title = {{ExoClock} {Project}. {III}. 450 {New} {Exoplanet} {Ephemerides} from {Ground} and {Space} {Observations}},
	volume = {265},
	issn = {0067-0049},
	url = {https://ui.adsabs.harvard.edu/abs/2023ApJS..265....4K},
	doi = {10.3847/1538-4365/ac9da4},
	urldate = {2024-03-26},
	journal = {ApJS},
	author = {Kokori, A. and Tsiaras, A. and Edwards, B. and Jones, A. and Pantelidou, G. and Tinetti, G. and Bewersdorff, L. and Iliadou, A. and Jongen, Y. and Lekkas, G. and Nastasi, A. and Poultourtzidis, E. and Sidiropoulos, C. and Walter, F. and Wünsche, A. and Abraham, R. and Agnihotri, V. K. and Albanesi, R. and Arce-Mansego, E. and Arnot, D. and Audejean, M. and Aumasson, C. and Bachschmidt, M. and Baj, G. and Barroy, P. R. and Belinski, A. A. and Bennett, D. and Benni, P. and Bernacki, K. and Betti, L. and Biagini, A. and Bosch, P. and Brandebourg, P. and Brát, L. and Bretton, M. and Brincat, S. M. and Brouillard, S. and Bruzas, A. and Bruzzone, A. and Buckland, R. A. and Caló, M. and Campos, F. and Carreño, A. and Carrion Rodrigo, J. A. and Casali, R. and Casalnuovo, G. and Cataneo, M. and Chang, C. -M. and Changeat, L. and Chowdhury, V. and Ciantini, R. and Cilluffo, M. and Coliac, J. -F. and Conzo, G. and Correa, M. and Coulon, G. and Crouzet, N. and Crow, M. V. and Curtis, I. A. and Daniel, D. and Dauchet, B. and Dawes, S. and Deldem, M. and Deligeorgopoulos, D. and Dransfield, G. and Dymock, R. and Eenmäe, T. and Esseiva, N. and Evans, P. and Falco, C. and Farfán, R. G. and Fernández-Lajús, E. and Ferratfiat, S. and Ferreira, S. L. and Ferretti, A. and Fiołka, J. and Fowler, M. and Futcher, S. R. and Gabellini, D. and Gainey, T. and Gaitan, J. and Gajdoš, P. and García-Sánchez, A. and Garlitz, J. and Gillier, C. and Gison, C. and Gonzales, J. and Gorshanov, D. and Grau Horta, F. and Grivas, G. and Guerra, P. and Guillot, T. and Haswell, C. A. and Haymes, T. and Hentunen, V. -P. and Hills, K. and Hose, K. and Humbert, T. and Hurter, F. and Hynek, T. and Irzyk, M. and Jacobsen, J. and Jannetta, A. L. and Johnson, K. and Jóźwik-Wabik, P. and Kaeouach, A. E. and Kang, W. and Kiiskinen, H. and Kim, T. and Kivila, Ü. and Koch, B. and Kolb, U. and Kučáková, H. and Lai, S. -P. and Laloum, D. and Lasota, S. and Lewis, L. A. and Liakos, G. -I. and Libotte, F. and Lomoz, F. and Lopresti, C. and Majewski, R. and Malcher, A. and Mallonn, M. and Mannucci, M. and Marchini, A. and Mari, J. -M. and Marino, A. and Marino, G. and Mario, J. -C. and Marquette, J. -B. and Martínez-Bravo, F. A. and Mašek, M. and Matassa, P. and Michel, P. and Michelet, J. and Miller, M. and Miny, E. and Molina, D. and Mollier, T. and Monteleone, B. and Montigiani, N. and Morales-Aimar, M. and Mortari, F. and Morvan, M. and Mugnai, L. V. and Murawski, G. and Naponiello, L. and Naudin, J. -L. and Naves, R. and Néel, D. and Neito, R. and Neveu, S. and Noschese, A. and Öğmen, Y. and Ohshima, O. and Orbanic, Z. and Pace, E. P. and Pantacchini, C. and Paschalis, N. I. and Pereira, C. and Peretto, I. and Perroud, V. and Phillips, M. and Pintr, P. and Pioppa, J. -B. and Plazas, J. and Poelarends, A. J. and Popowicz, A. and Purcell, J. and Quinn, N. and Raetz, M. and Rees, D. and Regembal, F. and Rocchetto, M. and Rocci, P. -F. and Rockenbauer, M. and Roth, R. and Rousselot, L. and Rubia, X. and Ruocco, N. and Russo, E. and Salisbury, M. and Salvaggio, F. and Santos, A. and Savage, J. and Scaggiante, F. and Sedita, D. and Shadick, S. and Silva, A. F. and Sioulas, N. and Školník, V. and Smith, M. and Smolka, M. and Solmaz, A. and Stanbury, N. and Stouraitis, D. and Tan, T. -G. and Theusner, M. and Thurston, G. and Tifner, F. P. and Tomacelli, A. and Tomatis, A. and Trnka, J. and Tylšar, M. and Valeau, P. and Vignes, J. -P. and Villa, A. and Vives Sureda, A. and Vora, K. and Vrašt'ák, M. and Walliang, D. and Wenzel, B. and Wright, D. E. and Zambelli, R. and Zhang, M. and Zíbar, M.},
	month = mar,
	year = {2023},
	eid= {4},
}

@article{gaia_collaboration_gaia_2021,
	title = {Gaia {Early} {Data} {Release} 3. {Summary} of the contents and survey properties},
	volume = {649},
	issn = {0004-6361},
	url = {https://ui.adsabs.harvard.edu/abs/2021A&A...649A...1G},
	doi = {10.1051/0004-6361/202039657},
	urldate = {2024-03-26},
	journal = {A\&A},
	author = {{Gaia Collaboration} and Brown, A. G. A. and Vallenari, A. and Prusti, T. and de Bruijne, J. H. J. and Babusiaux, C. and Biermann, M. and Creevey, O. L. and Evans, D. W. and Eyer, L. and Hutton, A. and Jansen, F. and Jordi, C. and Klioner, S. A. and Lammers, U. and Lindegren, L. and Luri, X. and Mignard, F. and Panem, C. and Pourbaix, D. and Randich, S. and Sartoretti, P. and Soubiran, C. and Walton, N. A. and Arenou, F. and Bailer-Jones, C. A. L. and Bastian, U. and Cropper, M. and Drimmel, R. and Katz, D. and Lattanzi, M. G. and van Leeuwen, F. and Bakker, J. and Cacciari, C. and Castañeda, J. and De Angeli, F. and Ducourant, C. and Fabricius, C. and Fouesneau, M. and Frémat, Y. and Guerra, R. and Guerrier, A. and Guiraud, J. and Jean-Antoine Piccolo, A. and Masana, E. and Messineo, R. and Mowlavi, N. and Nicolas, C. and Nienartowicz, K. and Pailler, F. and Panuzzo, P. and Riclet, F. and Roux, W. and Seabroke, G. M. and Sordo, R. and Tanga, P. and Thévenin, F. and Gracia-Abril, G. and Portell, J. and Teyssier, D. and Altmann, M. and Andrae, R. and Bellas-Velidis, I. and Benson, K. and Berthier, J. and Blomme, R. and Brugaletta, E. and Burgess, P. W. and Busso, G. and Carry, B. and Cellino, A. and Cheek, N. and Clementini, G. and Damerdji, Y. and Davidson, M. and Delchambre, L. and Dell'Oro, A. and Fernández-Hernández, J. and Galluccio, L. and García-Lario, P. and Garcia-Reinaldos, M. and González-Núñez, J. and Gosset, E. and Haigron, R. and Halbwachs, J. -L. and Hambly, N. C. and Harrison, D. L. and Hatzidimitriou, D. and Heiter, U. and Hernández, J. and Hestroffer, D. and Hodgkin, S. T. and Holl, B. and Janßen, K. and Jevardat de Fombelle, G. and Jordan, S. and Krone-Martins, A. and Lanzafame, A. C. and Löffler, W. and Lorca, A. and Manteiga, M. and Marchal, O. and Marrese, P. M. and Moitinho, A. and Mora, A. and Muinonen, K. and Osborne, P. and Pancino, E. and Pauwels, T. and Petit, J. -M. and Recio-Blanco, A. and Richards, P. J. and Riello, M. and Rimoldini, L. and Robin, A. C. and Roegiers, T. and Rybizki, J. and Sarro, L. M. and Siopis, C. and Smith, M. and Sozzetti, A. and Ulla, A. and Utrilla, E. and van Leeuwen, M. and van Reeven, W. and Abbas, U. and Abreu Aramburu, A. and Accart, S. and Aerts, C. and Aguado, J. J. and Ajaj, M. and Altavilla, G. and Álvarez, M. A. and Álvarez Cid-Fuentes, J. and Alves, J. and Anderson, R. I. and Anglada Varela, E. and Antoja, T. and Audard, M. and Baines, D. and Baker, S. G. and Balaguer-Núñez, L. and Balbinot, E. and Balog, Z. and Barache, C. and Barbato, D. and Barros, M. and Barstow, M. A. and Bartolomé, S. and Bassilana, J. -L. and Bauchet, N. and Baudesson-Stella, A. and Becciani, U. and Bellazzini, M. and Bernet, M. and Bertone, S. and Bianchi, L. and Blanco-Cuaresma, S. and Boch, T. and Bombrun, A. and Bossini, D. and Bouquillon, S. and Bragaglia, A. and Bramante, L. and Breedt, E. and Bressan, A. and Brouillet, N. and Bucciarelli, B. and Burlacu, A. and Busonero, D. and Butkevich, A. G. and Buzzi, R. and Caffau, E. and Cancelliere, R. and Cánovas, H. and Cantat-Gaudin, T. and Carballo, R. and Carlucci, T. and Carnerero, M. I. and Carrasco, J. M. and Casamiquela, L. and Castellani, M. and Castro-Ginard, A. and Castro Sampol, P. and Chaoul, L. and Charlot, P. and Chemin, L. and Chiavassa, A. and Cioni, M. -R. L. and Comoretto, G. and Cooper, W. J. and Cornez, T. and Cowell, S. and Crifo, F. and Crosta, M. and Crowley, C. and Dafonte, C. and Dapergolas, A. and David, M. and David, P. and de Laverny, P. and De Luise, F. and De March, R. and De Ridder, J. and de Souza, R. and de Teodoro, P. and de Torres, A. and del Peloso, E. F. and del Pozo, E. and Delbo, M. and Delgado, A. and Delgado, H. E. and Delisle, J. -B. and Di Matteo, P. and Diakite, S. and Diener, C. and Distefano, E. and Dolding, C. and Eappachen, D. and Edvardsson, B. and Enke, H. and Esquej, P. and Fabre, C. and Fabrizio, M. and Faigler, S. and Fedorets, G. and Fernique, P. and Fienga, A. and Figueras, F. and Fouron, C. and Fragkoudi, F. and Fraile, E. and Franke, F. and Gai, M. and Garabato, D. and Garcia-Gutierrez, A. and García-Torres, M. and Garofalo, A. and Gavras, P. and Gerlach, E. and Geyer, R. and Giacobbe, P. and Gilmore, G. and Girona, S. and Giuffrida, G. and Gomel, R. and Gomez, A. and Gonzalez-Santamaria, I. and González-Vidal, J. J. and Granvik, M. and Gutiérrez-Sánchez, R. and Guy, L. P. and Hauser, M. and Haywood, M. and Helmi, A. and Hidalgo, S. L. and Hilger, T. and Hładczuk, N. and Hobbs, D. and Holland, G. and Huckle, H. E. and Jasniewicz, G. and Jonker, P. G. and Juaristi Campillo, J. and Julbe, F. and Karbevska, L. and Kervella, P. and Khanna, S. and Kochoska, A. and Kontizas, M. and Kordopatis, G. and Korn, A. J. and Kostrzewa-Rutkowska, Z. and Kruszyńska, K. and Lambert, S. and Lanza, A. F. and Lasne, Y. and Le Campion, J. -F. and Le Fustec, Y. and Lebreton, Y. and Lebzelter, T. and Leccia, S. and Leclerc, N. and Lecoeur-Taibi, I. and Liao, S. and Licata, E. and Lindstrøm, E. P. and Lister, T. A. and Livanou, E. and Lobel, A. and Madrero Pardo, P. and Managau, S. and Mann, R. G. and Marchant, J. M. and Marconi, M. and Marcos Santos, M. M. S. and Marinoni, S. and Marocco, F. and Marshall, D. J. and Martin Polo, L. and Martín-Fleitas, J. M. and Masip, A. and Massari, D. and Mastrobuono-Battisti, A. and Mazeh, T. and McMillan, P. J. and Messina, S. and Michalik, D. and Millar, N. R. and Mints, A. and Molina, D. and Molinaro, R. and Molnár, L. and Montegriffo, P. and Mor, R. and Morbidelli, R. and Morel, T. and Morris, D. and Mulone, A. F. and Munoz, D. and Muraveva, T. and Murphy, C. P. and Musella, I. and Noval, L. and Ordénovic, C. and Orrù, G. and Osinde, J. and Pagani, C. and Pagano, I. and Palaversa, L. and Palicio, P. A. and Panahi, A. and Pawlak, M. and Peñalosa Esteller, X. and Penttilä, A. and Piersimoni, A. M. and Pineau, F. -X. and Plachy, E. and Plum, G. and Poggio, E. and Poretti, E. and Poujoulet, E. and Prša, A. and Pulone, L. and Racero, E. and Ragaini, S. and Rainer, M. and Raiteri, C. M. and Rambaux, N. and Ramos, P. and Ramos-Lerate, M. and Re Fiorentin, P. and Regibo, S. and Reylé, C. and Ripepi, V. and Riva, A. and Rixon, G. and Robichon, N. and Robin, C. and Roelens, M. and Rohrbasser, L. and Romero-Gómez, M. and Rowell, N. and Royer, F. and Rybicki, K. A. and Sadowski, G. and Sagristà Sellés, A. and Sahlmann, J. and Salgado, J. and Salguero, E. and Samaras, N. and Sanchez Gimenez, V. and Sanna, N. and Santoveña, R. and Sarasso, M. and Schultheis, M. and Sciacca, E. and Segol, M. and Segovia, J. C. and Ségransan, D. and Semeux, D. and Shahaf, S. and Siddiqui, H. I. and Siebert, A. and Siltala, L. and Slezak, E. and Smart, R. L. and Solano, E. and Solitro, F. and Souami, D. and Souchay, J. and Spagna, A. and Spoto, F. and Steele, I. A. and Steidelmüller, H. and Stephenson, C. A. and Süveges, M. and Szabados, L. and Szegedi-Elek, E. and Taris, F. and Tauran, G. and Taylor, M. B. and Teixeira, R. and Thuillot, W. and Tonello, N. and Torra, F. and Torra, J. and Turon, C. and Unger, N. and Vaillant, M. and van Dillen, E. and Vanel, O. and Vecchiato, A. and Viala, Y. and Vicente, D. and Voutsinas, S. and Weiler, M. and Wevers, T. and Wyrzykowski, Ł. and Yoldas, A. and Yvard, P. and Zhao, H. and Zorec, J. and Zucker, S. and Zurbach, C. and Zwitter, T.},
	month = may,
	year = {2021},
	eid= {A1},
}

@article{boldt-christmas_optimising_2024,
	title = {Optimising spectroscopic observations of transiting exoplanets},
	volume = {683},
	copyright = {© The Authors 2024},
	issn = {0004-6361, 1432-0746},
	url = {https://www.aanda.org/articles/aa/abs/2024/03/aa47398-23/aa47398-23.html},
	doi = {10.1051/0004-6361/202347398},
	language = {en},
	urldate = {2024-04-02},
	journal = {A\&A},
	author = {Boldt-Christmas, Linn and Lesjak, Fabio and Wehrhahn, Ansgar and Piskunov, Nikolai and Rains, Adam D. and Nortmann, Lisa and Kochukhov, Oleg},
	month = mar,
	year = {2024},
	eid= {A244},
}

@ARTICLE{foley_exoplanet_2024,
       author = {{Foley}, Bradford J.},
        title = "{Exoplanet Geology: What Can We Learn from Current and Future Observations?}",
      journal = {Reviews in Mineralogy and Geochemistry},
         year = 2024,
        month = jul,
       volume = {90},
       number = {1},
        pages = {559-594},
          doi = {10.2138/rmg.2024.90.15},
archivePrefix = {arXiv},
       eprint = {2404.15433},
 primaryClass = {astro-ph.EP},
       adsurl = {https://ui.adsabs.harvard.edu/abs/2024RvMG...90..559F}
}

@article{polyansky_exomol_2018,
	title = {{ExoMol} molecular line lists {XXX}: a complete high-accuracy line list for water},
	volume = {480},
	issn = {0035-8711},
	shorttitle = {{ExoMol} molecular line lists {XXX}},
	url = {https://doi.org/10.1093/mnras/sty1877},
	doi = {10.1093/mnras/sty1877},
	number = {2},
	urldate = {2024-05-11},
	journal = {MNRAS},
	author = {Polyansky, Oleg L and Kyuberis, Aleksandra A and Zobov, Nikolai F and Tennyson, Jonathan and Yurchenko, Sergei N and Lodi, Lorenzo},
	month = oct,
	year = {2018},
	eid= {2597--2608},
}

@article{rothman_hitran2012_2013,
	series = {{HITRAN2012} special issue},
	title = {The {HITRAN2012} molecular spectroscopic database},
	volume = {130},
	issn = {0022-4073},
	url = {https://www.sciencedirect.com/science/article/pii/S0022407313002859},
	doi = {10.1016/j.jqsrt.2013.07.002},
	urldate = {2024-05-11},
	journal = {JQSRT},
	author = {Rothman, L. S. and Gordon, I. E. and Babikov, Y. and Barbe, A. and Chris Benner, D. and Bernath, P. F. and Birk, M. and Bizzocchi, L. and Boudon, V. and Brown, L. R. and Campargue, A. and Chance, K. and Cohen, E. A. and Coudert, L. H. and Devi, V. M. and Drouin, B. J. and Fayt, A. and Flaud, J. -M. and Gamache, R. R. and Harrison, J. J. and Hartmann, J. -M. and Hill, C. and Hodges, J. T. and Jacquemart, D. and Jolly, A. and Lamouroux, J. and Le Roy, R. J. and Li, G. and Long, D. A. and Lyulin, O. M. and Mackie, C. J. and Massie, S. T. and Mikhailenko, S. and Müller, H. S. P. and Naumenko, O. V. and Nikitin, A. V. and Orphal, J. and Perevalov, V. and Perrin, A. and Polovtseva, E. R. and Richard, C. and Smith, M. A. H. and Starikova, E. and Sung, K. and Tashkun, S. and Tennyson, J. and Toon, G. C. and Tyuterev, Vl. G. and Wagner, G.},
	month = nov,
	year = {2013},
	eid= {4--50},
}

@article{hargreaves_accurate_2020,
	title = {An {Accurate}, {Extensive}, and {Practical} {Line} {List} of {Methane} for the {HITEMP} {Database}},
	volume = {247},
	issn = {0067-0049},
	url = {https://ui.adsabs.harvard.edu/abs/2020ApJS..247...55H},
	doi = {10.3847/1538-4365/ab7a1a},
	urldate = {2024-05-11},
	journal = {ApJS},
	author = {Hargreaves, Robert J. and Gordon, Iouli E. and Rey, Michael and Nikitin, Andrei V. and Tyuterev, Vladimir G. and Kochanov, Roman V. and Rothman, Laurence S.},
	month = apr,
	year = {2020},
	eid= {55},
}

@article{yurchenko_variationally_2011,
	title = {A variationally computed line list for hot {NH3}},
	volume = {413},
	issn = {0035-8711},
	url = {https://ui.adsabs.harvard.edu/abs/2011MNRAS.413.1828Y},
	doi = {10.1111/j.1365-2966.2011.18261.x},
	urldate = {2024-05-11},
	journal = {MNRAS},
	author = {Yurchenko, S. N. and Barber, R. J. and Tennyson, J.},
	month = may,
	year = {2011},
	eid= {1828--1834},
}

@article{sing_warm_2024,
	title = {A warm {Neptune}’s methane reveals core mass and vigorous atmospheric mixing},
	volume = {630},
	copyright = {2024 The Author(s)},
	issn = {1476-4687},
	url = {https://www.nature.com/articles/s41586-024-07395-z},
	doi = {10.1038/s41586-024-07395-z},
	language = {en},
	number = {8018},
	urldate = {2024-07-02},
	journal = {Nature},
	author = {Sing, David K. and Rustamkulov, Zafar and Thorngren, Daniel P. and Barstow, Joanna K. and Tremblin, Pascal and Alves de Oliveira, Catarina and Beck, Tracy L. and Birkmann, Stephan M. and Challener, Ryan C. and Crouzet, Nicolas and Espinoza, Néstor and Ferruit, Pierre and Giardino, Giovanna and Gressier, Amélie and Lee, Elspeth K. H. and Lewis, Nikole K. and Maiolino, Roberto and Manjavacas, Elena and Rauscher, Bernard J. and Sirianni, Marco and Valenti, Jeff A.},
	month = jun,
	year = {2024},
	eid= {831--835},
}

@article{smith_combined_2024,
	title = {A {Combined} {Ground}-based and {JWST} {Atmospheric} {Retrieval} {Analysis}: {Both} {IGRINS} and {NIRSpec} {Agree} that the {Atmosphere} of {WASP}-{77A} b {Is} {Metal}-poor},
	volume = {167},
	issn = {0004-6256},
	shorttitle = {A {Combined} {Ground}-based and {JWST} {Atmospheric} {Retrieval} {Analysis}},
	url = {https://ui.adsabs.harvard.edu/abs/2024AJ....167..110S},
	doi = {10.3847/1538-3881/ad17bf},
	urldate = {2024-07-16},
	journal = {AJ},
	author = {Smith, Peter C. B. and Line, Michael R. and Bean, Jacob L. and Brogi, Matteo and August, Prune and Welbanks, Luis and Desert, Jean-Michel and Lunine, Jonathan and Sanchez, Jorge and Mansfield, Megan and Pino, Lorenzo and Rauscher, Emily and Kempton, Eliza and Zalesky, Joseph and Fowler, Martin},
	month = mar,
	year = {2024},
	eid= {110},
}

@article{czesla_elusive_2024,
	title = {The elusive atmosphere of {WASP}-12 b. {High}-resolution transmission spectroscopy with {CARMENES}},
	volume = {683},
	issn = {0004-6361},
	url = {https://ui.adsabs.harvard.edu/abs/2024A&A...683A..67C},
	doi = {10.1051/0004-6361/202348107},
	urldate = {2024-08-12},
	journal = {A\&A},
	author = {Czesla, S. and Lampón, M. and Cont, D. and Lesjak, F. and Orell-Miquel, J. and Sanz-Forcada, J. and Nagel, E. and Nortmann, L. and Molaverdikhani, K. and López-Puertas, M. and Yan, F. and Quirrenbach, A. and Caballero, J. A. and Pallé, E. and Aceituno, J. and Amado, P. J. and Henning, Th. and Khalafinejad, S. and Montes, D. and Reiners, A. and Ribas, I. and Schweitzer, A.},
	month = mar,
	year = {2024},
	eid= {A67},
}

@article{cont_exploring_2024,
	title = {Exploring the ultra-hot {Jupiter} {WASP}-178b. {Constraints} on atmospheric chemistry and dynamics from a joint retrieval of {VLT}/{CRIRES}+ and space photometric data},
	volume = {688},
	issn = {0004-6361},
	url = {https://ui.adsabs.harvard.edu/abs/2024A&A...688A.206C},
	doi = {10.1051/0004-6361/202450064},
	urldate = {2024-09-30},
	journal = {A\&A},
	author = {Cont, D. and Nortmann, L. and Yan, F. and Lesjak, F. and Czesla, S. and Lavail, A. and Reiners, A. and Piskunov, N. and Hatzes, A. and Boldt-Christmas, L. and Kochukhov, O. and Marquart, T. and Nagel, E. and Rains, A. D. and Rengel, M. and Seemann, U. and Shulyak, D.},
	month = aug,
	year = {2024},
	eid= {A206},
}

@article{murphy_evidence_2024,
	title = {Evidence for morning-to-evening limb asymmetry on the cool low-density exoplanet {WASP}-107 b},
	copyright = {2024 The Author(s), under exclusive licence to Springer Nature Limited},
	issn = {2397-3366},
	url = {https://www.nature.com/articles/s41550-024-02367-9},
	doi = {10.1038/s41550-024-02367-9},
	language = {en},
	urldate = {2024-10-24},
	journal = {Nat. Astron.},
	author = {Murphy, Matthew M. and Beatty, Thomas G. and Schlawin, Everett and Bell, Taylor J. and Line, Michael R. and Greene, Thomas P. and Parmentier, Vivien and Rauscher, Emily and Welbanks, Luis and Fortney, Jonathan J. and Rieke, Marcia},
	month = sep,
	year = {2024},
	eid= {1--13},
}

@article{maire_workshop_2023,
	title = {Workshop {Summary}: {Exoplanet} {Orbits} and {Dynamics}},
	volume = {135},
	issn = {0004-6280},
	shorttitle = {Workshop {Summary}},
	url = {https://ui.adsabs.harvard.edu/abs/2023PASP..135j6001M},
	doi = {10.1088/1538-3873/acff88},
	urldate = {2024-11-12},
	journal = {PASP},
	author = {Maire, Anne-Lise and Delrez, Laetitia and Pozuelos, Francisco J. and Becker, Juliette and Espinoza, Nestor and Lillo-Box, Jorge and Revol, Alexandre and Absil, Olivier and Agol, Eric and Almenara, José M. and Anglada-Escudé, Guillem and Beust, Hervé and Blunt, Sarah and Bolmont, Emeline and Bonavita, Mariangela and Brandner, Wolfgang and Mirek Brandt, G. and Brandt, Timothy D. and Brown, Garett and Cantero Mitjans, Carles and Charalambous, Carolina and Chauvin, Gaël and Correia, Alexandre C. M. and Cranmer, Miles and Defrère, Denis and Deleuil, Magali and Demory, Brice-Olivier and De Rosa, Robert J. and Desidera, Silvano and Dévora-Pajares, Martín and Díaz, Rodrigo F. and Do Ó, Clarissa and Ducrot, Elsa and Dupuy, Trent J. and Ferrer-Chávez, Rodrigo and Fontanive, Clémence and Gillon, Michaël and Giuppone, Cristian and Gkouvelis, Leonardos and de Oliveira Gomes, Gabriel and Gomes, Sérgio R. A. and Günther, Maximilian N. and Hadden, Sam and Han, Yinuo and Hernandez, David M. and Jehin, Emmanuel and Kane, Stephen R. and Kervella, Pierre and Kiefer, Flavien and Konopacky, Quinn M. and Langlois, Maud and Lanssens, Benjamin and Lazzoni, Cecilia and Lendl, Monika and Li, Yiting and Libert, Anne-Sophie and Lovos, Flavia and Miculán, Romina G. and Murray, Zachary and Pallé, Enric and Rein, Hanno and Rodet, Laetitia and Roisin, Arnaud and Sahlmann, Johannes and Siverd, Robert and Stalport, Manu and Carlos Suárez, Juan and Tamayo, Daniel and Teyssandier, Jean and Thuillier, Antoine and Timmermans, Mathilde and Triaud, Amaury H. M. J. and Trifonov, Trifon and Valente, Ema F. S. and Van Grootel, Valérie and Vasist, Malavika and Wang, Jason J. and Wyatt, Mark C. and Xuan, Jerry and Young, Steven and Zimmerman, Neil T.},
	month = oct,
	year = {2023},
	eid= {106001},
}

@article{welbanks_high_2024,
	title = {A high internal heat flux and large core in a warm {Neptune} exoplanet},
	volume = {630},
	copyright = {2024 The Author(s), under exclusive licence to Springer Nature Limited},
	issn = {1476-4687},
	url = {https://www.nature.com/articles/s41586-024-07514-w},
	doi = {10.1038/s41586-024-07514-w},
	language = {en},
	number = {8018},
	urldate = {2024-11-25},
	journal = {Nature},
	author = {Welbanks, Luis and Bell, Taylor J. and Beatty, Thomas G. and Line, Michael R. and Ohno, Kazumasa and Fortney, Jonathan J. and Schlawin, Everett and Greene, Thomas P. and Rauscher, Emily and McGill, Peter and Murphy, Matthew and Parmentier, Vivien and Tang, Yao and Edelman, Isaac and Mukherjee, Sagnick and Wiser, Lindsey S. and Lagage, Pierre-Olivier and Dyrek, Achrène and Arnold, Kenneth E.},
	month = jun,
	year = {2024},
	eid= {836--840},
}

@article{pinhas_signatures_2017,
	title = {On signatures of clouds in exoplanetary transit spectra},
	volume = {471},
	issn = {0035-8711},
	url = {https://doi.org/10.1093/mnras/stx1849},
	doi = {10.1093/mnras/stx1849},
	number = {4},
	urldate = {2024-12-28},
	journal = {MNRAS},
	author = {Pinhas, Arazi and Madhusudhan, Nikku},
	month = nov,
	year = {2017},
	eid= {4355--4373},
}

@article{nortmann_crires_2025,
	title = {{CRIRES}+ transmission spectroscopy of {WASP}-127 b - {Detection} of the resolved signatures of a supersonic equatorial jet and cool poles in a hot planet},
	volume = {693},
	copyright = {© The Authors 2025},
	issn = {0004-6361, 1432-0746},
	url = {https://www.aanda.org/articles/aa/abs/2025/01/aa50438-24/aa50438-24.html},
	doi = {10.1051/0004-6361/202450438},
	language = {en},
	urldate = {2025-01-24},
	journal = {A\&A},
	author = {Nortmann, L. and Lesjak, F. and Yan, F. and Cont, D. and Czesla, S. and Lavail, A. and Rains, A. D. and Nagel, E. and Boldt-Christmas, L. and Hatzes, A. and Reiners, A. and Piskunov, N. and Kochukhov, O. and Heiter, U. and Shulyak, D. and Rengel, M. and Seemann, U.},
	month = jan,
	year = {2025},
	eid= {A213},
}

@article{lesjak_retrieving_2025,
	title = {Retrieving wind properties from the ultra-hot dayside of {WASP}-189 b with {CRIRES}+},
	volume = {693},
	copyright = {© The Authors 2025},
	issn = {0004-6361, 1432-0746},
	url = {https://www.aanda.org/articles/aa/abs/2025/01/aa51391-24/aa51391-24.html},
	doi = {10.1051/0004-6361/202451391},
	language = {en},
	urldate = {2025-01-24},
	journal = {A\&A},
	author = {Lesjak, F. and Nortmann, L. and Cont, D. and Yan, F. and Reiners, A. and Piskunov, N. and Hatzes, A. and Boldt-Christmas, L. and Czesla, S. and Lavail, A. and Nagel, E. and Rains, A. D. and Rengel, M. and Seemann, U. and Shulyak, D.},
	month = jan,
	year = {2025},
	eid= {A72},
}

@article{bourrier_dream_2023,
	title = {{DREAM}: {I}. {Orbital} architecture orrery},
	volume = {669},
	issn = {0004-6361},
	shorttitle = {{DREAM}},
	url = {https://ui.adsabs.harvard.edu/abs/2023A&A...669A..63B},
	doi = {10.1051/0004-6361/202245004},
	urldate = {2025-01-31},
	journal = {A\&A},
	author = {Bourrier, V. and Attia, M. and Mallonn, M. and Marret, A. and Lendl, M. and Konig, P. -C. and Krenn, A. and Cretignier, M. and Allart, R. and Henry, G. and Bryant, E. and Leleu, A. and Nielsen, L. and Hebrard, G. and Hara, N. and Ehrenreich, D. and Seidel, J. and dos Santos, L. and Lovis, C. and Bayliss, D. and Cegla, H. M. and Dumusque, X. and Boisse, I. and Boucher, A. and Bouchy, F. and Pepe, F. and Lavie, B. and Rey Cerda, J. and Ségransan, D. and Udry, S. and Vrignaud, T.},
	month = jan,
	year = {2023},
	eid= {A63},
}

@article{hejazi_elemental_2023,
	title = {Elemental {Abundances} of the {Super}-{Neptune} {WASP}-107b’s {Host} {Star} {Using} {High}-resolution, {Near}-infrared {Spectroscopy}},
	volume = {949},
	issn = {0004-637X},
	url = {https://dx.doi.org/10.3847/1538-4357/accb97},
	doi = {10.3847/1538-4357/accb97},
	language = {en},
	number = {2},
	urldate = {2025-02-05},
	journal = {ApJ},
	author = {Hejazi, Neda and Crossfield, Ian J. M. and Nordlander, Thomas and Mansfield, Megan and Souto, Diogo and Marfil, Emilio and Coria, David R. and Brande, Jonathan and Polanski, Alex S. and Hand, Joseph E. and Wienke, Kate F.},
	month = jun,
	year = {2023},
	eid= {79},
}

@article{cheverall_feasibility_2024,
	title = {Feasibility of {High}-resolution {Transmission} {Spectroscopy} for {Low}-velocity {Exoplanets}},
	volume = {167},
	issn = {1538-3881},
	url = {https://dx.doi.org/10.3847/1538-3881/ad380c},
	doi = {10.3847/1538-3881/ad380c},
	language = {en},
	number = {6},
	urldate = {2025-02-06},
	journal = {AJ},
	author = {Cheverall, Connor J. and Madhusudhan, Nikku},
	month = may,
	year = {2024},

	eid= {272},
}

@article{prinoth_high-resolution_2024,
	title = {High-resolution {Transmission} {Spectroscopy} of {Warm} {Jupiters}: {An} {ESPRESSO} {Sample} with {Predictions} for {ANDES}},
	volume = {168},
	issn = {0004-6256},
	shorttitle = {High-resolution {Transmission} {Spectroscopy} of {Warm} {Jupiters}},
	url = {https://ui.adsabs.harvard.edu/abs/2024AJ....168..133P},
	doi = {10.3847/1538-3881/ad5a7f},
	urldate = {2025-02-06},
	journal = {AJ},
	author = {Prinoth, Bibiana and Sedaghati, Elyar and Seidel, Julia V. and Hoeijmakers, H. Jens and Brahm, Rafael and Thorsbro, Brian and Jordán, Andrés},
	month = sep,
	year = {2024},
	eid= {133},
}

@article{pino_diagnosing_2018,
	title = {Diagnosing aerosols in extrasolar giant planets with cross-correlation function of water bands},
	volume = {619},
	issn = {0004-6361},
	url = {https://ui.adsabs.harvard.edu/abs/2018A&A...619A...3P},
	doi = {10.1051/0004-6361/201832986},
	urldate = {2025-02-10},
	journal = {A\&A},
	author = {Pino, Lorenzo and Ehrenreich, David and Allart, Romain and Lovis, Christophe and Brogi, Matteo and Malik, Matej and Nascimbeni, Valerio and Pepe, Francesco and Piotto, Giampaolo},
	month = oct,
	year = {2018},
	eid= {A3},
}

@article{hood_prospects_2020,
	title = {Prospects for {Characterizing} the {Haziest} {Sub}-{Neptune} {Exoplanets} with {High}-resolution {Spectroscopy}},
	volume = {160},
	issn = {0004-6256},
	url = {https://ui.adsabs.harvard.edu/abs/2020AJ....160..198H},
	doi = {10.3847/1538-3881/abb46b},
	urldate = {2025-02-10},
	journal = {AJ},
	author = {Hood, Callie E. and Fortney, Jonathan J. and Line, Michael R. and Martin, Emily C. and Morley, Caroline V. and Birkby, Jayne L. and Rustamkulov, Zafar and Lupu, Roxana E. and Freedman, Richard S.},
	month = nov,
	year = {2020},
	eid= {198},
}

@article{gandhi_seeing_2020,
	title = {Seeing above the clouds with high-resolution spectroscopy},
	volume = {498},
	issn = {0035-8711},
	url = {https://ui.adsabs.harvard.edu/abs/2020MNRAS.498..194G},
	doi = {10.1093/mnras/staa2424},
	urldate = {2025-02-10},
	journal = {MNRAS},
	author = {Gandhi, Siddharth and Brogi, Matteo and Webb, Rebecca K.},
	month = oct,
	year = {2020},
	eid= {194--204},
}

@article{dash_constraints_2024,
	title = {Constraints on atmospheric water abundance and cloud deck pressure in the warm {Neptune} {GJ} 3470 b via {CARMENES} transmission spectroscopy},
	volume = {530},
	issn = {0035-8711},
	url = {https://ui.adsabs.harvard.edu/abs/2024MNRAS.530.3100D},
	doi = {10.1093/mnras/stae997},
	urldate = {2025-02-10},
	journal = {MNRAS},
	author = {Dash, Spandan and Brogi, Matteo and Gandhi, Siddharth and Lafarga, Marina and Meech, Annabella and Bello-Arufe, Aaron and Wheatley, Peter J.},
	month = may,
	year = {2024},
	eid= {3100--3116},
}

@article{molliere_detecting_2019,
	title = {Detecting isotopologues in exoplanet atmospheres using ground-based high-dispersion spectroscopy},
	volume = {622},
	copyright = {© ESO 2019},
	issn = {0004-6361, 1432-0746},
	url = {https://www.aanda.org/articles/aa/abs/2019/02/aa34169-18/aa34169-18.html},
	doi = {10.1051/0004-6361/201834169},
	language = {en},
	urldate = {2025-02-10},
	journal = {A\&A},
	author = {Mollière, P. and Snellen, I. a. G.},
	month = feb,
	year = {2019},
	eid= {A139},
}

@article{basilicata_gaps_2024,
	title = {The {GAPS} {Programme} at {TNG} - {LV}. {Multiple} molecular species in the atmosphere of {HAT}-{P}-11 b and review of the {HAT}-{P}-11 planetary system},
	volume = {686},
	copyright = {© The Authors 2024},
	issn = {0004-6361, 1432-0746},
	url = {https://www.aanda.org/articles/aa/abs/2024/06/aa47659-23/aa47659-23.html},
	doi = {10.1051/0004-6361/202347659},
	language = {en},
	urldate = {2025-02-11},
	journal = {A\&A},
	author = {Basilicata, M. and Giacobbe, P. and Bonomo, A. S. and Scandariato, G. and Brogi, M. and Singh, V. and Paola, A. Di and Mancini, L. and Sozzetti, A. and Lanza, A. F. and Cubillos, P. E. and Damasso, M. and Desidera, S. and Biazzo, K. and Bignamini, A. and Borsa, F. and Cabona, L. and Carleo, I. and Ghedina, A. and Guilluy, G. and Maggio, A. and Mainella, G. and Micela, G. and Molinari, E. and Molinaro, M. and Nardiello, D. and Pedani, M. and Pino, L. and Poretti, E. and Southworth, J. and Stangret, M. and Turrini, D.},
	month = jun,
	year = {2024},
	eid= {A127},
}

@article{parker_limits_2025,
	title = {Limits on the atmospheric metallicity and aerosols of the sub-{Neptune} {GJ} 3090 b from high-resolution {CRIRES}+ spectroscopy},
	volume = {538},
	issn = {0035-8711},
	url = {https://doi.org/10.1093/mnras/staf469},
	doi = {10.1093/mnras/staf469},
	number = {4},
	urldate = {2025-04-10},
	journal = {MNRAS},
	author = {Parker, Luke T and Mendonça, João M and Diamond-Lowe, Hannah and Birkby, Jayne L and Meech, Annabella and Vaughan, Sophia R and Brogi, Matteo and Fisher, Chloe and Buchhave, Lars A and Bello-Arufe, Aaron and Kreidberg, Laura and Dittmann, Jason},
	month = apr,
	year = {2025},
	eid= {3263--3283},
}

@article{mukherjee_jwst_2025,
	title = {A {JWST} {Panchromatic} {Thermal} {Emission} {Spectrum} of the {Warm} {Neptune} {Archetype} {GJ} 436b},
	volume = {982},
	issn = {2041-8205},
	url = {https://dx.doi.org/10.3847/2041-8213/adba46},
	doi = {10.3847/2041-8213/adba46},
	language = {en},
	number = {2},
	urldate = {2025-05-12},
	journal = {ApJL},
	author = {Mukherjee, Sagnick and Schlawin, Everett and Bell, Taylor J. and Fortney, Jonathan J. and Beatty, Thomas G. and Greene, Thomas P. and Ohno, Kazumasa and Murphy, Matthew M. and Parmentier, Vivien and Line, Michael R. and Welbanks, Luis and Wiser, Lindsey S. and Rieke, Marcia J.},
	month = mar,
	year = {2025},

	eid= {L39},
}

@article{dubey_quantified_2025,
	title = {Quantified {Estimation} of {Molecular} {Detections} across {Different} {Classes} of {Neptunian} {Atmospheres} {Using} {Cross}-correlation {Spectroscopy}: {Prospects} for {Future} {Extremely} {Large} {Telescopes} with {High}-resolution {Spectrographs}},
	volume = {278},
	issn = {0067-0049},
	shorttitle = {Quantified {Estimation} of {Molecular} {Detections} across {Different} {Classes} of {Neptunian} {Atmospheres} {Using} {Cross}-correlation {Spectroscopy}},
	url = {https://ui.adsabs.harvard.edu/abs/2025ApJS..278...19D},
	doi = {10.3847/1538-4365/adbf05},
	urldate = {2025-05-12},
	journal = {ApJS},
	author = {Dubey, Dwaipayan and Majumdar, Liton and Beichman, Charles and Blake, Geoffrey A. and Vasisht, Gautam and Henning, Thomas},
	month = may,
	year = {2025},
	eid= {19},
}

@article{shields_habitability_2016,
	series = {The habitability of planets orbiting {M}-dwarf stars},
	title = {The habitability of planets orbiting {M}-dwarf stars},
	volume = {663},
	issn = {0370-1573},
	url = {https://www.sciencedirect.com/science/article/pii/S0370157316303179},
	doi = {10.1016/j.physrep.2016.10.003},
	urldate = {2025-06-13},
	journal = {Physics Reports},
	author = {Shields, Aomawa L. and Ballard, Sarah and Johnson, John Asher},
	month = dec,
	year = {2016},
	eid= {1--38},
}

@article{wardenier_modelling_2023,
	title = {Modelling the effect of {3D} temperature and chemistry on the cross-correlation signal of transiting ultra-hot {Jupiters}: a study of five chemical species on {WASP}-76b},
	volume = {525},
	issn = {0035-8711},
	shorttitle = {Modelling the effect of {3D} temperature and chemistry on the cross-correlation signal of transiting ultra-hot {Jupiters}},
	url = {https://doi.org/10.1093/mnras/stad2586},
	doi = {10.1093/mnras/stad2586},
	number = {4},
	urldate = {2025-07-01},
	journal = {MNRAS},
	author = {Wardenier, Joost P and Parmentier, Vivien and Line, Michael R and Lee, Elspeth K H},
	month = nov,
	year = {2023},
	eid= {4942--4961},
}

@article{savel_peering_2025,
	title = {Peering into the {Black} {Box}: {Forward} {Modeling} of the {Uncertainty} {Budget} of {High}-resolution {Spectroscopy} of {Exoplanet} {Atmospheres}},
	volume = {169},
	issn = {0004-6256},
	shorttitle = {Peering into the {Black} {Box}},
	url = {https://ui.adsabs.harvard.edu/abs/2025AJ....169..135S},
	doi = {10.3847/1538-3881/ada27e},
	urldate = {2025-07-01},
	journal = {AJ},
	author = {Savel, Arjun B. and Bedell, Megan and Kempton, Eliza M. -R. and Smith, Peter C. B. and Bean, Jacob L. and Zhao, Lily L. and Wong, Kaze W. K. and Sanchez, Jorge A. and Line, Michael R.},
	month = mar,
	year = {2025},
	eid= {135},
}

@article{finnerty_keck_2023,
	title = {Keck {Planet} {Imager} and {Characterizer} {Emission} {Spectroscopy} of {WASP}-33b},
	volume = {166},
	issn = {0004-6256},
	url = {https://ui.adsabs.harvard.edu/search/fq=%7B!type%3Daqp%20v%3D%24fq_database%7D&fq_database=(database%3Aastronomy%20OR%20database%3Aphysics)&q=Keck%2FKPIC%20WASP-33b&sort=date%20desc%2C%20bibcode%20desc&p_=0},
	doi = {10.3847/1538-3881/acda91},
	language = {en},
	number = {1},
	urldate = {2025-07-01},
	journal = {AJ},
	author = {Finnerty, Luke and Schofield, Tobias and Sappey, Ben and Xuan, Jerry W. and Ruffio, Jean-Baptiste and Wang, Jason J. and Delorme, Jacques-Robert and Blake, Geoffrey A. and Buzard, Cam and Fitzgerald, Michael P. and Baker, Ashley and Bartos, Randall and Bond, Charlotte Z. and Calvin, Benjamin and Cetre, Sylvain and Doppmann, Greg and Echeverri, Daniel and Jovanovic, Nemanja and Liberman, Joshua and López, Ronald A. and Martin, Emily C. and Mawet, Dimitri and Morris, Evan and Pezzato, Jacklyn and Phillips, Caprice L. and Ragland, Sam and Skemer, Andrew and Venenciano, Taylor and Wallace, J. Kent and Wallack, Nicole L. and Wang, Ji and Wizinowich, Peter},
	month = jul,
	year = {2023},
	eid= {31},
}

@article{gandhi_spatially_2022,
	title = {Spatially resolving the terminator: variation of {Fe}, temperature, and winds in {WASP}-76 b across planetary limbs and orbital phase},
	volume = {515},
	issn = {0035-8711},
	shorttitle = {Spatially resolving the terminator},
	url = {https://ui.adsabs.harvard.edu/abs/2022MNRAS.515..749G},
	doi = {10.1093/mnras/stac1744},
	urldate = {2025-07-05},
	journal = {MNRAS},
	author = {Gandhi, Siddharth and Kesseli, Aurora and Snellen, Ignas and Brogi, Matteo and Wardenier, Joost P. and Parmentier, Vivien and Welbanks, Luis and Savel, Arjun B.},
	month = sep,
	year = {2022},
	eid= {749--766},
}

@article{teske_jwst_2025,
	title = {{JWST} {COMPASS}: {NIRSpec}/{G395H} {Transmission} {Observations} of {TOI}-776 c, a 2 {R}⊕ {M} {Dwarf} {Planet}},
	volume = {169},
	issn = {1538-3881},
	shorttitle = {{JWST} {COMPASS}},
	url = {https://dx.doi.org/10.3847/1538-3881/adb975},
	doi = {10.3847/1538-3881/adb975},
	language = {en},
	number = {5},
	urldate = {2025-07-05},
	journal = {AJ},
	author = {Teske, Johanna and Batalha, Natasha E. and Wallack, Nicole L. and Kirk, James and Wogan, Nicholas F. and Gordon, Tyler A. and Alam, Munazza K. and Aguichine, Artyom and Wolfgang, Angie and Wakeford, Hannah R. and Scarsdale, Nicholas and Adams Redai, Jea and Moran, Sarah E. and López-Morales, Mercedes and Meech, Annabella and Gao, Peter and Batalha, Natalie M. and Alderson, Lili and Gagnebin, Anna},
	month = apr,
	year = {2025},

	eid= {249},
}

@book{nettelmann_exoplanetary_2021,
	series = {2514-3433},
	title = {Exoplanetary {Interiors}},
	isbn = {978-0-7503-1472-5},
	url = {https://doi.org/10.1088/2514-3433/abfa8fch16},
	publisher = {IOP Publishing},
	author = {Nettelmann, Nadine and Valencia, Diana},
	year = {2021},
	doi = {10.1088/2514-3433/abfa8fch16},
}

@article{eso_cpl_development_team_esorex_2015,
	title = {{EsoRex}: {ESO} {Recipe} {Execution} {Tool}},
	shorttitle = {{EsoRex}},
	url = {https://ui.adsabs.harvard.edu/abs/2015ascl.soft04003E},
	urldate = {2025-07-05},
	journal = {Astrophysics Source Code Library},
	author = {{ESO CPL Development Team}},
	month = apr,
	year = {2015},
	eid= {ascl:1504.003},
}

@article{brogi_carbon_2014,
	title = {Carbon monoxide and water vapor in the atmosphere of the non-transiting exoplanet {HD} 179949 b},
	volume = {565},
	issn = {0004-6361},
	url = {https://ui.adsabs.harvard.edu/abs/2014A&A...565A.124B},
	doi = {10.1051/0004-6361/201423537},
	urldate = {2025-07-05},
	journal = {A\&A},
	author = {Brogi, M. and de Kok, R. J. and Birkby, J. L. and Schwarz, H. and Snellen, I. A. G.},
	month = may,
	year = {2014},
	eid= {A124},
}

@article{schwarz_evidence_2015,
	title = {Evidence against a strong thermal inversion in {HD} 209458b from high-dispersion spectroscopy},
	volume = {576},
	issn = {0004-6361},
	url = {https://ui.adsabs.harvard.edu/abs/2015A&A...576A.111S},
	doi = {10.1051/0004-6361/201425170},
	urldate = {2025-07-05},
	journal = {A\&A},
	author = {Schwarz, Henriette and Brogi, Matteo and de Kok, Remco and Birkby, Jayne and Snellen, Ignas},
	month = apr,
	year = {2015},
	eid= {A111},
}

@ARTICLE{snellen_exoplanet_2025,
       author = {{Snellen}, Ignas A.~G.},
        title = "{Exoplanet Atmospheres at High Spectral Resolution}",
      journal = {\araa},
         year = 2025,
        month = aug,
       volume = {63},
       number = {1},
        pages = {83-125},
          doi = {10.1146/annurev-astro-052622-031342},
archivePrefix = {arXiv},
       eprint = {2505.08926},
 primaryClass = {astro-ph.EP},
       adsurl = {https://ui.adsabs.harvard.edu/abs/2025ARA&A..63...83S}
}

@article{gustafsson_grid_2008,
	title = {A grid of {MARCS} model atmospheres for late-type stars. {I}. {Methods} and general properties},
	volume = {486},
	issn = {0004-6361},
	url = {https://ui.adsabs.harvard.edu/abs/2008A&A...486..951G},
	doi = {10.1051/0004-6361:200809724},
	urldate = {2025-07-05},
	journal = {A\&A},
	author = {Gustafsson, B. and Edvardsson, B. and Eriksson, K. and Jørgensen, U. G. and Nordlund, Å. and Plez, B.},
	month = aug,
	year = {2008},
	eid= {951--970},
}

@article{grevesse_solar_2007,
	title = {The {Solar} {Chemical} {Composition}},
	volume = {130},
	issn = {0038-6308},
	url = {https://ui.adsabs.harvard.edu/abs/2007SSRv..130..105G},
	doi = {10.1007/s11214-007-9173-7},
	urldate = {2025-07-05},
	journal = {Space Sci. Rev.},
	author = {Grevesse, N. and Asplund, M. and Sauval, A. J.},
	month = jun,
	year = {2007},
	eid= {105--114},
}

@ARTICLE{krishnamurthy_continuous_2025,
       author = {{Krishnamurthy}, Vigneshwaran and {Carteret}, Yann and {Piaulet-Ghorayeb}, Caroline and {Splinter}, Jared and {Doshi}, Dhvani and {Radica}, Michael and {Coulombe}, Louis-Philippe and {Allart}, Romain and {Bourrier}, Vincent and {Cowan}, Nicolas B. and {Doyon}, Ren{\'e} and {Lafreni{\`e}re}, David and {Albert}, Lo{\"\i}c and {Benneke}, Bj{\"o}rn and {Dang}, Lisa and {Jayawardhana}, Ray and {Johnstone}, Doug and {Kaltenegger}, Lisa and {Langeveld}, Adam B. and {Pelletier}, Stefan and {Rowe}, Jason F. and {Roy}, Pierre-Alexis and {Taylor}, Jake and {Turner}, Jake D.},
        title = "{Continuous helium absorption from both the leading and trailing tails of WASP-107 b}",
      journal = {Nat. Astron.},
         year = 2026,
        month = feb,
       volume = {10},
        pages = {258-270},
          doi = {10.1038/s41550-025-02710-8},
archivePrefix = {arXiv},
       eprint = {2505.20588},
 primaryClass = {astro-ph.EP},
       adsurl = {https://ui.adsabs.harvard.edu/abs/2026NatAs..10..258K}
}

@article{dholakia_general_2025,
	title = {A {General}, {Differentiable} {Transit} {Model} for {Ellipsoidal} {Occulters}: {Derivation}, {Application}, and {Forecast} of {Planetary} {Oblateness} and {Obliquity} {Constraints} with {JWST}},
	volume = {987},
	issn = {0004-637X},
	shorttitle = {A {General}, {Differentiable} {Transit} {Model} for {Ellipsoidal} {Occulters}},
	url = {https://dx.doi.org/10.3847/1538-4357/addb4e},
	doi = {10.3847/1538-4357/addb4e},
	language = {en},
	number = {2},
	urldate = {2025-07-05},
	journal = {AJ},
	author = {Dholakia, Shashank and Dholakia, Shishir and Pope, Benjamin J. S.},
	month = jul,
	year = {2025},

	eid= {150},
}

@article{murphy_panchromatic_2025,
	title = {A {Panchromatic} {Characterization} of the {Evening} and {Morning} {Atmosphere} of {WASP}-107 b: {Composition} and {Cloud} {Variations}, and {Insight} into the {Effect} of {Stellar} {Contamination}},
	volume = {170},
	issn = {1538-3881},
	shorttitle = {A {Panchromatic} {Characterization} of the {Evening} and {Morning} {Atmosphere} of {WASP}-107 b},
	url = {https://dx.doi.org/10.3847/1538-3881/addf38},
	doi = {10.3847/1538-3881/addf38},
	language = {en},
	number = {1},
	urldate = {2025-07-05},
	journal = {AJ},
	author = {Murphy, Matthew M. and Beatty, Thomas G. and Schlawin, Everett and Bell, Taylor J. and Radica, Michael and Kennedy, Thomas D. and Mehta, Nishil and Welbanks, Luis and Line, Michael R. and Parmentier, Vivien and Greene, Thomas P. and Mukherjee, Sagnick and Fortney, Jonathan J. and Ohno, Kazumasa and Wiser, Lindsey and Arnold, Kenneth and Rauscher, Emily and Edelman, Isaac R. and Rieke, Marcia J.},
	month = jun,
	year = {2025},

	eid= {61},
}

@article{cassisi_effective_2019,
	title = {Effective temperature – radius relationship of {M} dwarfs},
	volume = {626},
	copyright = {© ESO 2019},
	issn = {0004-6361, 1432-0746},
	url = {https://www.aanda.org/articles/aa/abs/2019/06/aa35468-19/aa35468-19.html},
	doi = {10.1051/0004-6361/201935468},
	language = {en},
	urldate = {2025-07-05},
	journal = {A\&A},
	author = {Cassisi, S. and Salaris, M.},
	month = jun,
	year = {2019},
	
	eid= {A32},
}

@article{parsons_scatter_2018,
	title = {The scatter of the {M} dwarf mass-radius relationship},
	volume = {481},
	issn = {0035-8711},
	url = {https://ui.adsabs.harvard.edu/abs/2018MNRAS.481.1083P},
	doi = {10.1093/mnras/sty2345},
	urldate = {2025-07-05},
	journal = {MNRAS},
	author = {Parsons, S. G. and Gänsicke, B. T. and Marsh, T. R. and Ashley, R. P. and Breedt, E. and Burleigh, M. R. and Copperwheat, C. M. and Dhillon, V. S. and Green, M. J. and Hermes, J. J. and Irawati, P. and Kerry, P. and Littlefair, S. P. and Rebassa-Mansergas, A. and Sahman, D. I. and Schreiber, M. R. and Zorotovic, M.},
	month = nov,
	year = {2018},
	eid= {1083--1096},
}

@article{maguire_assessing_2024,
	title = {Assessing methods for telluric removal on atmospheric retrievals of high-resolution optical exoplanetary transmission spectra},
	volume = {692},
	issn = {0004-6361},
	url = {https://ui.adsabs.harvard.edu/abs/2024A&A...692A...8M},
	doi = {10.1051/0004-6361/202451784},
	urldate = {2025-07-06},
	journal = {A\&A},
	author = {Maguire, Cathal and Sedaghati, Elyar and Gibson, Neale P. and Smette, Alain and Pino, Lorenzo},
	month = dec,
	year = {2024},
	eid= {A8},
}

@article{meech_applications_2022,
	title = {Applications of a {Gaussian} process framework for modelling of high-resolution exoplanet spectra},
	volume = {512},
	issn = {0035-8711},
	url = {https://ui.adsabs.harvard.edu/abs/2022MNRAS.512.2604M},
	doi = {10.1093/mnras/stac662},
	urldate = {2025-07-06},
	journal = {MNRAS},
	author = {Meech, Annabella and Aigrain, Suzanne and Brogi, Matteo and Birkby, Jayne L.},
	month = may,
	year = {2022},
	eid= {2604--2617},
}

@ARTICLE{hong_velocity_2025,
       author = {{Hong}, Kevin S. and {Finnerty}, Luke and {Fitzgerald}, Michael P.},
        title = "{Velocity Shift and Signal-to-Noise Ratio Limits for High-resolution Spectroscopy of Hot Jupiters Using Keck/KPIC}",
      journal = {\aj},
         year = 2025,
        month = aug,
       volume = {170},
       number = {2},
          eid = {110},
        pages = {110},
          doi = {10.3847/1538-3881/ade3c4},
archivePrefix = {arXiv},
       eprint = {2505.09781},
 primaryClass = {astro-ph.EP},
       adsurl = {https://ui.adsabs.harvard.edu/abs/2025AJ....170..110H}
}

@article{powell_transit_2019,
	title = {Transit {Signatures} of {Inhomogeneous} {Clouds} on {Hot} {Jupiters}: {Insights} from {Microphysical} {Cloud} {Modeling}},
	volume = {887},
	issn = {0004-637X},
	shorttitle = {Transit {Signatures} of {Inhomogeneous} {Clouds} on {Hot} {Jupiters}},
	url = {https://dx.doi.org/10.3847/1538-4357/ab55d9},
	doi = {10.3847/1538-4357/ab55d9},
	language = {en},
	number = {2},
	urldate = {2025-07-06},
	journal = {ApJ},
	author = {Powell, Diana and Louden, Tom and Kreidberg, Laura and Zhang, Xi and Gao, Peter and Parmentier, Vivien},
	month = dec,
	year = {2019},
	eid= {170},
}

@article{kataria_atmospheric_2016,
	title = {The {Atmospheric} {Circulation} of a {Nine}-hot-{Jupiter} {Sample}: {Probing} {Circulation} and {Chemistry} over a {Wide} {Phase} {Space}},
	volume = {821},
	issn = {0004-637X},
	shorttitle = {The {Atmospheric} {Circulation} of a {Nine}-hot-{Jupiter} {Sample}},
	url = {https://ui.adsabs.harvard.edu/abs/2016ApJ...821....9K},
	doi = {10.3847/0004-637X/821/1/9},
	urldate = {2025-07-06},
	journal = {ApJ},
	author = {Kataria, Tiffany and Sing, David K. and Lewis, Nikole K. and Visscher, Channon and Showman, Adam P. and Fortney, Jonathan J. and Marley, Mark S.},
	month = apr,
	year = {2016},
	eid= {9},
}

@article{lavvas_aerosol_2017,
	title = {Aerosol {Properties} of the {Atmospheres} of {Extrasolar} {Giant} {Planets}},
	volume = {847},
	issn = {0004-637X},
	url = {https://ui.adsabs.harvard.edu/abs/2017ApJ...847...32L},
	doi = {10.3847/1538-4357/aa88ce},
	urldate = {2025-07-07},
	journal = {ApJ},
	author = {Lavvas, P. and Koskinen, T.},
	month = sep,
	year = {2017},
	eid= {32},
}

@article{miguel_exploring_2014,
	title = {Exploring {Atmospheres} of {Hot} {Mini}-{Neptunes} and {Extrasolar} {Giant} {Planets} {Orbiting} {Different} {Stars} with {Application} to {HD} 97658b, {WASP}-12b, {CoRoT}-2b, {XO}-1b, and {HD} 189733b},
	volume = {780},
	issn = {0004-637X},
	url = {https://ui.adsabs.harvard.edu/abs/2014ApJ...780..166M},
	doi = {10.1088/0004-637X/780/2/166},
	urldate = {2025-07-07},
	journal = {ApJ},
	author = {Miguel, Y. and Kaltenegger, L.},
	month = jan,
	year = {2014},
	eid= {166},
}

@article{soni_signature_2024,
	title = {Signature of {Vertical} {Mixing} in {Hydrogen}-dominated {Exoplanet} {Atmospheres}},
	volume = {977},
	issn = {0004-637X},
	url = {https://ui.adsabs.harvard.edu/abs/2024ApJ...977...52S},
	doi = {10.3847/1538-4357/ad891f},
	urldate = {2025-07-07},
	journal = {ApJ},
	author = {Soni, Vikas and Acharyya, Kinsuk},
	month = dec,
	year = {2024},
	eid= {52},
}

@article{hoeijmakers_search_2015,
	title = {A search for {TiO} in the optical high-resolution transmission spectrum of {HD} 209458b: {Hindrance} due to inaccuracies in the line database},
	volume = {575},
	copyright = {© ESO, 2015},
	issn = {0004-6361, 1432-0746},
	shorttitle = {A search for {TiO} in the optical high-resolution transmission spectrum of {HD} 209458b},
	url = {https://www.aanda.org/articles/aa/abs/2015/03/aa24794-14/aa24794-14.html},
	doi = {10.1051/0004-6361/201424794},
	language = {en},
	urldate = {2025-07-07},
	journal = {A\&A},
	author = {Hoeijmakers, H. J. and Kok, R. J. de and Snellen, I. a. G. and Brogi, M. and Birkby, J. L. and Schwarz, H.},
	month = mar,
	year = {2015},
	eid= {A20},
}

@article{tannock_146-248_2022,
	title = {A 1.46-2.48 μm spectroscopic atlas of a {T6} dwarf (1060 {K}) atmosphere with {IGRINS}: first detections of {H2S} and {H2}, and verification of {H2O}, {CH4}, and {NH3} line lists},
	volume = {514},
	issn = {0035-8711},
	shorttitle = {A 1.46-2.48 μm spectroscopic atlas of a {T6} dwarf (1060 {K}) atmosphere with {IGRINS}},
	url = {https://ui.adsabs.harvard.edu/abs/2022MNRAS.514.3160T},
	doi = {10.1093/mnras/stac1412},
	urldate = {2025-07-07},
	journal = {MNRAS},
	author = {Tannock, Megan E. and Metchev, Stanimir and Hood, Callie E. and Mace, Gregory N. and Fortney, Jonathan J. and Morley, Caroline V. and Jaffe, Daniel T. and Lupu, Roxana},
	month = aug,
	year = {2022},
	eid= {3160--3178},
}

@article{bowesman_high-resolution_2021,
	title = {A high-resolution line list for {AlO}},
	volume = {508},
	issn = {0035-8711},
	url = {https://ui.adsabs.harvard.edu/abs/2021MNRAS.508.3181B},
	doi = {10.1093/mnras/stab2525},
	urldate = {2025-07-07},
	journal = {MNRAS},
	author = {Bowesman, Charles A. and Shuai, Meiyin and Yurchenko, Sergei N. and Tennyson, Jonathan},
	month = dec,
	year = {2021},
	eid= {3181--3193},
}

@article{hedges_effect_2016,
	title = {Effect of pressure broadening on molecular absorption cross sections in exoplanetary atmospheres},
	volume = {458},
	issn = {0035-8711},
	url = {https://ui.adsabs.harvard.edu/abs/2016MNRAS.458.1427H},
	doi = {10.1093/mnras/stw278},
	urldate = {2025-07-07},
	journal = {MNRAS},
	author = {Hedges, Christina and Madhusudhan, Nikku},
	month = may,
	year = {2016},
	eid= {1427--1449},
}

@article{cont_retrieving_2025,
	title = {Retrieving day- and nightside atmospheric properties of the ultra-hot {Jupiter} {TOI}-2109b: {Detection} of {Fe} and {CO} emission lines and evidence for inefficient heat transport},
	volume = {698},
	issn = {0004-6361},
	shorttitle = {Retrieving day- and nightside atmospheric properties of the ultra-hot {Jupiter} {TOI}-2109b},
	url = {https://ui.adsabs.harvard.edu/abs/2025A&A...698A..31C},
	doi = {10.1051/0004-6361/202554572},
	urldate = {2025-07-07},
	journal = {A\&A},
	author = {Cont, D. and Nortmann, L. and Lesjak, F. and Yan, F. and Shulyak, D. and Lavail, A. and Stangret, M. and Pallé, E. and Amado, P. J. and Caballero, J. A. and Hatzes, A. and Henning, Th. and Piskunov, N. and Quirrenbach, A. and Reiners, A. and Ribas, I. and Agüí Fernández, J. F. and Akın, C. and Boldt-Christmas, L. and Chaturvedi, P. and Czesla, S. and Hahlin, A. and Heng, K. and Kochukhov, O. and Marquart, T. and Molaverdikhani, K. and Montes, D. and Morello, G. and Nagel, E. and Orell-Miquel, J. and Rains, A. D. and Rengel, M. and Schweitzer, A. and Sánchez-López, A. and Seemann, U.},
	month = may,
	year = {2025},
	eid= {A31},
}

@article{ryabchikova_major_2015,
	title = {A major upgrade of the {VALD} database},
	volume = {90},
	issn = {0031-8949},
	url = {https://ui.adsabs.harvard.edu/abs/2015PhyS...90e4005R},
	doi = {10.1088/0031-8949/90/5/054005},
	urldate = {2025-07-07},
	journal = {Physica Scripta},
	author = {Ryabchikova, T. and Piskunov, N. and Kurucz, R. L. and Stempels, H. C. and Heiter, U. and Pakhomov, Yu and Barklem, P. S.},
	month = may,
	year = {2015},
	eid= {054005},
}

@article{rossiter_detection_1924,
	title = {On the detection of an effect of rotation during eclipse in the velocity of the brigher component of beta {Lyrae}, and on the constancy of velocity of this system.},
	volume = {60},
	issn = {0004-637X},
	url = {https://ui.adsabs.harvard.edu/abs/1924ApJ....60...15R},
	doi = {10.1086/142825},
	urldate = {2025-07-07},
	journal = {ApJ},
	author = {Rossiter, R. A.},
	month = jul,
	year = {1924},
	eid= {15--21},
}

@article{mclaughlin_results_1924,
	title = {Some results of a spectrographic study of the {Algol} system.},
	volume = {60},
	issn = {0004-637X},
	url = {https://ui.adsabs.harvard.edu/abs/1924ApJ....60...22M},
	doi = {10.1086/142826},
	urldate = {2025-07-07},
	journal = {ApJ},
	author = {McLaughlin, D. B.},
	month = jul,
	year = {1924},
	eid= {22--31},
}

@article{shen_eccentricity_2008,
	title = {On the {Eccentricity} {Distribution} of {Exoplanets} from {Radial} {Velocity} {Surveys}},
	volume = {685},
	issn = {0004-637X},
	url = {https://ui.adsabs.harvard.edu/abs/2008ApJ...685..553S},
	doi = {10.1086/590548},
	urldate = {2025-07-14},
	journal = {ApJ},
	author = {Shen, Yue and Turner, Edwin L.},
	month = sep,
	year = {2008},
	eid= {553--559},
}

@article{zakamska_observational_2011,
	title = {Observational biases in determining extrasolar planet eccentricities in single-planet systems},
	volume = {410},
	issn = {0035-8711},
	url = {https://ui.adsabs.harvard.edu/abs/2011MNRAS.410.1895Z},
	doi = {10.1111/j.1365-2966.2010.17570.x},
	urldate = {2025-07-14},
	journal = {MNRAS},
	author = {Zakamska, Nadia L. and Pan, Margaret and Ford, Eric B.},
	month = jan,
	year = {2011},
	eid= {1895--1910},
}

@article{lucy_spectroscopic_1971,
	title = {Spectroscopic binaries with circular orbits.},
	volume = {76},
	issn = {0004-6256},
	url = {https://ui.adsabs.harvard.edu/abs/1971AJ.....76..544L},
	doi = {10.1086/111159},
	urldate = {2025-07-14},
	journal = {AJ},
	author = {Lucy, L. B. and Sweeney, M. A.},
	month = aug,
	year = {1971},
	eid= {544--556},
}

@article{yurchenko_data_2025,
	title = {Data challenges and prospects of high-resolution spectroscopy of exoplanets},
	copyright = {2025 Springer Nature Limited},
	issn = {2522-5820},
	url = {https://www.nature.com/articles/s42254-025-00839-z},
	doi = {10.1038/s42254-025-00839-z},
	language = {en},
	urldate = {2025-07-17},
	journal = {Nat. Rev. Phys.},
	author = {Yurchenko, Sergei N. and Tennyson, Jonathan and Brogi, Matteo},
	month = jul,
	year = {2025},
	eid= {1--15},
}

@article{seidel_vertical_2025,
	title = {Vertical structure of an exoplanet’s atmospheric jet stream},
	volume = {639},
	copyright = {2025 The Author(s), under exclusive licence to Springer Nature Limited},
	issn = {1476-4687},
	url = {https://www.nature.com/articles/s41586-025-08664-1},
	doi = {10.1038/s41586-025-08664-1},
	language = {en},
	number = {8056},
	urldate = {2025-07-21},
	journal = {Nature},
	author = {Seidel, Julia V. and Prinoth, Bibiana and Pino, Lorenzo and dos Santos, Leonardo A. and Chakraborty, Hritam and Parmentier, Vivien and Sedaghati, Elyar and Wardenier, Joost P. and Farret Jentink, Casper and Zapatero Osorio, Maria Rosa and Allart, Romain and Ehrenreich, David and Lendl, Monika and Roccetti, Giulia and Damasceno, Yuri and Bourrier, Vincent and Lillo-Box, Jorge and Hoeijmakers, H. Jens and Pallé, Enric and Santos, Nuno and Suárez Mascareño, Alejandro and Sousa, Sergio G. and Tabernero, Hugo M. and Pepe, Francesco A.},
	month = mar,
	year = {2025},
	eid= {902--908},
}

@article{allart_spectrally_2018,
	title = {Spectrally resolved helium absorption from the extended atmosphere of a warm {Neptune}-mass exoplanet},
	volume = {362},
	issn = {0036-8075},
	url = {https://ui.adsabs.harvard.edu/abs/2018Sci...362.1384A},
	doi = {10.1126/science.aat5879},
	urldate = {2025-07-21},
	journal = {Science},
	author = {Allart, R. and Bourrier, V. and Lovis, C. and Ehrenreich, D. and Spake, J. J. and Wyttenbach, A. and Pino, L. and Pepe, F. and Sing, D. K. and Lecavelier des Etangs, A.},
	month = dec,
	year = {2018},
	eid= {1384--1387},
}

@article{anderson_discoveries_2017,
	title = {The discoveries of {WASP}-91b, {WASP}-105b and {WASP}-107b: {Two} warm {Jupiters} and a planet in the transition region between ice giants and gas giants},
	volume = {604},
	copyright = {© ESO, 2017},
	issn = {0004-6361, 1432-0746},
	shorttitle = {The discoveries of {WASP}-91b, {WASP}-105b and {WASP}-107b},
	url = {https://www.aanda.org/articles/aa/abs/2017/08/aa30439-17/aa30439-17.html},
	doi = {10.1051/0004-6361/201730439},
	language = {en},
	urldate = {2025-07-21},
	journal = {A\&A},
	author = {Anderson, D. R. and Cameron, A. Collier and Delrez, L. and Doyle, A. P. and Gillon, M. and Hellier, C. and Jehin, E. and Lendl, M. and Maxted, P. F. L. and Madhusudhan, N. and Pepe, F. and Pollacco, D. and Queloz, D. and Ségransan, D. and Smalley, B. and Smith, A. M. S. and Triaud, A. H. M. J. and Turner, O. D. and Udry, S. and West, R. G.},
	month = aug,
	year = {2017},
	eid= {A110},
}

@article{tsai_photochemically_2023,
	title = {Photochemically produced {SO2} in the atmosphere of {WASP}-39b},
	volume = {617},
	copyright = {2023 The Author(s)},
	issn = {1476-4687},
	url = {https://www.nature.com/articles/s41586-023-05902-2},
	doi = {10.1038/s41586-023-05902-2},
	language = {en},
	number = {7961},
	urldate = {2025-07-21},
	journal = {Nature},
	author = {Tsai, Shang-Min and Lee, Elspeth K. H. and Powell, Diana and Gao, Peter and Zhang, Xi and Moses, Julianne and Hébrard, Eric and Venot, Olivia and Parmentier, Vivien and Jordan, Sean and Hu, Renyu and Alam, Munazza K. and Alderson, Lili and Batalha, Natalie M. and Bean, Jacob L. and Benneke, Björn and Bierson, Carver J. and Brady, Ryan P. and Carone, Ludmila and Carter, Aarynn L. and Chubb, Katy L. and Inglis, Julie and Leconte, Jérémy and Line, Michael and López-Morales, Mercedes and Miguel, Yamila and Molaverdikhani, Karan and Rustamkulov, Zafar and Sing, David K. and Stevenson, Kevin B. and Wakeford, Hannah R. and Yang, Jeehyun and Aggarwal, Keshav and Baeyens, Robin and Barat, Saugata and de Val-Borro, Miguel and Daylan, Tansu and Fortney, Jonathan J. and France, Kevin and Goyal, Jayesh M. and Grant, David and Kirk, James and Kreidberg, Laura and Louca, Amy and Moran, Sarah E. and Mukherjee, Sagnick and Nasedkin, Evert and Ohno, Kazumasa and Rackham, Benjamin V. and Redfield, Seth and Taylor, Jake and Tremblin, Pascal and Visscher, Channon and Wallack, Nicole L. and Welbanks, Luis and Youngblood, Allison and Ahrer, Eva-Maria and Batalha, Natasha E. and Behr, Patrick and Berta-Thompson, Zachory K. and Blecic, Jasmina and Casewell, S. L. and Crossfield, Ian J. M. and Crouzet, Nicolas and Cubillos, Patricio E. and Decin, Leen and Désert, Jean-Michel and Feinstein, Adina D. and Gibson, Neale P. and Harrington, Joseph and Heng, Kevin and Henning, Thomas and Kempton, Eliza M.-R. and Krick, Jessica and Lagage, Pierre-Olivier and Lendl, Monika and Lothringer, Joshua D. and Mansfield, Megan and Mayne, N. J. and Mikal-Evans, Thomas and Palle, Enric and Schlawin, Everett and Shorttle, Oliver and Wheatley, Peter J. and Yurchenko, Sergei N.},
	month = may,
	year = {2023},
	eid= {483--487},
}

@article{changeat_cloud_2025,
	title = {Cloud and haze parameterization in atmospheric retrievals: {Insights} from {Titan}'s {Cassini} data and {JWST} observations of hot {Jupiters}},
	volume = {699},
	copyright = {© The Authors 2025},
	issn = {0004-6361, 1432-0746},
	shorttitle = {Cloud and haze parameterization in atmospheric retrievals},
	url = {https://www.aanda.org/articles/aa/abs/2025/07/aa53186-24/aa53186-24.html},
	doi = {10.1051/0004-6361/202453186},
	language = {en},
	urldate = {2025-07-22},
	journal = {A\&A},
	author = {Changeat, Q. and Bardet, D. and Chubb, K. and Dyrek, A. and Edwards, B. and Ohno, K. and Venot, O.},
	month = jul,
	year = {2025},
	
	eid= {A219},
}

@article{rengel_about_2022,
	title = {About the atomic and molecular databases in the planetary community – {A} contribution in the {Laboratory} {Astrophysics} {Data} {WG} {IAU} 2022 {GA} session},
	volume = {18},
	issn = {1743-9213, 1743-9221},
	url = {https://www.cambridge.org/core/journals/proceedings-of-the-international-astronomical-union/article/about-the-atomic-and-molecular-databases-in-the-planetary-community-a-contribution-in-the-laboratory-astrophysics-data-wg-iau-2022-ga-session/81A1F203521985D865276F6CF8C32BF2},
	doi = {10.1017/S1743921323000169},
	language = {en},
	number = {S371},
	urldate = {2025-07-22},
	journal = {Proceedings of the International Astronomical Union},
	author = {Rengel, M.},
	month = aug,
	year = {2022},
	eid= {87--91},
}

@article{grasser_peering_2024,
	title = {Peering above the clouds of the warm {Neptune} {GJ} 436 b with {CRIRES}+},
	volume = {688},
	copyright = {© The Authors 2024},
	issn = {0004-6361, 1432-0746},
	url = {https://www.aanda.org/articles/aa/abs/2024/08/aa49932-24/aa49932-24.html},
	doi = {10.1051/0004-6361/202449932},
	language = {en},
	urldate = {2025-09-21},
	journal = {A\&A},
	author = {Grasser, Natalie and Snellen, Ignas A. G. and Landman, Rico and Picos, Darío González and Gandhi, Siddharth},
	month = aug,
	year = {2024},
	eid= {A191},
}

@misc{piskunov_tsd_2025,
	url = {http://arxiv.org/abs/2509.12737},
	doi = {10.48550/arXiv.2509.12737},
	urldate = {2025-09-21},
	author = {Piskunov, Nikolai and Rains, Adam D. and Boldt-Christmas, Linn},
	month = sep,
	year = {2025},
	note = {arXiv:2509.12737 [astro-ph]},
}

@article{allart_wasp-127b_2020,
	title = {{WASP}-127b: a misaligned planet with a partly cloudy atmosphere and tenuous sodium signature seen by {ESPRESSO}},
	volume = {644},
	issn = {0004-6361},
	shorttitle = {{WASP}-127b},
	url = {https://ui.adsabs.harvard.edu/abs/2020A&A...644A.155A},
	doi = {10.1051/0004-6361/202039234},
	urldate = {2025-11-11},
	journal = {A\&A},
	author = {Allart, R. and Pino, L. and Lovis, C. and Sousa, S. G. and Casasayas-Barris, N. and Zapatero Osorio, M. R. and Cretignier, M. and Palle, E. and Pepe, F. and Cristiani, S. and Rebolo, R. and Santos, N. C. and Borsa, F. and Bourrier, V. and Demangeon, O. D. S. and Ehrenreich, D. and Lavie, B. and Lendl, M. and Lillo-Box, J. and Micela, G. and Oshagh, M. and Sozzetti, A. and Tabernero, H. and Adibekyan, V. and Allende Prieto, C. and Alibert, Y. and Amate, M. and Benz, W. and Bouchy, F. and Cabral, A. and Dekker, H. and D'Odorico, V. and Di Marcantonio, P. and Dumusque, X. and Figueira, P. and Genova Santos, R. and González Hernández, J. I. and Lo Curto, G. and Manescau, A. and Martins, C. J. A. P. and Mégevand, D. and Mehner, A. and Molaro, P. and Nunes, N. J. and Poretti, E. and Riva, M. and Suárez Mascareño, A. and Udry, S. and Zerbi, F.},
	month = dec,
	year = {2020},
	eid= {A155},
}

@article{lafarga_hot_2023,
	title = {The hot {Neptune} {WASP}-166 b with {ESPRESSO} - {III}. {A} blue-shifted tentative water signal constrains the presence of clouds},
	volume = {521},
	issn = {0035-8711},
	url = {https://ui.adsabs.harvard.edu/abs/2023MNRAS.521.1233L},
	doi = {10.1093/mnras/stad480},
	urldate = {2025-11-11},
	journal = {MNRAS},
	author = {Lafarga, M. and Brogi, M. and Gandhi, S. and Cegla, H. M. and Seidel, J. V. and Doyle, L. and Allart, R. and Buchschacher, N. and Lendl, M. and Lovis, C. and Sosnowska, D.},
	month = may,
	year = {2023},
	eid = {1233--1252},
}

@ARTICLE{palle_andes_2025,
       author = {{Palle}, Enric and {Biazzo}, Katia and {Bolmont}, Emeline and {Molli{\`e}re}, Paul and {Poppenhaeger}, Katja and {Birkby}, Jayne and {Brogi}, Matteo and {Chauvin}, Gael and {Chiavassa}, Andrea and {Hoeijmakers}, Jens and {Lellouch}, Emmanuel and {Lovis}, Christophe and {Maiolino}, Roberto and {Nortmann}, Lisa and {Parviainen}, Hannu and {Pino}, Lorenzo and {Turbet}, Martin and {Weder}, Jesse and {Albrecht}, Simon and {Antoniucci}, Simone and {Barros}, Susana C. and {Beaudoin}, Andre and {Benneke}, Bjorn and {Boisse}, Isabelle and {Bonomo}, Aldo S. and {Borsa}, Francesco and {Brandeker}, Alexis and {Brandner}, Wolfgang and {Buchhave}, Lars A. and {Cheffot}, Anne-Laure and {Deborde}, Robin and {Debras}, Florian and {Doyon}, Rene and {Di Marcantonio}, Paolo and {Giacobbe}, Paolo and {Gonz{\'a}lez Hern{\'a}ndez}, Jonay I. and {Helled}, Ravit and {Kreidberg}, Laura and {Machado}, Pedro and {Maldonado}, Jesus and {Marconi}, Alessandro and {Martins}, B.~L. Canto and {Miceli}, Adriano and {Mordasini}, Christoph and {N'Diaye}, Mamadou and {Niedzielski}, Andrzej and {Nisini}, Brunella and {Origlia}, Livia and {Peroux}, Celine and {Pietrow}, Alexander G.~M. and {Pinna}, Enrico and {Rauscher}, Emily and {Reffert}, Sabine and {Rodr{\'\i}guez-L{\'o}pez}, Cristina and {Rousselot}, Philippe and {Sanna}, Nicoletta and {Santos}, Nuno C. and {Simonnin}, Adrien and {Su{\'a}rez Mascare{\~n}o}, Alejandro and {Zanutta}, Alessio and {Zapatero-Osorio}, Maria Rosa and {Zechmeister}, Mathias},
        title = "{Ground-breaking exoplanet science with the ANDES spectrograph at the ELT}",
      journal = {Experimental Astronomy},
         year = 2025,
        month = jun,
       volume = {59},
       number = {3},
          eid = {29},
          doi = {10.1007/s10686-025-10000-4},
}

\end{document}